\documentclass[prd,nofootinbib,superscriptaddress,floatfix]{revtex4-1}
\pdfoutput=1
\makeatletter
\DeclareRobustCommand*{\bfseries}{%
  \not@math@alphabet\bfseries\mathbf
  \fontseries\bfdefault\selectfont
  \boldmath
}
\makeatother

\usepackage{calc}
\usepackage{rotating}
\usepackage[english]{babel}
\usepackage{graphicx}
\usepackage{booktabs}
\usepackage{subfig}
\usepackage{float}
\usepackage{amsmath}
\usepackage{amssymb}
\usepackage{amsthm}
\usepackage{latexsym}
\usepackage{dcolumn}
\newcommand{\newc}{\newcommand*}

\long\def\begincomment#1\endcomment{%
        \begingroup\sf\baselineskip12pt#1\endgroup}
\newc{\etal}{\textrm{et al.}} 
\newc{\eg}{\textrm{e.g.}} 
\newc{\ie}{\textrm{i.e.}}
\newc{\etc}{\textrm{etc}}
\newc\vs{\textrm{vs.}}
\newc{\cl}{\rm {CL}}

\newc{\ev}{\ensuremath{\,\mathrm{eV}}}
\newc{\kev}{\ensuremath{\,\mathrm{keV}}}
\newc{\mev}{\ensuremath{\,\mathrm{MeV}}}
\newc{\gev}{\ensuremath{\,\mathrm{GeV}}}
\newc{\tev}{\ensuremath{\,\mathrm{TeV}}}

\newc{\MeV}{\mev} 
\newc{\TeV}{\tev}
\newc{\invpb}{\ensuremath{/\text{pb}}}
\newc{\invfb}{\ensuremath{/\text{fb}}}

\newc\nb{\ensuremath{\,\mathrm{nb}}} \newc\pb{\ensuremath{\,\mathrm{pb}}} \newc\fb{\ensuremath{\,\mathrm{fb}}}

\newc\pc{\ensuremath{\,\mathrm{pc}}}
\newc\kpc{\ensuremath{\,\mathrm{kpc}}}
\newc\mpc{\ensuremath{\,\mathrm{Mpc}}}

\newc\ps{\ensuremath{\,\mathrm{ps}}} 

\newc\cmeter{\ensuremath{\,\mathrm{cm}}} 
\newc\meter{\ensuremath{\,\mathrm{m}}} 
\newc\kmeter{\ensuremath{\,\mathrm{km}}}

\newc\second{\ensuremath{\,\mathrm{s}}}
\newc\msecond{\ensuremath{\,\mathrm{ms}}}
\newc\nsecond{\ensuremath{\,\mathrm{ns}}}
\newc\psecond{\ensuremath{\,\mathrm{ps}}}

\newc{\chisq}{\ensuremath{\chi^2 - \chi^2_{\mathrm{min}}}}
\newc{\chisqmin}{\ensuremath{\chi^2_{\mathrm{min}}}}
\newc{\Delchisq}{\ensuremath{\Delta\chi^2}}

\newc{\like}{\ensuremath{\mathcal{L}}}
\newc\lsim{\ensuremath{\mathrel{\rlap{\lower4pt\hbox{\hskip1pt$\sim$}}\raise1pt\hbox{$<$}}}}
\newc\gsim{\ensuremath{\mathrel{\rlap{\lower4pt\hbox{\hskip1pt$\sim$}}\raise1pt\hbox{$>$}}}}

\newc{\VEV}[1]{\ensuremath{\langle #1 \rangle}}

\newc{\dl}{\ensuremath{\stackrel{\leftarrow}{D}}}
\newc{\dr}{\ensuremath{\stackrel{\rightarrow}{D}}}

\newc{\bcenter}{\begin{center}}    \newc{\ecenter}{\end{center}}
\newc{\bfl}{\begin{flushleft}}    \newc{\efl}{\end{flushleft}}
\newc{\bfr}{\begin{flushright}}    \newc{\efr}{\end{flushright}}

\newc{\bi}{\begin{itemize}}
\newc{\ei}{\end{itemize}}
\newc{\bed}{\begin{description}}
\newc{\eed}{\end{description}}
\newc{\ben}{\begin{enumerate}}
\newc{\een}{\end{enumerate}}

\newc{\be}{\begin{equation}}
\newc{\ee}{\end{equation}}
\newc{\bea}{\begin{eqnarray}}
\newc{\eea}{\end{eqnarray}}

\newc{\alphas}{\ensuremath{\alpha_s}}
\newc{\alphatwo}{\ensuremath{\alpha_2}}
\newc{\alphaone}{\ensuremath{\alpha_1}}
\newc{\alphai}[1]{\ensuremath{\alpha_{#1}}}
\newc{\alphaem}{\ensuremath{\alpha_{\mathrm{em}}}}

\newc{\alphaeff}{\ensuremath{\alpha_{\mathrm{eff}}}}

\newc{\sineff}{\ensuremath{\sin \theta_{\mathrm{eff}}}}
\newc{\sinsqeff}{\ensuremath{\sin^2 \theta_{\mathrm{eff}}}}
\newc{\dalphahad}{\ensuremath{\Delta \alpha_{\mathrm{had}}}}

\newc{\yt}{\ensuremath{h_t}} \newc{\yb}{\ensuremath{h_b}} \newc{\ytau}{\ensuremath{h_{\tau}}}

\newc\mz{\ensuremath{m_Z}} 
\newc\mw{\ensuremath{m_W}}
\newc\mZ{\mz}        \newc\mW{\mw}

\newc\mhsm{\ensuremath{ m_{H_{\mathrm{SM}}}}}

\newc{\mtop}{\ensuremath{ m_t}}               \newc{\mtpole}{\ensuremath{ M_t}}
\newc{\mbottom}{\ensuremath{ m_b}} 
\newc{\mtau}{\ensuremath{ m_{\tau}}}
\newc{\mt}{\mtpole}
\newc{\mb}{\mbottom} 

\newc{\llbar}{\ensuremath{\ell\bar{\ell}}}
\newc{\tauptaum}{\ensuremath{ \tau^+\tau^-}}

\newc{\qqbar}{\ensuremath{ q\bar{q}}} \newc{\ppbar}{\ensuremath{ p\bar{p}}}
\newc{\bbbar}{\ensuremath{ b\bar{b}}} \newc{\ttbar}{\ensuremath{ t\bar{t}}}
\newc{\ffbar}{\ensuremath{ f\bar{f}}} \newc{\tautaubar}{\ensuremath{ \tau\bar{\tau}}}

\newc{\mchi}{\ensuremath{m_\neutone}}
\newc{\squark}{\ensuremath{\tilde{q}}}
\newc{\slepton}{\ensuremath{\tilde{l}}}
\newc{\gluino}{\ensuremath{\tilde{g}}} 
\newc{\mgluino}{\ensuremath{{m_{\gluino}}}}
\newc{\sthw}{\ensuremath{ \sin\theta_W}}              \newc{\cthw}{\ensuremath{\cos\theta_W}}
\newc{\tanthw}{\ensuremath{ \tan\theta_W}}              \newc{\cotthw}{\ensuremath{\cot\theta_W}}

\newc{\ssqthw}{\ensuremath{\sin^2 \theta_W}}

\newc{\msbar}{\ensuremath{\overline{MS}}} \newc{\drbar}{\ensuremath{\overline{DR}}}

\newc{\mtmtsmmsbar}{\ensuremath{ m_t(m_t)^{\msbar}_{{\mathrm{SM}}}}}
\newc{\mtmtsmdrbar}{\ensuremath{ m_t(m_t)^{\drbar}_{{\mathrm{SM}}}}}
\newc{\mtmtmssmdrbar}{\ensuremath{ m_t(m_t)^{\drbar}_{{\mathrm{SUSY}}}}}

\newc{\mbmbmsbar}{\ensuremath{ m_b(m_b)^{\msbar} }}

\newc{\mbmbsmmsbar}{\ensuremath{ m_b(m_b)^{\msbar}_{{\mathrm{SM}}}}}
\newc{\mbmzsmmsbar}{\ensuremath{ m_b(\mz)^{\msbar}_{{\mathrm{SM}}}}}
\newc{\mbmzsmdrbar}{\ensuremath{ m_b(\mz)^{\drbar}_{{\mathrm{SM}}}}}
\newc{\mbmzmssmdrbar}{\ensuremath{ m_b(\mz)^{\drbar}_{{\mathrm{SUSY}}}}}

\newc{\mtaumzsmmsbar}{\ensuremath{ m_{\tau}(\mz)^{\msbar}_{{\mathrm{SM}}}}}
\newc{\mtaumzsmdrbar}{\ensuremath{ m_{\tau}(\mz)^{\drbar}_{{\mathrm{SM}}}}}
\newc{\mtaumzmssmdrbar}{\ensuremath{ m_{\tau}(\mz)^{\drbar}_{{\mathrm{SUSY}}}}}

\newc{\alphasmzms}{\ensuremath{\alpha_s(M_Z)^{\overline{MS}}}}
\newc{\alphaimzms}[1]{\ensuremath{\alpha_{#1}(M_Z)^{\overline{MS}}}}

\newc{\alphaemmz}{\ensuremath{\alpha_{\mathrm{em}}(M_Z)^{\overline{MS}}}}

\newc{\mzero}{\ensuremath{{m_0}}}
\newc{\mhalf}{\ensuremath{ m_{1/2}}}
\newc{\tanb}{\ensuremath{\tan\beta}}
\newc{\azero}{\ensuremath{ A_0}}
\newc{\signmu}{\ensuremath{\rm{sgn}\,\mu}}

\newc{\mgut}{\ensuremath{ M_{\rm GUT}}}
\newc{\mplanck}{\ensuremath{ M_{\rm P}}}      \newc{\mpl}{\ensuremath{ M_{\rm Pl}}}
\newc{\msusy}{\ensuremath{ M_{\rm SUSY}}}      \newc{\ms}{\ensuremath{ M_{\rm S}}}

 \newc{\hu}{\ensuremath{ H_u}}       \newc{\hd}{\ensuremath{ H_d}}
 \newc{\mhu}{\ensuremath{ m_{H_u}}}       \newc{\mhd}{\ensuremath{ m_{H_d}}}
 \newc{\mhuew}{\ensuremath{ m^{\ast}_{H_u}}}       \newc{\mhdew}{\ensuremath{ m^{\ast}_{H_d}}}
 \newc{\mhuewsq}{\ensuremath{ m^{\ast\, 2}_{H_u}}}       \newc{\mhdewsq}{\ensuremath{ m^{\ast\, 2}_{H_d}}}
 \newc{\mhl}{\ensuremath{m_\hl}} 
 \newc{\mglu}{\ensuremath{m_{\tilde g}}} 
 \newc{\mul}{\ensuremath{m_{\tilde{u}_L}}} 
 \newc{\mtone}{\ensuremath{m_{\tilde{t}_1}}} 
\newc{\sigsip}{\ensuremath{\sigma^{\rm SI}_{p}}}	\newc{\sigsin}{\ensuremath{\sigma^{\rm SI}_{n}}}
\newc{\sigsdp}{\ensuremath{\sigma^{\rm SD}_{p}}}	\newc{\sigsdn}{\ensuremath{\sigma^{\rm SD}_{n}}}
\newc{\sigsi}{\ensuremath{\sigma^{\rm SI}}}	\newc{\sigsd}{\ensuremath{\sigma^{\rm SD}}}

\newc{\abund}{\ensuremath{ \Omega h^2}}
\newc{\omegadm}{\ensuremath{ \Omega_{{\rm DM}}}}     \newc{\abunddm}{\ensuremath{ \Omega_{{\rm DM}} h^2}} 
\newc{\omegam}{\ensuremath{ \Omega_{{\rm m}}}}       \newc{\abundm}{\ensuremath{ \Omega_{{\rm m}} h^2}}
\newc{\omegab}{\ensuremath{ \Omega_{{\rm b}}}}	\newc{\abundb}{\ensuremath{ \Omega_{{\rm b}} h^2}}
\newc{\omegatot}{\ensuremath{ \Omega_{{\rm TOT}}}}
\newc{\omegacdm}{\ensuremath{ \Omega_{{\rm CDM}}}}   \newc{\abundcdm}{\ensuremath{ \Omega_{{\rm CDM}} h^2}}
\newc{\omegalambda}{\ensuremath{ \Omega_{\Lambda}}} \newc{\abundlambda}{\ensuremath{ \Omega_{\Lambda} h^2}}
\newc{\omegarad}{\ensuremath{ \Omega_{{\rm rad}}}}  \newc{\abundrad}{\ensuremath{ \Omega_{{\rm rad}} h^2}}

\newc{\rhocrit}{\ensuremath{ \rho_{\rm crit}}}
\newc{\rhochi}{\ensuremath{ \rho_{\chi}}}

\newc{\abunchi}{\ensuremath{\Omega_\chi h^2}}
\newc{\abundlsp}{\ensuremath{\Omega_{\rm LSP}h^2}}

\newcommand*{\abundchi}{\ensuremath{\Omega_\chi h^2}}% For multiple citations with one key
\newc{\amu}{\ensuremath{ a_{\mu}}}        \newc{\amususy}{\ensuremath{ a_{\mu}^{\mathrm{SUSY}}}}
\newc{\amuexpt}{\ensuremath{ a_{\mu}^{\mathrm{expt}}}}        \newc{\amusm}{\ensuremath{ a_{\mu}^{\mathrm{SM}}}}
\newc\deltaamu{\ensuremath{\Delta a_{\mu}}} \newc{\deltaamususy}{\ensuremath{\delta a_{\mu}^{\mathrm{SUSY}}}}
\newc\gmtwo{\ensuremath{ (g-2)_{\mu}}} 
\newc{\deltagmtwomususy}{\ensuremath{\delta\left(g-2\right)_{\mu}^{\mathrm{SUSY}}}}
\newc{\deltagmtwomu}{\ensuremath{\delta\left(g-2\right)_{\mu}}}

\newc\BR{\ensuremath{\rm BR}}

\newc\bsgamma{\ensuremath{ b\rightarrow s \gamma }}
\newc\bxsgamma{\ensuremath{\overline{B}\rightarrow X_{s}\gamma}}

\newc\brbsgamma{\ensuremath{\BR\left(\bsgamma\right)}}
\newc\brbxsgamma{\ensuremath{\BR\left(\bxsgamma\right)}}

\newc\bsmumu{\ensuremath{B_s\to\mu^+\mu^-}}
\newc\brbsmumu{\ensuremath{\BR\left(B_s\to\mu^+\mu^-\right)}}

\newc\bdmmumu{\ensuremath{\overline{B}_d\to\mu^+\mu^-}}

\newc\bbbarmix{\ensuremath{\overline{B}_s\mbox{-}B_s}}      % B_s mixing
\newc\delmbs{\ensuremath{\Delta M_{B_s}}}

\newc{\butaunu}{\ensuremath{\left(B_u \rightarrow \tau \nu\right)}}
\newc{\brbutaunu}{\ensuremath{\BR\left(B_u \rightarrow \tau \nu\right)}}

\newcommand*{\reftable}[1]{Table~\ref{#1}}

     \newcommand*{\refsec}[1]{Sec.~\ref{#1}}
\newcommand*{\refref}[1]{Ref.\cite{#1}}

\newcommand*{\neutone}{\ensuremath{\chi}}

\newcommand*{\mhhat}{\ensuremath{\hat{m}_h}}
\newcommand*{\stau}{\ensuremath{\tilde{\tau}}}

\newcommand*{\alphaT}{\ensuremath{\alpha_T}}

\newcommand*{\razor}{\textrm{razor}}

\newcommand*{\razorexp}{\ensuremath{\cms\ \razor\ 4.4\invfb} analysis}

\newcommand*{\bayesfits}{\textrm{The BayesFITS Group}}
\newcommand*{\softsusy}{SOFTSUSY}
\newcommand*{\feynhiggs}{FeynHiggs}
\newcommand*{\micromegas}{MicrOMEGAs}

\newcommand*{\susyhit}{\text{SUSY-HIT}}

\newcommand*{\cms}{\text{CMS}}

\newcommand*{\pythia}{\text{PYTHIA}}

\newcommand*{\superiso}{\text{SuperIso Relic}}
\newcommand*{\stauc}{\text{\stau-coannihilation}}

\let\oldcite\cite
\renewcommand*{\cite}{~\oldcite}

\newcommand*{\hl}{\ensuremath{h}}

\newcommand*{\ha}{\ensuremath{A}}
\newcommand*{\mha}{\ensuremath{m_\ha}}
\newcommand*{\mh}{\ensuremath{m_h}}

\restylefloat{figure}

\begin{document}
%%%%%%%%%%%%%%%%%%%%%%%%%%%%%%%%%%%%%%%%%%%%%%%%%%%%%%%%%%%%%%%%%%%%%%%%%%%%%%%%

% Title
\title{Constrained MSSM favoring new territories:\\ The impact of new LHC limits and a 125
GeV Higgs boson}
%lr \title{Treating the CMSSM and the NUHM with an LHC Razor}

% Authors & affiliations
\author{Andrew Fowlie}
\email{A.Fowlie@sheffield.ac.uk}
\affiliation{Department of Physics and Astronomy, University of
  Sheffield, Sheffield S3 7RH, England}

\author{Malgorzata Kazana}
\email{Malgorzata.Kazana@fuw.edu.pl}
\affiliation{National Centre for Nuclear Research, Ho{\. z}a
  69, 00-681 Warsaw, Poland}

\author{Kamila Kowalska}
\email{Kamila.Kowalska@fuw.edu.pl}
\affiliation{National Centre for Nuclear Research, Ho{\. z}a
  69, 00-681 Warsaw, Poland}

\author{Shoaib Munir}
\email{Shoaib.Munir@fuw.edu.pl}
\affiliation{National Centre for Nuclear Research, Ho{\. z}a
  69, 00-681 Warsaw, Poland}

\author{Leszek Roszkowski}
\email{L.Roszkowski@sheffield.ac.uk}
\altaffiliation{On leave of absence from the University of Sheffield.}
\affiliation{National Centre for Nuclear Research, Ho{\. z}a 69, 
00-681 Warsaw, Poland}

\author{Enrico Maria Sessolo}
\email{Enrico-Maria.Sessolo@fuw.edu.pl}
\affiliation{National Centre for Nuclear Research, Ho{\. z}a 69, 00-681 Warsaw, Poland}

\author{Sebastian Trojanowski}
\email{Sebastian.Trojanowski@fuw.edu.pl}
\affiliation{National Centre for Nuclear Research, Ho{\. z}a 69, 00-681 Warsaw, Poland}

\author{Yue-Lin Sming Tsai}
\email{Sming.Tsai@fuw.edu.pl}
\affiliation{National Centre for Nuclear Research, Ho{\. z}a 69, 00-681 Warsaw, Poland}

% Appears in brackets after all authors
\collaboration{\bayesfits}

% Date
\date{\today}

%%%%%%%%%%%%%%%%%%%%%%%%%%%%%%%%%%%%%%%%%%%%%%%%%%%%%%%%%%%%%%%%%%%%%%%%%%%%%%%%
% Abstract
%%%%%%%%%%%%%%%%%%%%%%%%%%%%%%%%%%%%%%%%%%%%%%%%%%%%%%%%%%%%%%%%%%%%%%%%%%%%%%%%
\begin{abstract}
  We present an updated and extended global analysis of the
  Constrained MSSM (CMSSM) taking into account new limits on
  supersymmetry from $\sim5\invfb$ data sets at the LHC. In particular,
  in the case of the razor limit obtained by the \cms\ Collaboration we
  simulate detector efficiency for the experimental analysis and
  derive an approximate but accurate likelihood function. We discuss the impact on the global fit 
  of a possible Higgs boson with mass near 125\gev,
  as implied by recent data, and of a new improved limit on \brbsmumu. We identify high posterior probability regions of
  the CMSSM parameters as the stau-coannihilation and the $A$-funnel
  region, with the importance of the latter now being much larger due
  to the combined effect of the above three LHC results and of dark
  matter relic density. We also find that the focus point region is
  now disfavored. Ensuing implications for superpartner masses favor
  even larger values than before, and even lower ranges for dark
  matter spin-independent cross section, $\sigsip\lsim10^{-9}\pb$. We
  also find that relatively minor variations in applying experimental constraints can induce a large shift in
  the location of the best-fit point. This puts into question the
  robustness of applying the usual $\chi^2$ approach to the CMSSM.  We
  discuss the goodness-of-fit and find that, while it is difficult to
  calculate a $p$-value, the \gmtwo\ constraint makes, nevertheless,
  the overall fit of the CMSSM poor. We consider a scan
  without this constraint, and we allow $\mu$ to be either positive or
  negative.  We find that the global fit improves enormously for both
  signs of $\mu$, with a slight preference for $\mu<0$ caused by a
  better fit to \brbsgamma\ and \brbsmumu.
\end{abstract}
%%%%%%%%%%%%%%%%%%%%%%%%%%%%%%%%%%%%%%%%%%%%%%%%%%%%%%%%%%%%%%%%%%%%%%%%%%%%%%%%
% Title
\maketitle
%%%%%%%%%%%%%%%%%%%%%%%%%%%%%%%%%%%%%%%%%%%%%%%%%%%%%%%%%%%%%%%%%%%%%%%%%%%%%%%%

%%%%%%%%%%%%%%%%%%%%%%%%%%%%%%%%%%%%%%%%%%%%%%%%%%%%%%%%%%%%%%%%%%%%%%%%%%%%%%%%
\section{\label{sec:introduction}Introduction}
%%%%%%%%%%%%%%%%%%%%%%%%%%%%%%%%%%%%%%%%%%%%%%%%%%%%%%%%%%%%%%%%%%%%%%%%%%%%%%%%

The experimental collaborations ATLAS and \cms\ at the Large
Hadron Collider (LHC) have each so far collected around 5\invfb\ of data and
have analyzed a large part of it to set new improved limits on several
models of new physics beyond the Standard Model (SM), including
low-energy supersymmetry (SUSY). 
In particular, lower limits on the soft masses
\mzero\ and \mhalf\ of the Constrained Minimal Supersymmetric Standard Model
(CMSSM)\cite{hep-ph/9312272} have been pushed further up by a recent
\cms\ analysis of all-hadronic final states, which applied a
razor method to $4.4\invfb$ of
data\cite{CMS-PAS-SUS-12-005}. (In contrast, the other two free
parameters of the CMSSM, \azero\ and \tanb\ remain almost unaffected
by the above data.) This result considerably improved
previous limits by the same collaboration using the same method with
$0.8\invfb$ of data\cite{CMS-PAS-SUS-11-008}, as well as limits from the \alphaT\ method
using $1.1\invfb$ of data\cite{Chatrchyan:2011zy} and the MHT method
with the same dataset\cite{CMS-PAS-SUS-11-004}.  Much improved
lower limits on SUSY masses have also recently been produced by ATLAS, the
strongest of which have been obtained from searches with all-hadronic
final states\cite{ATLAS-CONF-2012-033,ATLAS-CONF-2012-037}. In
particular, the recent ``0-lepton" search with 2-to-6 jets has
resulted in a 95\% confidence level~(\cl) exclusion contour in the
CMSSM parameter space with 4.7\invfb\cite{ATLAS-CONF-2012-033} of data
which competes with the razor result in the same region of parameter space.

Furthermore, last year both ATLAS and \cms\ excluded all but two small
windows of SM (and SM-like) Higgs mass range, by combining their
searches in the $\gamma\gamma$, $bb$, $\tau\tau$, $WW$ and $ZZ$ final
states\cite{CMS-PAS-HIG-12-008,ATLAS-CONF-2012-019}. In December 2011
both collaborations also reported some excess of events in the
subdominant but background-clean $\gamma\gamma$ final
state\cite{ATLAS:2012ad,Chatrchyan:2012tw}.  In the $ZZ\rightarrow 4l$
final state a small excess has also been found but at a somewhat
smaller mass of around 119 \gev\cite{ATLAS:2012ac,Chatrchyan:2012dg}.
The Tevatron collaborations CDF and D0 also found some excess over a
broader mass range\cite{TEVNPH:2012ab}.  The hints of a possible Higgs
signal around 125\gev\ generated much excitement and
activity\cite{Baer:2011ab,*Buchmueller:2011ab,*Kadastik:2011aa,*Cao:2011sn,
  Baer:2012uy,*Akula:2011aa,*Aparicio:2012iw,
  Ellwanger:2011aa,*Ellwanger:2012ke,*Gunion:2012zd,*King:2012is,*Kang:2012tn,*Cao:2012fz,*Vasquez:2012hn,*Gabrielli:2012hd,
  Hall:2011aa,*Li:2011ab,*Arbey:2011ab,*Arbey:2011aa,*Draper:2011aa,*Moroi:2011aa,*Carena:2011aa,*Arvanitaki:2011ck,*Gozdz:2012xx,*FileviezPerez:2012iw,*Chang:2012gp,
  *Desai:2012qy,*Maiani:2012ij,*Cheng:2012np,*Kyae:2012ea,*Boudjema:2012cq,*Christensen:2012ei,*Gogoladze:2012ii,*Byakti:2012qk,*Ajaib:2012vc,*Ibe:2012dd,*Basak,
  Matsuzaki:2012gd,*Azatov:2012bz,*Blankenburg:2012ex,*LopezHonorez:2012kv,
  *Barger:2012hv,*Giardino:2012ww,*Ellis:2012rx,*Bai:2011wz,*arXiv:1204.1061,*Arhrib:2012yv,*Arhrib:2012ia,
  Balazs:2012qc,*Gosh,*Feng:2012jf,*Okada:2012gf,*Feng,*He,*Barroso:2012wz,*Carpenter:2012zr,*Ginzburg:2012hc,*Espinosa,*Jeong:2012ma,*Donkin:2012yn}.\footnote{On July 4th, 2012, the discovery at $4.9\sigma$ by
  CMS\cite{CMS:2012gu} and at $5.0\sigma$ by ATLAS\cite{ATLAS:2012gk} of a boson
  consistent with the SM Higgs, with mass near 125\gev, was
  announced. Particularly, the mass claimed by CMS, $125.3\pm0.6\gev$,
  is very close in central value and experimental error to the signal
  case considered in this paper. In light of this important discovery
  we shall discuss only the case of a SM-like Higgs boson with mass of
  125\gev.}

Another recent important highlight of experimental progress in
constraining SUSY and other frameworks of new physics has been the
new, much improved limit
$\brbsmumu<4.5\times10^{-9}$~(95\%~\cl)\cite{Aaij:2012ac}, which is
already approaching the SM value of $\left(3.2\pm0.2
\right)\times10^{-9}$\cite{Buras:2010wr}. Its effect on the CMSSM will
also be considerable, as we shall see below.

In a previous paper\cite{Fowlie:2011mb} by the BayesFITS group a
global statistical analysis of the CMSSM based on about 1.1\invfb\ of
data was presented. In addition to the usual set of relevant
constraints from the relic abundance of cold dark matter (DM) in the
Universe, direct mass limits from LEP and the Tevatron, flavor
physics, \etc., one of the most
restrictive limits, from the \alphaT\ analysis of
\cms\cite{Chatrchyan:2011zy} was applied. The analysis included some crucial
features. We generated approximate efficiency and likelihood maps in
order to reproduce the \cms\ \alphaT\ limit, as described in detail
in Ref.\cite{Fowlie:2011mb}. This allowed us to include the \cms\ exclusion
limit into the combined likelihood function along with the other
constraints, and to map out high probability regions of the CMSSM
parameter space. Furthermore, in Ref.\cite{Roszkowski:2012uf} the impact of
recent limits from FermiLAT on dwarf spheroidal galaxies was
investigated in order to derive implications of direct and indirect
detection of cold DM for the CMSSM, along with an extension of the
approximate likelihood maps for the \cms\ \alphaT\ result to
significantly larger CMSSM mass parameter ranges, and an update on
a number of our results from Ref.\cite{Fowlie:2011mb}.

One of the conclusions derived from previous global analyses, both
ours and the ones performed by other
groups\cite{Akrami:2009hp,*Strege:2011pk,*Bechtle:2012zk,Buchmueller:2011sw,Baer:2011ab,*Buchmueller:2011ab,*Kadastik:2011aa,*Cao:2011sn},
was that the dominant contribution to the total $\chi^2$ comes from
the anomalous magnetic moment of the muon \gmtwo. It seems obvious
that relaxing this particular constraint would in a natural way
improve the CMSSM fit because satisfying it requires quite low masses
of the scalars.

In fact, there exists a quite convincing argument to do so.  It has
been known for years that a significant discrepancy is observed
between the experimental measurement of the muon anomalous magnetic
moment, coming from the experiment E821 at Brookhaven National
Laboratory\cite{g2experiment}, and its theoretical predictions within
the SM framework. The discrepancy is at more than 3$\sigma$,
\deltagmtwomu = $28.7\pm8.0\times 10^{-10}$\cite{Davier:2011}, and is
usually interpreted as a strong indication of new physics beyond the
SM.

However, since the poor fit of the CMSSM is to such a large extent a
result of basically only one constraint, it is worth examining
whether it is as robust as the other most important constraints.
In fact, despite much effort, there seem to remain a number of issues
of which we only highlight a few here. The accuracy of theoretical predictions is strongly affected
by the nonperturbative effects related to the low-energy strong
interactions. The main leading-order (LO) contribution to \deltagmtwomu\
comes from the hadron vacuum polarization and is between 5 $\times
10^{-10}$ and 6 $\times 10^{-10}$. It can be related to the measured
hadronic cross section provided by the experiment and has been
calculated very precisely with a fractional accuracy of
0.7\%\cite{Jegerlehner:2009,*Davier:2010,*Hagiwara:2011,Jegerlehner:2011, Davier:2011}. On the other
hand, a next-to-leading order (NLO) contribution of the order of
$O(\alpha^3)$ that comes from the light-by-light scattering through
the hadronic vacuum, though one order of magnitude smaller than the
LO contribution, is much more poorly known (with a
fractional accuracy of 30\%), since it cannot be calculated accurately
based on the experimental data and is strongly model dependent. As a consequence, its contribution to \deltagmtwomususy\ is between 2.5 $\times 10^{-10}$ and 4 $\times 10^{-10}$\cite{Prades:2009}. 
Due to all those uncertainties one should be careful in interpreting
the effect of \deltagmtwomu\ on the searches for SUSY, in particular
the CMSSM.  We, therefore, also present here some global fits both in
the presence and in the absence of the \gmtwo\ constraint.

Relaxing the \gmtwo\ constraint has an important consequence. Since
the supersymmetric contribution to \deltagmtwomu\ is proportional to
\signmu, in order to satisfy the experimental limit one is forced to
choose $\signmu>0$, as has been the case in most of the previous global
fit analyses. However, with the \gmtwo\ constraint abandoned, the
justification to limit the Higgs/Higgsino mass parameter $\mu$ to positive values is no
longer there, since the other constraints are much less affected by
the sign of $\mu$. The analysis of the impact of the negative $\mu$ on the
global CMSSM fit was performed in, \eg, Ref.\cite{Allanach:2006cc,Roszkowski:2007fd} for the
data from the pre-LHC experiments, but with the \gmtwo\ constraint
taken into account.  

On the other hand, for negative $\mu$ the fit to \brbsgamma\ actually
improves considerably\cite{Roszkowski:2007fd} in the higher
superpartner mass ranges implied by new LHC
limits. This is because, in order to provide
a contribution from SUSY to the positive discrepancy between the
experimental and the SM values, one actually needs positive contributions from
both the charged Higgs/top and the chargino/stop loops, the latter of
which is inverse-proportional to the sign of $\mu$. 
As we shall see, considering both signs of $\mu$ and relaxing
the \gmtwo\ constraint will lead to a rather complex picture. In
particular, it will significantly improve the statistical fit of the
CMSSM.

In this paper we update our recent global analysis of the
CMSSM\cite{Fowlie:2011mb}. While we mainly focus on a Bayesian
approach and derive posterior probability density function (pdf) maps,
we also compute, for each case we consider, the lowest $\chi^2$ (best-fit point).

We find that it can often be difficult to robustly establish the
location of the best-fit point in the CMSSM parameter space, in
particular in the most studied case with $\mu>0$ and the \gmtwo\
constraint included. Basically, one can find a very good fit in either
a (relatively small) stau coannihilation (henceforth \stauc) region or
in a (much more extended) \ha-funnel region, where \ha\ is the
pseudoscalar Higgs, both at large \mhalf\ and not as large
\mzero. First, the lowest values of $\chi^2$ in both regions is
often very similar. Second, in the $A$-funnel region we find an
extended ``plateau'' of comparable, low values of $\chi^2$. As a
result, fairly small changes in the treatment of experimental
constraints (most notably the LHC lower mass limits via a likelihood
function), etc, may cause a large shift in the location of the
best-fit point, as we will present in detail below.  Our analysis here
confirms our earlier assertion spelled out in Ref.\cite{Fowlie:2011mb}
(page 17) and puts into question the robustness of results obtained
with the $\chi^2$ approach in the framework of the CMSSM.

The main new elements of this study are as follows:
\begin{itemize}
\item{the derivation of an approximate but accurate likelihood map
corresponding to the \cms\ razor limit based on $4.4\invfb$ of data;}
\item{studying the impact of a SM-like Higgs with the mass around 125\gev;}
\item{considering the effect of the recently updated
limit on $\brbsmumu$.}
\end{itemize}
All these three ingredients will play a major role in
shifting  high posterior probability regions from the previously
favored \stauc\ region, and to some degree also focus
point region, to mainly the \ha-funnel region. In particular, as we
discuss below, different ways of mimicking the \cms\ limit in the
likelihood map can have a major impact on both the location and also
the value of the best-fit point.

Also, motivated by the results of the previous scans and some
theoretical arguments, we move here beyond the usual CMSSM global
fit analysis and investigate the effects due to: 
\begin{itemize}
\item{relaxing the \gmtwo\ constraint; and}
\item{taking a negative sign of parameter $\mu$.}
\end{itemize}

This paper is organized as follows. In \refsec{sec:method} we
detail our methodology, including our statistical analysis, scanning
algorithm and our treatment of the likelihood from the \razorexp. In
\refsec{sec:results} we present the results from our scans and discuss their novel features. In \refsec{sec:statistics} we give a statistical discussion of our results and we summarize our findings in
\refsec{sec:summary}.

%%%%%%%%%%%%%%%%%%%%%%%%%%%%%%%%%%%%%%%%%%%%%%%%%%%%%%%%%%%%%%%%%%%%%%%%%%%%%%%%
\section{Method\label{sec:method}}
%%%%%%%%%%%%%%%%%%%%%%%%%%%%%%%%%%%%%%%%%%%%%%%%%%%%%%%%%%%%%%%%%%%%%%%%%%%%%%%%
%%%%%%%%%%%%%%%%%%%%%%%%%%%%%%%%%%%%%%%%%%%%%%%%%%%%%%%%%%%%%%%%%%%%%%%%%%%%%%%%
% Stat language
\subsection{The framework\label{subsec:framework}}
%%%%%%%%%%%%%%%%%%%%%%%%%%%%%%%%%%%%%%%%%%%%%%%%%%%%%%%%%%%%%%%%%%%%%%%%%%%%%%%%

Our aim is to map out the regions of the  parameter space of the SUSY model
under consideration that are in best agreement with all relevant
experimental constraints. To this end, we follow the strategy outlined
in Refs.\cite{deAustri:2006pe,Fowlie:2011mb}. Here we merely summarize its main features. 

In Bayesian statistics, for a theory described by some parameters $m$, experimental observables
$\xi(m)$ can be compared with data $d$ and a posterior probability density
function (pdf) $p(m|d)$ can be calculated through Bayes' Theorem
%%%%%%
\begin{equation}
p(m|d)=\frac{p(d|\xi(m))\pi(m)}{p(d)}\, ,
\label{Bayesth}
\end{equation}
%%%%%%
where the likelihood $p(d|\xi(m))\equiv\mathcal{L}$ gives the
probability density for obtaining $d$ from a measurement of $\xi$, the
prior $\pi(m)$ parametrizes assumptions about the theory prior to
performing the measurement and the evidence $p(d)\equiv\mathcal{Z}$
represents the assumptions on the data. As long as one considers only
one model the evidence is a constant in the theory parameters, and
thus a normalization factor, but, as we will see in Sec.~IVB, it is
a necessary element of model comparison.

The Bayesian approach yields a simple and
natural procedure for calculating the posterior pdf of any
limited subset of $r$ variables in the $n$-dimensional parameter space, $s_{i=1,..,r}\subset m$.
%lr *** limited subset of $r$ variables in the $n$-dimensional parameter space, $\psi_{i=1,..,r}\subset m$.
One just needs to marginalize, or integrate, over the remaining
parameters
%%%%%%
\begin{equation}
p(s_{i=1,..,r}|d)=\int p(m|d)d^{n-r}m\,.
%lr **** p(\psi_{i=1,..,r}|d)=\int p(m|d)d^{n-r}m\,,
\label{marginalization}
\end{equation}
%%%%%%
%lr **** where $n$ denotes the dimension of the full parameter space. 
To describe our
methodology for the Bayesian scan we use the same notation
as in  Ref.\cite{Fowlie:2011mb}.

The likelihood function is a central object in our statistical
analysis. We construct it using the prescription described in
\refref{deAustri:2006pe}. In particular, we model positive
measurements with a Gaussian function, and smear out the experimental
limits from negative searches using the theoretical error $\tau$.

As stated in the Introduction, in the current analysis we include
three new important ingredients provided by LHC data.  First, we
include the new exclusion limit on the (\mzero, \mhalf) plane of the
CMSSM, which has been obtained by the \cms\ Collaboration by applying
the razor method to 4.4\invfb\ of data (see subsection
\ref{subsec:razor_cut} for details).  Second, we consider the impact
of the new information from the Higgs boson experimental searches and
we assume a SM-like Higgs with the mass 125\gev.  Finally, we include
in the likelihood function the new, improved limit on $\brbsmumu$.

%%%%%%%%%%%%%%%%%%%%%%%%%%%%%%%%%%%%%%%%%%%%%%%%%%%%%%%%%%%%%%%%%%%%%%%%%%%%%%%%
% 
\subsection{The efficiency and likelihood  maps for the \cms\ razor $4.4\invfb$ analysis\label{subsec:razor_cut}}
%%%%%%%%%%%%%%%%%%%%%%%%%%%%%%%%%%%%%%%%%%%%%%%%%%%%%%%%%%%%%%%%%%%%%%%%%%%%%%%%

We derive our LHC likelihood for the \cms\
search\cite{Chatrchyan:2011ek,CMS-PAS-SUS-12-005} for $R$-parity
conserving SUSY in all-hadronic events performed with the \razor\
method summarized below.  The results based on the LHC data sample of
$4.4\invfb$ of integrated luminosity recorded at $\sqrt{s} = 7$\tev\
shows no excess of events over the SM predictions. Our aim is to
translate the analysis scheme into a simplified approach to obtain a
signal selection efficiency for a large number of points in the CMSSM
parameter space.

Studies by the LHC collaborations have shown that jets plus missing
energy constraints are relatively insensitive to the values of \tanb\
and \azero\cite{Allanach:2011ut,*Bechtle:2011dm}, because these
parameters have little effect on the squark and gluino masses. The
choice of \tanb\ is dictated by the requirement of the appropriate
radiative electroweak symmetry breaking (REWSB). The range of the
theoretically excluded region in the (\mzero, \mhalf) plane where
$\mu^2$ becomes negative and, consequently, REWSB does not occur,
strongly depends on the values of \tanb. We choose a value of
$\tanb=3$ which assures that the no-REWSB region does not appear
within the analyzed parameter range, and we fix $\azero = 0$ and
$\signmu = +1$ (or $-1$).

For \mzero\ in the range of $100-4000\gev$ and \mhalf\ in the range
of $100-2000\gev$ we generate a 2-dimensional grid of points in the
(\mzero, \mhalf) plane.  A scanning step of 50\gev\ is chosen in both
dimensions. For each point, we generate a mass spectrum and a  decay
table of supersymmetric particles, using the publicly available
packages \softsusy\cite{softsusy} and \susyhit\cite{susyhit},
respectively. The mass spectrum and the decay tables are then passed
to \pythia\ 6.4\cite{PYTHIA} for the event generation process. The
hadronized events are then passed to the fast detector simulator
PGS4\cite{PGS4}, which reconstructs the physical objects (photons,
electrons, muons, hadronically decaying taus, and hadronic jets).  We
updated the detector parameter-card following the recommendations of
the experimental collaboration on the CMS settings.

Our \razor\ analysis performed in this paper follows closely the one
of the \cms\ Collaboration\cite{CMS-PAS-SUS-12-005}.  All
reconstructed events are divided into six disjoint event samples
(boxes), dependent on the presence or absence of a lepton of a given
flavor: electron box (ELE), muon box (MU), three dilepton boxes
(ELE-MU, MU-MU, ELE-ELE), and hadronic box (HAD). For the analysis
described in this paper we limit ourselves to reconstructing the
hadronic box, which has been shown to yield an excellent approximation
of the overall bound with 4.4/fb\cite{CMS-PAS-SUS-12-005}.

At the preselection stage cuts are applied on the transverse energies
$E_T$ and the pseudorapidities $\eta$ of the reconstructed jets:
$E_T>40\gev$, $|\eta|<3$ for all jets, and $E_T>60\gev$ for two
leading jets. All jets appearing in a single event are grouped together to form two
megajets, which we label $jet_1$ and $jet_2$. The selection of the preferred jet combination is based on
the invariant mass of the dijet system. All possible combinations of jets are taken into
account and the one is chosen for which the invariant mass is minimal.

A pair of megajets should reconstruct the energy distribution of the
visible decay products in the center of mass (CM) frame. However, due to the presence of two unseen
lightest SUSY particles (LSP), it is possible to reconstruct this frame only approximately. The
idea of the \razor\ analysis is to
replace the CM frame with the
so called $R$-frame, defined as a longitudinally boosted frame in
which the energies of the visible products can be written in terms of
some Lorentz invariant scale, which correctly approximates the energy
distribution in the CM frame. The Lorentz boost factor of the
transformation between the CM and $R$ frames is given by
%%%
\begin{equation}
\beta_R=\frac{p_z^{jet_1}+p_z^{jet_2}}{E^{jet_1}+E^{jet_2}}
\end{equation}
%%%
and the longitudinal boost invariant mass scale $M_R$ is defined as
%%%
\begin{equation}
M_R=\sqrt{(E^{jet_1}+E^{jet_2})^2-(p_z^{jet_1}+p_z^{jet_2})^2}.
\end{equation}
With such a definition, $M_R$ approximates the peak in the energy
distribution of the visible decay products. 
%%%
%lr The transverse mass $M_T^R$ is also defined
One also defines the transverse mass $M_T^R$ as
%%%
\begin{equation}
M^R_T=\sqrt{\frac{E_T^{miss}(p_T^{jet_1}+p_T^{jet_2})-\vec{E}_T^{miss}\cdot(\vec{p}_T^{jet_1}+\vec{p}_T^{jet_2})}{2}},
\end{equation}
%%%
as well as the \razor\ dimensionless ratio,
%%%
\begin{equation}
R=\frac{M_T^R}{M_R}.
\end{equation}
%%%
The variable $R$ would peak around zero for the QCD multijets and
around 0.5 for the SUSY signal, constituting a good discriminator
allowing to reduce the magnitude of the QCD background. The events in
the hadronic box are required to satisfy the conditions: $M_R>400\gev$ 
and $0.18 < R^2 < 0.5$.

To construct a 2-dimensional pdf for the signal, all accepted events
are divided into 38 separate bins in the $(M_R, R^2)$ plane. The
corresponding numbers of the observed events ($o$), expected
background events ($b$) and errors on the expected background yield
($\delta b$) are given in Table~\ref{razor_bins}\cite{Pierini}. Note
that two bins, namely $1200\gev<M_R<1600\gev$, $0.3<R^2<0.4$ and
$1200\gev<M_R<1600\gev$, $0.4<R^2<0.5$, feature a bigger than
3$\sigma$ excess of the observed signal over the expected
background. We will come back to this issue while discussing the
probability distribution assigned to each bin.

%%%
\begin{table}[t]\footnotesize
\begin{tabular}{|c|@{}c@{}|@{}c@{}|@{}c@{}|@{}c@{}|@{}c@{}|}
\hline  
$M_R$, $R^2$ & 500-550, 0.3-0.4 & 500-550, 0.4-0.5 & 550-600, 0.3-0.4 & 550-600, 0.4-0.5 & 600-650, 0.3-0.4  \\
\hline  
observed & 246 & 112 & 124 & 85 & 86 \\
background &  259.5 $\pm$ 19.4 & 118.9 $\pm$ 14.4 & 162.8 $\pm$ 16.1 & 73.6 $\pm$ 12.0 & 104.8 $\pm$ 14.8\\
\hline
$M_R$, $R^2$ & 600-650, 0.4-0.5 & 650-700, 0.2-0.3 & 650-700, 0.3-0.4 & 650-700, 0.4-0.5 & 700-800, 0.2-0.3\\
\hline  
observed & 26 & 192 & 57 & 23 & 247 \\
background & 43.0 $\pm$ 9.2 &  209.8 $\pm$ 21.2 & 68.0 $\pm$ 11.6 & 26.0 $\pm$ 7.2 & 233.9 $\pm$ 27.2 \\
\hline
$M_R$, $R^2$  & 700-800, 0.3-0.4 & 700-800, 0.4-0.5 & 800-900, 0.2-0.3 & 800-900, 0.3-0.4 & 800-900, 0.4-0.5 \\
\hline  
observed & 65 & 27 & 92 & 24 & 6\\
background  & 74.1 $\pm$ 15.1 & 24.8 $\pm$ 8.2 & 104.3 $\pm$ 17.7 & 29.3 $\pm$ 9.4 & 8.5 $\pm$ 4.3\\
\hline
$M_R$, $R^2$  & 900-1000, 0.2-0.3 & 900-1000, 0.3-0.4 & 900-1000, 0.4-0.5 & 1000-1200, 0.18-0.2 & 1000-1200, 0.2-0.3 \\
\hline  
observed  & 50 & 13 & 3 & 20 & 31\\
background  & 48.6 $\pm$ 12.6 & 11.3 $\pm$ 5.6 & 2.7 $\pm$ 2.2 & 15.8 $\pm$ 5.8 & 33.1 $\pm$ 10.2 \\
\hline
$M_R$, $R^2$  & 1000-1200, 0.3-0.4 & 1000-1200, 0.4-0.5 & 1200-1600, 0.18-0.2 & 1200-1600, 0.2-0.3 & 1200-1600, 0.3-0.4  \\
\hline  
observed  & 5 & 3 & 10 & 13 & 8\\
background  & 6.3 $\pm$ 3.8 & 1.3 $\pm$ 1.3 & 4.8 $\pm$ 2.9 & 9.3 $\pm$ 4.9 & 1.2 $\pm$ 1.2 \\
\hline
$M_R$, $R^2$ & 1200-1600, 0.4-0.5 & 1600-2000, 0.18-0.2 & 1600-2000, 0.2-0.3 & 1600-2000, 0.3-0.4 & 1600-2000, 0.4-0.5 \\
\hline  
observed  & 3 & 0 & 0 & 1 & 0\\
background & 0.4 $\pm$ 0.4 & 0.5 $\pm$ 0.5 & 0.6 $\pm$ 0.6 & 0.4 $\pm$ 0.4 & 0.3 $\pm$ 0.3 \\
\hline
$M_R$, $R^2$ & 2000-2800, 0.18-0.2 & 2000-2800, 0.2-0.3 & 2000-2800, 0.3-0.4 & 2000-2800, 0.4-0.5 & 2800-3500, 0.18-0.2 \\
observed  & 0 & 0 & 0 &  0 & 0 \\
background  &  0.4 $\pm$ 0.4 & 0.4 $\pm$ 0.4 & 0.3 $\pm$ 0.3 & 0.3 $\pm$ 0.3 & 0.3 $\pm$ 0.3\\
\cline{1-6}
$M_R$, $R^2$ & 2800-3500, 0.2-0.3 & 2800-3500, 0.3-0.4 & 2800-3500, 0.4-0.5 \\ 
\cline{1-4} 
observed & 0 & 0 & 0\\ 
background  & 0.3 $\pm$ 0.3 & 0.3 $\pm$ 0.3 & 0.3 $\pm$ 0.3 \\
\cline{1-4}
\end{tabular}
\caption {Bins used in the \razor\ analysis with the corresponding numbers of observed ($o$) and expected background events ($b\pm\delta b$).} \label{razor_bins} 
\end{table}
%%%

The efficiency $\epsilon$ of the detector is defined as the fraction
of events that passed all the cuts. The signal for the $i$-th bin is
than computed in the usual way,
%%%%
\begin{equation}
s_i=\epsilon_i\times\sigma\times\int L,
\end{equation}
%%%%
where $\int L$ is an integrated luminosity, here $\int L = 4.4
\invfb$, and $\sigma$ is the total cross section for the
production of supersymmetric particles at $\sqrt{s}=7$ TeV. The
probability of observing $o_i$ events in the $i$-th bin, given the
known number of the expected events $s_i$, and the number of the expected SM background events
$b_i$, is given by a counting-experiment likelihood (Poisson distribution) 
convolved with an additional function that takes care of the predicted error on the background yields ($\delta b_i$)
\begin{equation}
\mathcal{L}_i(o_i, s_i, b_i)=\int P(o_i|s_i,\bar{b}_i)F(\bar{b}_i|b_i,\delta b_i)d\bar{b}_i,
\end{equation}
where
\begin{equation}
P(o_i|s_i,b_i)=\frac{e^{-(s_i+b_i)}(s_i+b_i)^{o_i}}{o_i!}\,.\label{poisson}
\end{equation}
The values for $o_i$, $b_i$ and $\delta b_i$ are given in
Table~\ref{razor_bins}.  For the bins where the number of the observed
events does not exceed the predicted background by more than 3$\sigma$
(in fact, in our case it never exceeds 2$\sigma$), we use for the
function $F$ a standard Gaussian distribution
\begin{equation}
F(\bar{b}_i|b_i,\delta b_i)=\frac{1}{\delta b_i\cdot\sqrt{2 \pi}}\,
\exp\left[-\frac{1}{2}\left(\frac{\bar{b}_i - b_i}{\delta
      b_i}\right)^2\right].\label{gauss} 
\end{equation}
On the other hand, one should be a bit more careful when dealing with
the two bins in which the excess of events over the predicted
background is more than 3$\sigma$. The Poisson distribution
(\ref{poisson}) peaks at $o_i\approx s_i+(b_i\pm\delta b_i)$. If the
number of the observed events in a given bin is not much bigger than
the predicted background yield, the maximal likelihood one can obtain
in that bin will correspond to the background-only hypothesis (with
$s_i=0$). Any nonzero signal will suppress the likelihood, allowing to
exclude points on the (\mzero, \mhalf) plane almost independently of
the exact value of the signal. The likelihood map corresponding to
such a case will show a smoothly dropping likelihood function. On the
contrary, if the number of observed events is much bigger than the
predicted background yield, then the relation $o_i\approx
s_i+(b_i\pm\delta b_i)$ can hold only for the nonzero signal (contributions from the background with more than 3$\sigma$ error will
be suppressed by the Gaussian factor). For such points, the Poisson
likelihood (\ref{poisson}) will be enhanced by the nonzero signal,
which will result in the presence of spurious peaks in the likelihood map.
%In order to accurately reproduce the
%95\%~$CL_s$ exclusion bound published by CMS, it becomes necessary to
In order to reduce the statistical significance of event excess over the
background, instead of the Gaussian distribution a correct
way to proceed is to model the background uncertainties with a distribution that
assigns higher probabilities to the distribution right tail, for
example, with a log-normal distribution
\begin{equation}
F(\bar{b}_i|b_i,\delta b_i)=\frac{1}{\bar{b}_i\cdot\delta b_i\cdot
  \sqrt{2 \pi}}\, \exp\left[-\frac{(\ln\bar{b}_i-b_i)^2}{2\cdot\delta
    b_i^2}\right].\label{lognormal} 
\end{equation}
With such a distribution 
one can accommodate larger deviations from the background's central value, 
which allow to maximalize the total likelihood with a smaller number of signal
events and, at the same time, do not suppress the likelihood through convolution.  
%higher background values 
%(without supressing the total likelihood with a Gaussian factor) 
%and reduce the signal necessary to maximalize the total likelihood.
Such an approach allows us to eliminate the spurious
peaks in the likelihood map. Note that the presence of the
peaks would affect the contribution from the razor limit to the total $\chi^2$, and consequently to \chisqmin\ and the location of the best-fit point, even at large \mzero\ and \mhalf, far above the
region excluded by the razor limit (in the large mass region this contribution amounts to approximately three units of $\chi^2$). Such a situation would be clearly unphysical.

On the other hand, note that the procedure that we have adopted here
is not unique, even though it correctly reproduces the CMS limit. We
will discuss the impact of some other approaches to modeling the \razor\
exclusion limits in Sec.~\ref{sec:bestfit}.

The final total likelihood is obtained as a product of likelihoods for each separate bin
\begin{equation}
\mathcal{L}_{\razor}=\prod_{i=1}^{38}\mathcal{L}_i.
\end{equation}

\begin{figure}[t]
\centering
\subfloat[$\mu > 0$.]
{%
\label{fig:LikeMap-a}%
% razor 4.4/fb likelihood plot
\includegraphics[width=0.39\linewidth]{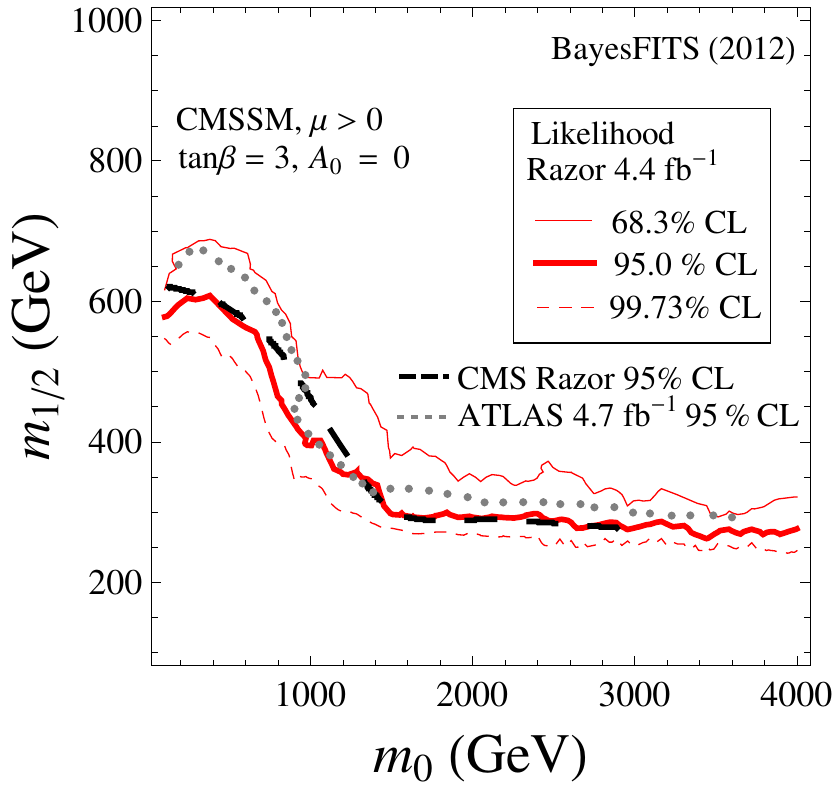}
}
\hspace{8pt}
\subfloat[$\mu < 0$.]
{% 
\label{fig:LikeMap-b}%
% razor 4.4/fb SB profile likelihood
%\includegraphics[width=0.39\linewidth]{razorNUHM4fb_c.pdf}
\includegraphics[width=0.39\linewidth]{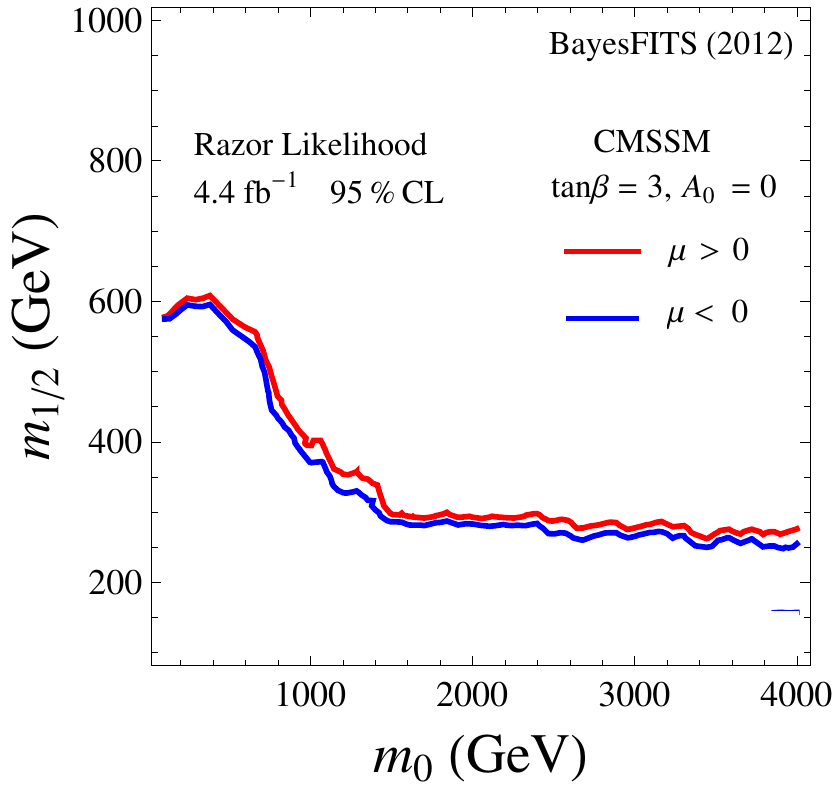}
}%
\caption[]{\subref{fig:LikeMap-a} Our approximation of the CMS razor
  4.4/fb likelihood map as described in the text. \tanb\ and \azero\
  are fixed to the values in the legend. The thick solid line 
  shows the 95.0\%~CL ($2\sigma$) bound. It approximates the \cms\ $95\%$~CL exclusion
  contour, shown by the dashed black line. The thin solid line and the thin dashed line show our calculations of the 68.3\%~CL ($1\sigma$) the 99.73\%~CL ($3\sigma$) exclusion bound, respectively. The dotted gray line shows the ATLAS 95\%~CL exclusion bound. \subref{fig:LikeMap-b} Our calculation of the CMS razor 95\%~CL exclusion line for $\mu>0$ (red) and $\mu<0$ (blue).}
\label{fig:likemap}
\end{figure}

We obtain the 95{\%}~CL exclusion limits using the $\Delta\chi^2$
statistics test and validate our result against the official CMS
plot\cite{CMS-PAS-SUS-12-005}.  We present in
Fig.~\ref{fig:likemap}\subref{fig:LikeMap-a} the 68.3\% (1$\sigma$),
95.0\% (2$\sigma$) and 99.73\%~CL ($3\sigma$) limits obtained from our
likelihood. For comparison we also show the official CMS exclusion
limit. We find a very good agreement, provided we rescale our signal
by a factor 1.8, which is a reasonable assumption given that
\pythia\ calculates the $pp$ cross section at only the leading
order\footnote{The cross section, and consequently the number of
  expected supersymmetric events, changes by over ten orders of
  magnitude over the (\mzero, \mhalf) plane. The resulting likelihood
  function is, therefore, not sensitive to next-to-leading order
  corrections to the cross section. Even if $\sigma_\text{NLO} \sim
  \sigma_\text{LO}$, the corrections would only slightly shift the
  isocontours of cross section and likelihood on the (\mzero, \mhalf)
  plane.}, and PGS4 might present some deficit in the efficiency
reconstruction.

The approximate efficiency maps derived above allow us to evaluate a
likelihood function, so that we can find the regions of the SUSY model's
parameter space that are in best agreement with the CMS \razor\
limit. Marked in the figure is also the 95\%~\cl\ limit from ATLAS,
which at low \mzero\ is actually a bit stronger. We note here that the
ATLAS limit was expected to be lower than the \razor\ one in the
(\mzero, \mhalf) plane. The actual limit being somewhat higher than
expected is a result of downwards fluctuation in the number of
background events. Given the fact that the two limits are actually
comparable within the experimental resolution around the region where
they are located, we will henceforth only show the CMS limit in our
figures.

We also verify the influence of selecting the negative sign of $\mu$ on our likelihood
distribution. While the  independence of the exclusion limit of \tanb\
and \azero\ in the analysis with all-hadronic final states is a well-known 
fact, it was never investigated before in the case of $\mu <
0$. The results of such a scan are presented in
Fig.~\ref{fig:likemap}\subref{fig:LikeMap-b}, where we show our
derived razor 95\%~CL bound. It appears clear that the position of the
line in the (\mzero, \mhalf) plane is almost insensitive to the sign of the
parameter $\mu$.

%%%%%%%%%%%%%%%%%%%%%%%%%   F   I   G   U   R   E   %%%%%%%%%%%%%%%%%%%%%%%%%%%%
% Likelihood maps

%%%%%%%%%%%%%%%%%%%%%%%%%%%%%%%%%%%%%%%%%%%%%%%%%%%%%%%%%%%%%%%%%%%%%%%%%%%%%%%

\subsection{The Higgs likelihood \label{subsec:higgs}}

In this paper we investigate the impact of the Higgs discovery at the
LHC on the CMSSM. In the CMSSM, 
so long as $\mha\gg\mz$, the lightest Higgs boson is to a very
good accuracy SM-like, \ie, its couplings to $ZZ$ and $WW$ are almost
the same as those of the SM Higgs (the
so-called decoupling regime)\cite{Djouadi:2005gj}. This has been a conclusion of many
previous studies, and has been also carefully checked in Ref.\cite{Roszkowski:2006mi}
%*** arXiv:hep-ph/0611173 On the detectability of the CMSSM light Higgs
%boson at the Tevatron Leszek Roszkowski (Sheffield and CERN), Roberto
%Ruiz de Austri (Autonoma Madrid), Roberto Trotta (Oxford) Journal-ref:
%JHEP 0704:084,2007, check/quote also djouadi's big review? *** 
with
experimental constraints available at that time (among which the
constraints on \mzero\ and \mhalf\ were clearly weaker than those
available now). We will show in Sec.~\ref{sec:cmssmresults} that this assumption is justified {\it a posteriori}, given the present constraints.
While the results from the LHC on the Higgs boson do
indicate that the discovered boson is indeed SM-like, here we will
assume that it is the lightest Higgs boson of the CMSSM that has
actually been discovered. Note that in our analysis we will be using
information about the Higgs mass but will not be applying constraints on
its couplings, in particular on the one to $\gamma\gamma$.

In setting up the Higgs likelihood function one has to take into
account an appreciable theoretical error on the light Higgs mass
calculation in the MSSM which comes primarily from neglecting
higher-order loop corrections, renormalization scheme differences, \etc., which is
estimated to be around $2-3$\gev\cite{Heinemeyer:2011aa}.  One therefore
has to distinguish between the ``true'' value of the Higgs mass \mhhat\ which would
result from an exact calculation (and which we identify with the
physical mass), and the value of the Higgs mass, denoted here by
\mhl, calculated within a given approximation encoded in one or
another spectrum calculator.\footnote{In our numerical scans we use \softsusy\
  version 3.2.4\cite{softsusy_old} but one should be aware that all
  available Higgs mass codes presently have similar (or larger)
  theoretical errors.}

The Higgs
mass can  initially be measured  with only a limited precision. We assume that the
mass of a SM-like Higgs is measured at $\mhhat=125~\gev$ with a Gaussian
experimental uncertainty of $\sigma=2~\gev$,
%%%%
\begin{equation}
p(d | \mhhat) = \exp\left[-(125\gev-\mhhat)^2/2\sigma^2\right].
\end{equation}
%%%%

Since we have only an imperfect Higgs mass calculation,
we assume that the Higgs masses calculated with \softsusy\ are
Gaussian-distributed around the ``true'' Higgs masses, that is
%%%%
\begin{equation}
p(\mhhat | \mh) = \exp\left[-(\mhhat - \mh)^2/2\tau^2\right],
\end{equation}
%%%%
with a theoretical error of $\tau = 2
\gev$.\footnote{Alternatively we could take a linear,
  rather than Gaussian distribution, which would be much more
  conservative.}
Our likelihood is defined as a convolution of the two
functions\cite{deAustri:2006pe}, 
%%%%
\begin{equation}
\label{eq:convolution}
\like(\mh) = \int p(d | \mhhat) \times p(\mhhat | \mh)\,\mathrm{d}\mhhat.
\end{equation}
%%%%
We choose to add the experimental and theoretical errors in quadrature, finally obtaining
%%%%
\begin{equation}
\like_{\mhl\simeq125\gev}(\mhl) = \exp\left[-(125\gev - \mhl)^2/2(\tau^2+\sigma^2)\right].
\end{equation}

%%%%%%%%%%%%%%%%%%%%%%%%%%%%%%%%%%%%%%%%%%%%%%%%%%%%%%%%%%%%%%%%%%%%%%%%%%%%%%%%
\section{\label{sec:results}Results}
%%%%%%%%%%%%%%%%%%%%%%%%%%%%%%%%%%%%%%%%%%%%%%%%%%%%%%%%%%%%%%%%%%%%%%%%%%%%%%%%

%%%%%%%%%%%%%%%%%%%%%%%%%%%%   T   A   B   L   E   %%%%%%%%%%%%%%%%%%%%%%%%%%%%
% Table showing our prior ranges
\begin{table}[t]
%\input{./Tables/priors}
%%%%%%%%%%%%%%%%%%%%%%%%%%%%%%%%%%%%%%%%%%%%%%%%%%%%%%%%%%%%%%%%%%%%%%%%%%%%%%%%
\begin{tabular}{|l|l|l|l|}
\hline 
%%%%%%%%%%%%%%%%%%%%%%%%%%%%%%%%%%%%%%%%%%%%%%%%%%%%%%%%%%%%%%%%%%%%%%%%%%%%%%%%
CMSSM parameter & Description & Prior Range & Prior Distribution \\
%%%%%%%%%%%%%%%%%%%%%%%%%%%%%%%%%%%%%%%%%%%%%%%%%%%%%%%%%%%%%%%%%%%%%%%%%%%%%%%%
\hline %\midrule
%%%%%%%%%%%%%%%%%%%%%%%%%%%%%%%%%%%%%%%%%%%%%%%%%%%%%%%%%%%%%%%%%%%%%%%%%%%%%%%%
\mzero        	& Universal scalar mass          & 100, 4000 	& Log\\
\mhalf		& Universal gaugino mass         & 100, 2000 	& Log\\
\azero        	& Universal trilinear coupling   & -7000, 7000	& Linear\\
\tanb	        & Ratio of Higgs vevs            & 3, 62 	& Linear\\
\signmu		& Sign of Higgs parameter        & +1 or $-1$ 		& Fixed\footnote{The sign of parameter $\mu$ is fixed for a given scan.}\\
%%%%%%%%%%%%%%%%%%%%%%%%%%%%%%%%%%%%%%%%%%%%%%%%%%%%%%%%%%%%%%%%%%%%%%%%%%%%%%%%
%\hline %\midrule
%\multicolumn{4}{|l|}{additionally in NUHM} \\
%\hline %\cmidrule(r){1-4}
%%%%%%%%%%%%%%%%%%%%%%%%%%%%%%%%%%%%%%%%%%%%%%%%%%%%%%%%%%%%%%%%%%%%%%%%%%%%%%%%
%\mhu        	& GUT-scale soft mass of $\hu$          & 100, 4000 	& Log\\
%\mhd		& GUT-scale soft mass of $\hd$          & 100, 4000 	& Log\\
%%%%%%%%%%%%%%%%%%%%%%%%%%%%%%%%%%%%%%%%%%%%%%%%%%%%%%%%%%%%%%%%%%%%%%%%%%%%%%%%
\hline 
%%%%%%%%%%%%%%%%%%%%%%%%%%%%%%%%%%%%%%%%%%%%%%%%%%%%%%%%%%%%%%%%%%%%%%%%%%%%%%%%
Nuisance & Description & Central value $\pm$ std. dev. & Prior Distribution \\
%%%%%%%%%%%%%%%%%%%%%%%%%%%%%%%%%%%%%%%%%%%%%%%%%%%%%%%%%%%%%%%%%%%%%%%%%%%%%%%%
\hline 
%%%%%%%%%%%%%%%%%%%%%%%%%%%%%%%%%%%%%%%%%%%%%%%%%%%%%%%%%%%%%%%%%%%%%%%%%%%%%%%%
\mt           	& Top quark pole mass 	& $172.9\pm1.1$ 	& Gaussian\\
\mbmbsmmsbar 	& Bottom quark mass	& $4.19\pm0.12$ 	& Gaussian\\
\alphasmzms	& Strong coupling	& $0.1184\pm0.0007$   	& Gaussian\\
1/\alphaemmz 	& Reciprocal of electromagnetic coupling  	& $127.916\pm 0.015$ 	& Gaussian\\
%%%%%%%%%%%%%%%%%%%%%%%%%%%%%%%%%%%%%%%%%%%%%%%%%%%%%%%%%%%%%%%%%%%%%%%%%%%%%%%%
\hline 
\end{tabular}
\caption{Priors for the parameters of the CMSSM  and for the SM nuisance
  parameters used in our scans. Masses and \azero\ are in
  GeV.}
\label{tab:priors}
\end{table}
%%%%%%%%%%%%%%%%%%%%%%%%%%%%%%%%%%%%%%%%%%%%%%%%%%%%%%%%%%%%%%%%%%%%%%%%%%%%%%%%

In this section we will present our numerical results. We scanned the
parameter space of the CMSSM over the ranges given in
\reftable{tab:priors}. Note that, compared to \refref{Fowlie:2011mb},
we doubled the ranges of \mzero\ and \mhalf, which are now the same as
in \refref{Roszkowski:2012uf}, and we enlarged the range of \azero\
from $(-2\tev,2\tev)$ to $(-7\tev,7\tev)$ in order to approach
$\mhl\sim125$ \gev. As before, we applied a log prior to the mass
parameters \mzero\ and \mhalf, and a linear one to \azero\ and \tanb.
We performed our scans for $\mu>0$ and $\mu<0$ separately, for each
case with and without the \gmtwo\ constraint.

In the current analysis we have improved our treatment of the SM nuisance parameters. In our
previous analyses, we sampled the nuisance parameters from finite
linear intervals (linear priors), and included Gaussian likelihood
functions that described their experimental measurements. In this
analysis, we sample the nuisance parameters directly from Gaussian
priors that describe their experimental measurements and do not include them into the likelihood function. This improves
our algorithm's efficiency and is a more intuitive method.

%%%%%%%%%%%%%%%%%%%%%%%%%%%%   T   A   B   L   E   %%%%%%%%%%%%%%%%%%%%%%%%%%%%
% Table showing all experimental constraints in our scans
\begin{table}[t]
%lr \input{./Tables/exp_constraints}
%%%%%%%%%%%%%%%%%%%%%%%%%%%%%%%%%%%%%%%%%%%%%%%%%%%%%%%%%%%%%%%%%%%%%%%%%%%%%%%%
\begin{tabular}{|l|l|l|l|l|l|}
\hline %\toprule
%%%%%%%%%%%%%%%%%%%%%%%%%%%%%%%%%%%%%%%%%%%%%%%%%%%%%%%%%%%%%%%%%%%%%%%%%%%%%%%%
Measurement & Mean or Range & Exp.~Error & Th.~Error & Likelihood Distribution & Ref.\\
%%%%%%%%%%%%%%%%%%%%%%%%%%%%%%%%%%%%%%%%%%%%%%%%%%%%%%%%%%%%%%%%%%%%%%%%%%%%%%%%
\hline %\cmidrule(r){1-6}
%lr \multicolumn{6}{|l|}{\razorexp} \\
%lr \hline %\cmidrule(r){1-6}
%%%%%%%%%%%%%%%%%%%%%%%%%%%%%%%%%%%%%%%%%%%%%%%%%%%%%%%%%%%%%%%%%%%%%%%%%%%%%%%%
\razorexp  	& See text 	& See text  	& $0$ 	& Poisson &\cite{CMS-PAS-SUS-12-005}\\ 
%%%%%%%%%%%%%%%%%%%%%%%%%%%%%%%%%%%%%%%%%%%%%%%%%%%%%%%%%%%%%%%%%%%%%%%%%%%%%%%%
%lr \hline %\cmidrule(r){1-6}
%lr \multicolumn{6}{|l|}{Non-LHC} \\
%lr \hline %\cmidrule(r){1-6}
%%%%%%%%%%%%%%%%%%%%%%%%%%%%%%%%%%%%%%%%%%%%%%%%%%%%%%%%%%%%%%%%%%%%%%%%%%%%%%%%
SM-like Higgs mass $\mhl$	& $125$ 	& $2$   	& $2$
& Gaussian &
\cite{CMS-PAS-HIG-12-008,ATLAS-CONF-2012-019,Heinemeyer:2011aa}
\\
\abundchi 			& $0.1120$ 	& $0.0056$  	& $10\%$ 		& Gaussian &  \cite{Komatsu:2010fb}\\
\sinsqeff 			& $0.23116$     & $0.00013$ 	& $0.00015$             & Gaussian &  \cite{Nakamura:2010zzi}\\
\mw                     	& $80.399$      & $0.023$   	& $0.015$               & Gaussian &  \cite{Nakamura:2010zzi}\\
\deltagmtwomususy $\times 10^{10}$ 	& $28.7 $  	& $8.0$ 	& $1.0$ 		& Gaussian &  \cite{Nakamura:2010zzi,Miller:2007kk} \\
\brbxsgamma $\times 10^{4}$ 		& $3.60$   	& $0.23$ 	& $0.21$ 		& Gaussian &  \cite{Nakamura:2010zzi}\\
\brbutaunu $\times 10^{4}$          & $1.66$  	& $0.66$ 	& $0.38$ 		& Gaussian &  \cite{Asner:2010qj}\\
\delmbs                         & $17.77$ 	& $0.12$ 	& $2.40$  		& Gaussian &  \cite{Nakamura:2010zzi}\\
\brbsmumu			& $< 4.5 \times 10^{-9}$  	& $0$ & $14\%$    	& Upper limit -- Error Fn &  \cite{Aaij:2012ac}\\
%%%%%%%%%%%%%%%%%%%%%%%%%%%%%%%%%%%%%%%%%%%%%%%%%%%%%%%%%%%%%%%%%%%%%%%%%%%%%%%%
\hline %\bottomrule
\end{tabular}\caption{The experimental measurements that we apply to
  constrain the CMSSM's parameters. Masses are
  in GeV. 
%lr The numbers in parenthesis in the list of LEP and Tevatron experimental
%lr   measurements are weaker experimental bounds, which we use for
%some   sparticle mass hierarchies.
} 
\label{tab:exp_constraints}
\end{table}
%%%%%%%%%%%%%%%%%%%%%%%%%%%%%%%%%%%%%%%%%%%%%%%%%%%%%%%%%%%%%%%%%%%%%%%%%%%%%%%%
The experimental constraints applied in our scans are listed in
\reftable{tab:exp_constraints}. In comparison with our previous papers
\refref{Fowlie:2011mb,Roszkowski:2012uf}, the new upper limit on
\brbsmumu\ is used, which is evidently more constraining than the old
one.  Note also that LEP and Tevatron limits on the Higgs sector and
superpartner masses are not listed in \reftable{tab:exp_constraints}
because the subsequent LHC limits were generally stronger, and in any
case in this paper we consider only the case of the Higgs signal. The
\razor\ and Higgs limits are included as described in
Sec.~\ref{sec:method}.

In Ref.\cite{Roszkowski:2012uf} we showed that the
effect of the current limits from FermiLAT and XENON100 strongly
depends on a proper treatment of astrophysical uncertainties. If the
uncertainties are treated in a conservative way, both direct and indirect limits from DM
searches are not more constraining than the accelerator ones,
hence we ignore them in the present analysis.

We have developed a new numerical code, BayesFITS, similar in spirit
to the MasterCode\cite{mastercode} and Fittino\cite{fittino}
frameworks (which perform frequentist analyses), and to
SuperBayeS\cite{SuperBayeS} and PySUSY\footnote{Written by Andrew
  Fowlie, public release forthcoming, see
  \url{http://www.hepforge.org/projects}.} (which perform Bayesian
analyses). BayesFITS engages several external, publicly available
packages: for sampling it uses MultiNest\cite{Feroz:2008xx} with 4000
live points, evidence tolerance factor set to 0.5, and sampling
efficiency equal to 0.8.  The mass spectrum is computed with
\softsusy\ and written in the form of SUSY Les Houches Accord files,
which are then taken as input files to compute various observables. We
use \superiso\ v3.2\cite{superiso} to calculate \brbxsgamma,
\brbsmumu, \brbutaunu, and \deltagmtwomususy, and \feynhiggs\
2.8.6\cite{feynhiggs:00} to calculate the electroweak variables \mw,
\sinsqeff, and \delmbs.  The DM observables, such as the relic density
and direct detection cross sections, are calculated with \micromegas\
2.4.5\cite{micromegas}.

Below we will present the results of our scans as one-dimensional (1D)
or two-dimensional (2D) marginalized posterior pdf maps
of parameters and observables. In evaluating the
posterior pdf's, we marginalize over the given SUSY model's other parameters and
the SM's nuisance parameters, as mentioned above and described in detail in Refs.\cite{Fowlie:2011mb,Roszkowski:2012uf}.

%%%%%%%%%%%%%%%%%%%%%%%%%%%%%%%%%%%%%%%%%%%%%%%%%%%%%%%%%%%%%%%%%%%%%%%%%%%%%%%%
\subsection{\label{sec:cmssmresults}The CMSSM with \gmtwo}
%%%%%%%%%%%%%%%%%%%%%%%%%%%%%%%%%%%%%%%%%%%%%%%%%%%%%%%%%%%%%%%%%%%%%%%%%%%%%%%%

\begin{figure}[t]
\centering
\subfloat[]{%
\label{fig:Comparison_m0m12_lhc}%
\includegraphics[width=0.39\textwidth]{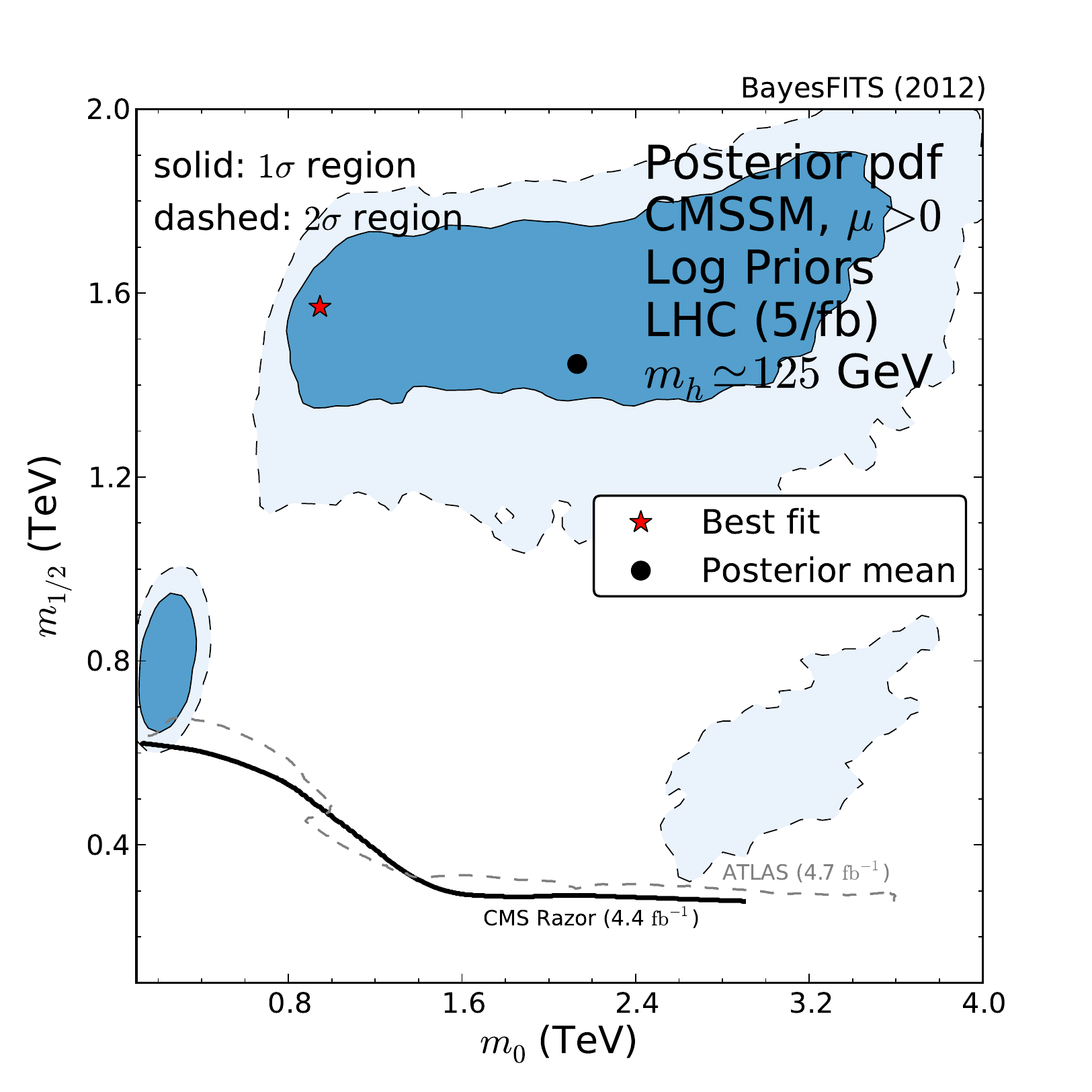}
}%
\hspace{1pt}%
\subfloat[]{%
\label{fig:Comparison_a0tanb_lhc}%
\includegraphics[width=0.39\textwidth]{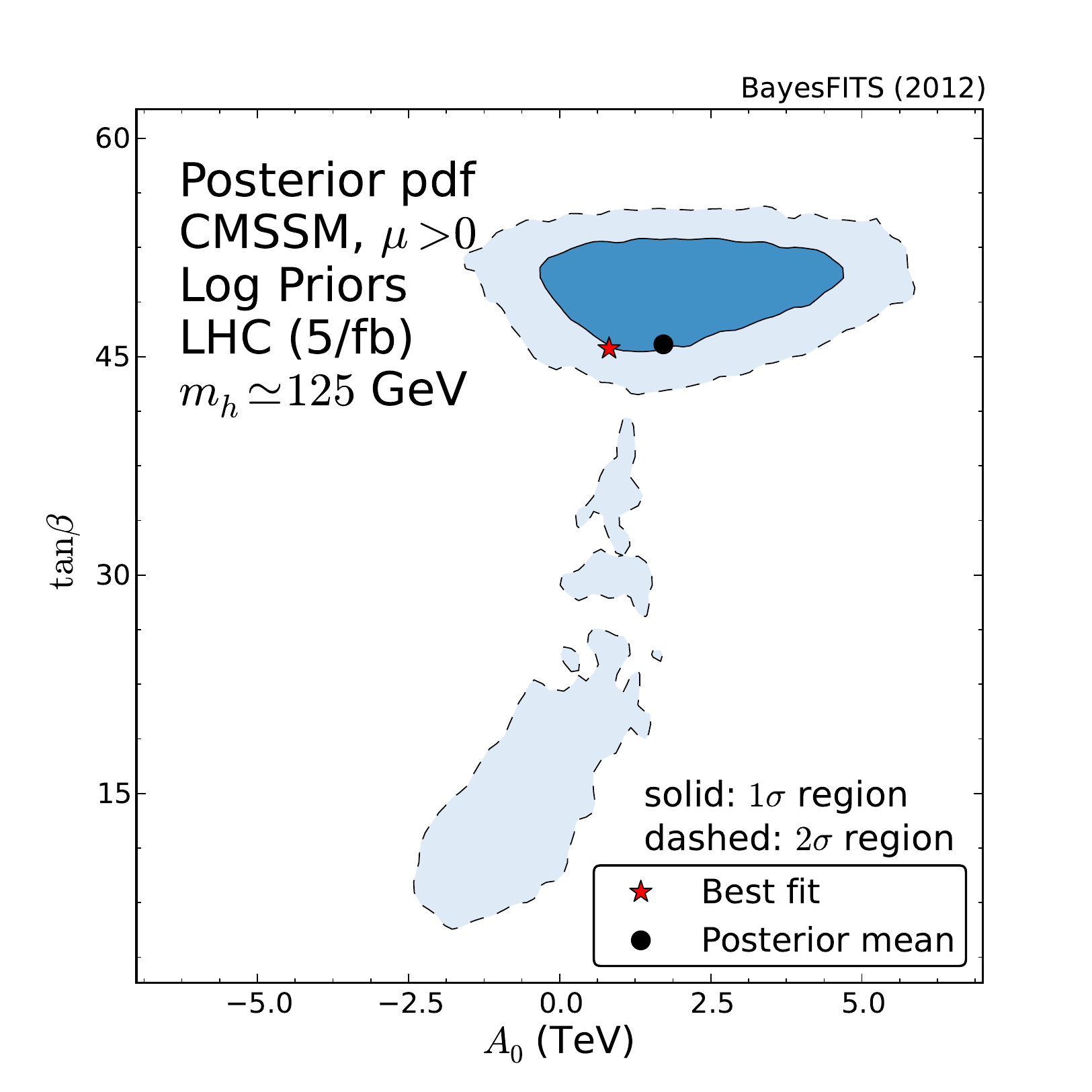}
}
\caption[]{Marginalized posterior pdf in \subref{fig:Comparison_m0m12_lhc} the (\mzero, \mhalf)
  plane and \subref{fig:Comparison_a0tanb_lhc} the (\azero, \tanb) plane of the CMSSM, constrained by the
  experiments listed in Table~\ref{tab:exp_constraints}. The solid black line shows the CMS razor 95\%~CL exclusion bound.
}
\label{fig:cmssm_params_lhc_mhl125}
\end{figure}

In
Figs.~\ref{fig:cmssm_params_lhc_mhl125}\subref{fig:Comparison_m0m12_lhc}
and~\ref{fig:cmssm_params_lhc_mhl125}\subref{fig:Comparison_a0tanb_lhc}
we show the marginalized posterior pdf in the (\mzero, \mhalf) plane
and in the (\azero, \tanb) plane, respectively. In these and the
following plots we show the Bayesian 68.3\% ($1\sigma$) credible
regions in dark blue, encircled by solid contours, and the 95\%
($2\sigma$) credible regions in light blue, encircled by dashed
contours.

The posterior presented in
Fig.~\ref{fig:cmssm_params_lhc_mhl125}\subref{fig:Comparison_m0m12_lhc}
features a bimodal behavior, with two well-defined $1\sigma$ credible
regions. One mode, smaller in size, which is located at small \mzero, is the \stau-coannihilation
region, whereas a much more extended mode lies in the
$A$-funnel region.  Although the bimodal behavior is superficially
similar to what was already observed in Ref.\cite{Fowlie:2011mb}, there are
substantial differences. Most notably, the high probability mode
which, in that paper and in Ref.\cite{Roszkowski:2012uf}, was spread over
the focus point (FP)/hyperbolic branch (HB) region at large \mzero\
and $\mhalf\ll\mzero$, has now moved up
to the $A$-funnel region.

The reason for the different behavior of the posterior with respect to
Ref.\cite{Fowlie:2011mb} is twofold.  On the one hand, we have found
that the highest density of points with the right Higgs mass can be
found at $\mhalf\gsim 1\tev$, which moves the posterior credible
regions up in the plane. On the other hand, some points with a large
\mhl\ can also be found in the FP/HB region but the scan tends to
ignore them in favor of points in the $A$-funnel region over which the
$b$-physics constraints are better satisfied.  The new upper bound on
\brbsmumu\ from LHCb also yields a substantial contribution.  The
approximately rectangular region bounded by $\mzero\sim 500-2000\gev$
and $m_{1/2}\sim 400-1000\gev$ is now cut out at the 95\%~CL. Notice
that in our previous papers\cite{Fowlie:2011mb,Roszkowski:2012uf} the
same part of parameter space was included in the 95\%~credible region.

%%%%%%%%%%%%%%%%%%%%%%%%%   F   I   G   U   R   E   %%%%%%%%%%%%%%%%%%%%%%%%%%%%
% 2 by 1 plot of nonLHC+lhc 2d pdf m0-m12 and  a0-tanbeta
%%%%%%%%%%%%%%%%%%%%%%%%%%%%%%%%%%%%%%%%%%%%%%%%%%%%%%%%%%%%%%%%%%%%%%%%%%%%%%%%

The new 4.4/fb razor exclusion bound reduces the size of the $1\sigma$
credible region of \stau-coannihilation at small \mzero, with respect
to what was observed in our previous
analyses\cite{Fowlie:2011mb,Roszkowski:2012uf}, where we used the
1.1/fb $\alpha_T$ likelihood.  The razor constraint also excludes more
of the FP/HB region.  We point out here that the improved exclusion
bound on \mhalf\ in the \stau-coannihilation region is mostly due to
the increased luminosity, while in the FP/HB region to switching from
the $\alpha_T$ to the razor search.  The razor bound with 0.8/fb
luminosity\cite{CMS-PAS-SUS-11-008} was better than the $\alpha_T$
bound in the FP/HB region, but worse in the \stau-coannihilation
region, where the improvement due to luminosity is more dramatic. As a
matter of fact, in the \stau-coannihilation the dominant cross section
is $pp\rightarrow\tilde{q}\tilde{q}$, while in the FP/HB region it is
$pp\rightarrow\tilde{g}\tilde{g}$. $M_R$ is in all effect an estimate
of the difference $m_{\tilde{g}(\tilde{q})}-m_{\chi}$. Since in the
CMSSM the gluino and LSP masses are correlated, the sensitivity in the
FP/HP region does not increase with luminosity as fast as in the
region at small \mzero. Finally, we note that, in
this case, the best-fit point is located on the left-hand side of the
$\ha$-funnel region. We postpone further discussion of \chisqmin\ and
the stability of the location of the best-fit point until
Sec.~\ref{sec:statistics}.

%%%%%%%%%%%%%%%%%%%%%%%%%   F   I   G   U   R   E   %%%%%%%%%%%%%%%%%%%%%%%%%%%%
% Plots in 2 by 2 subfigure environment
% msquar, mgluino, mchi, mchar1 1d plots
% non-LHC + razor 4.4/fb
\begin{figure}[p]
\centering
\subfloat[]{%
\label{fig:4masses-a}% squark
\includegraphics[width=0.37\textwidth]{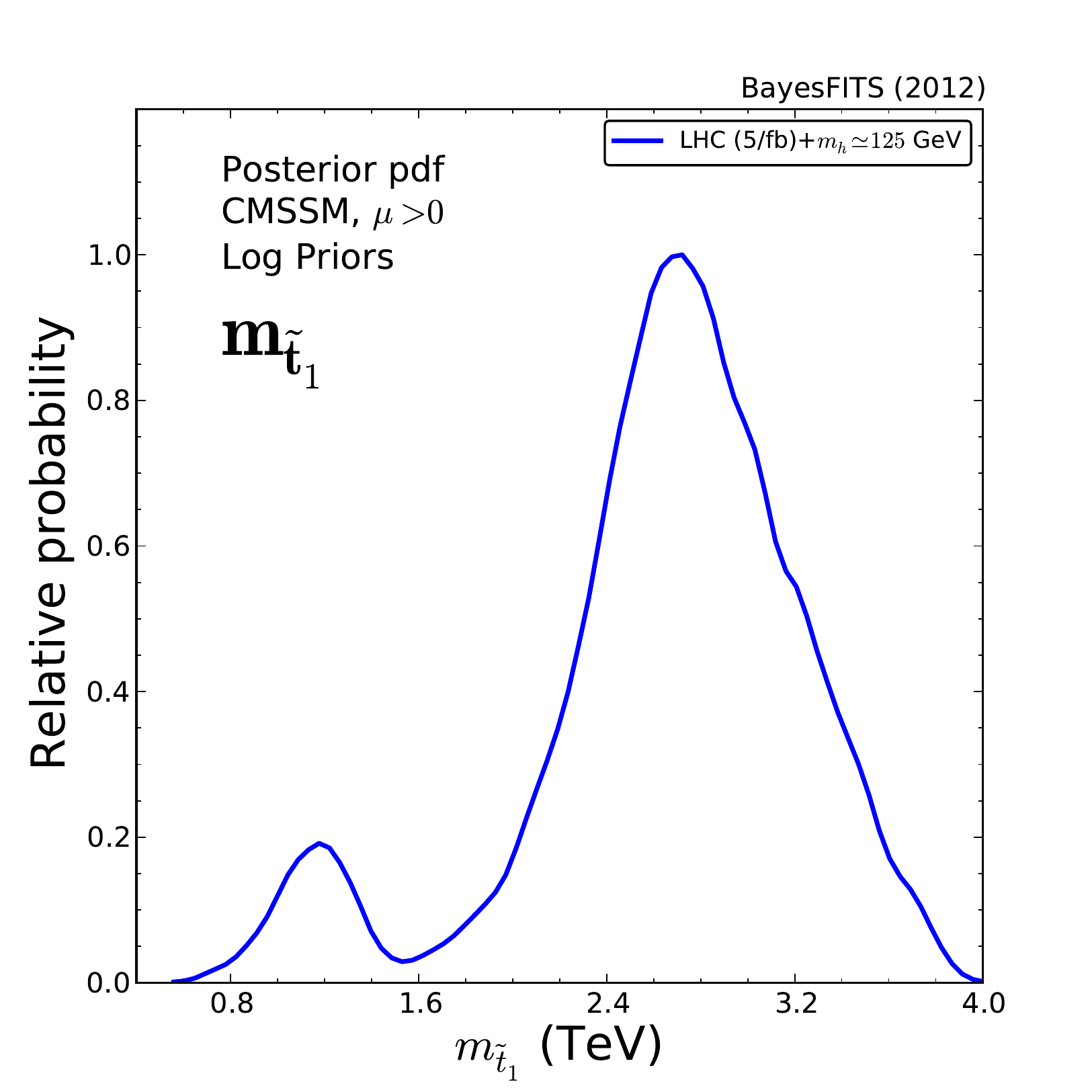}
}%
\hspace{1pt}%
\subfloat[]{%
\label{fig:4masses-b}% gluino
\includegraphics[width=0.37\textwidth]{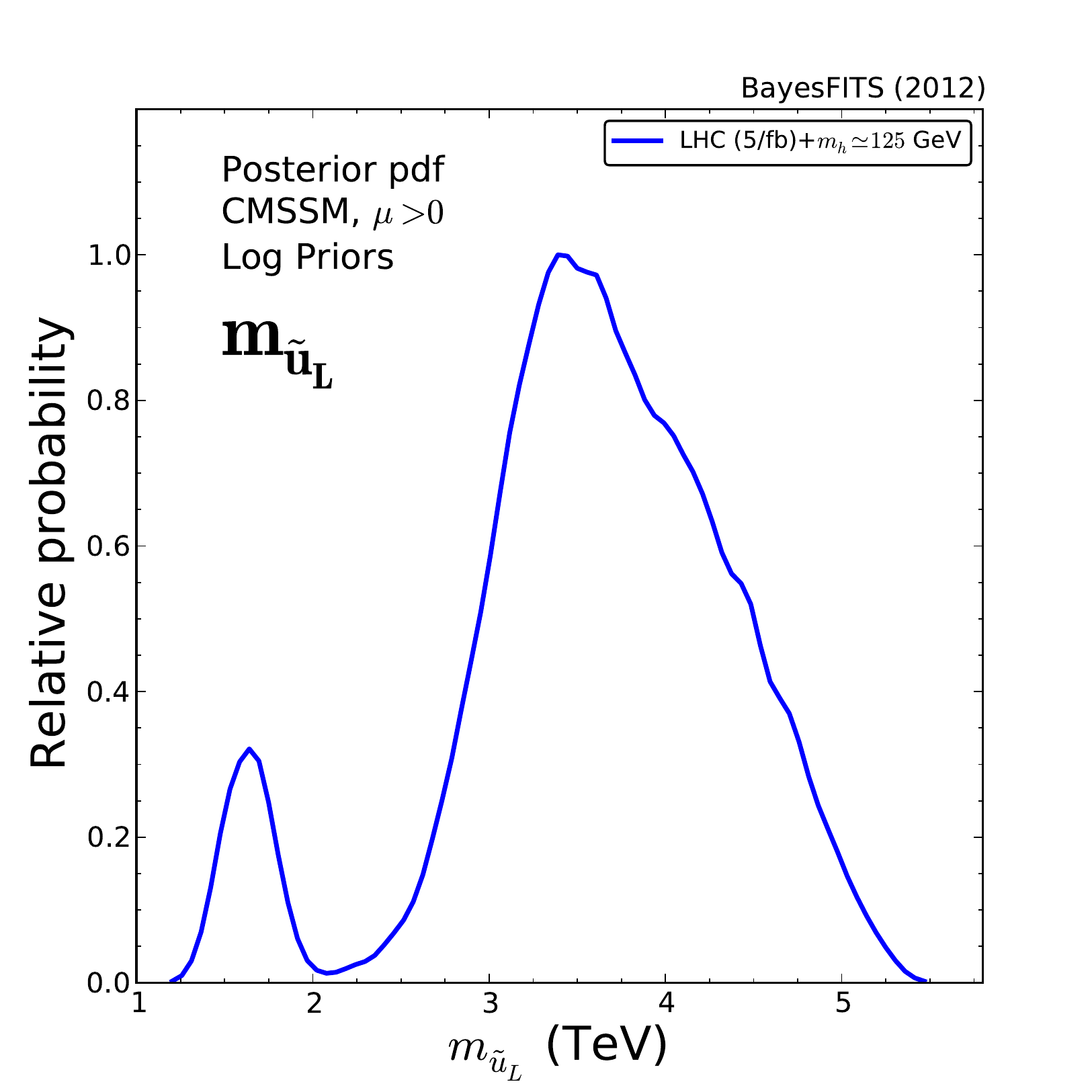}
}\\
\subfloat[]{%
\label{fig:4masses-c}% neut1
\includegraphics[width=0.37\textwidth]{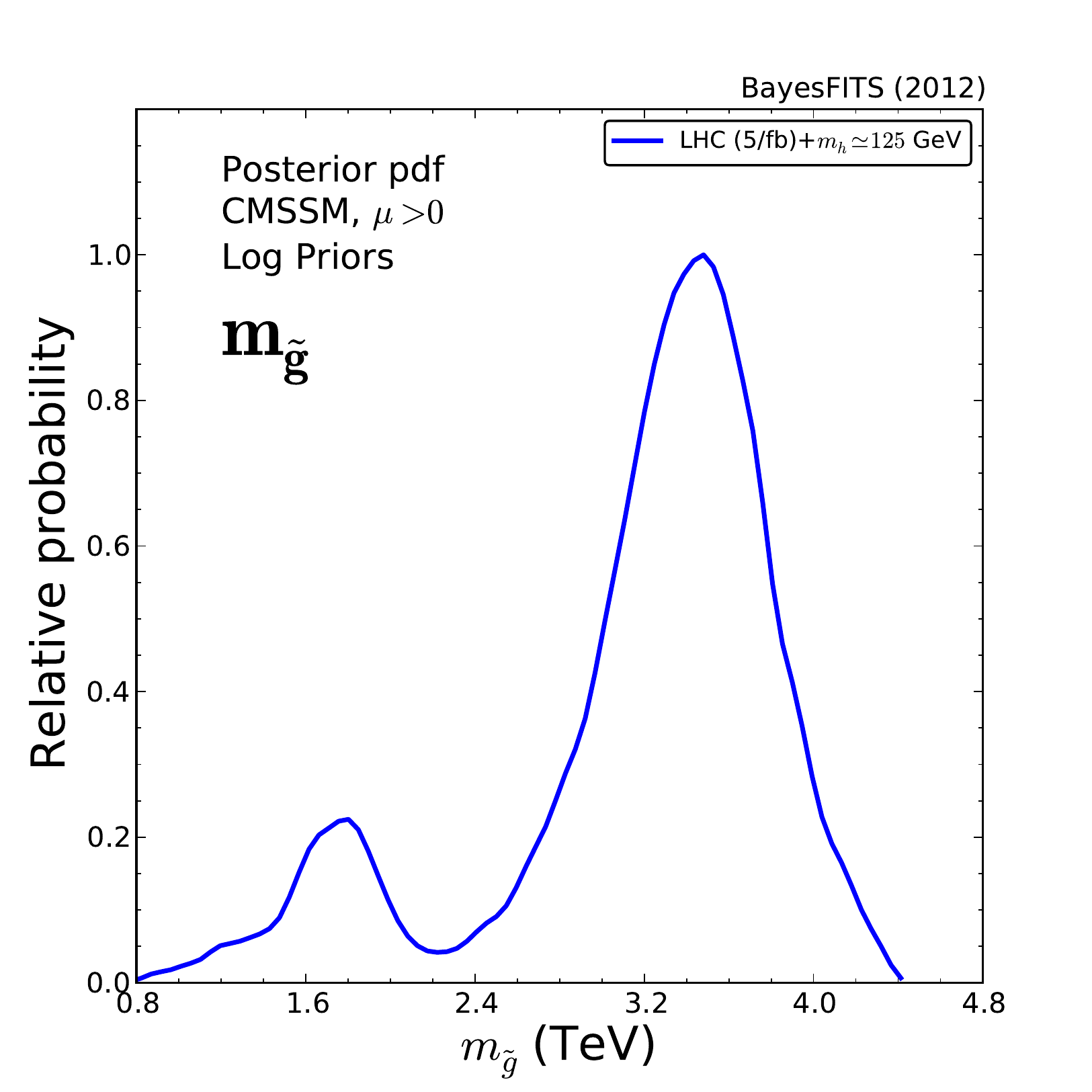}
}%
\hspace{1pt}%
\subfloat[]{%
\label{fig:4masses-d}% char1
\includegraphics[width=0.37\textwidth]{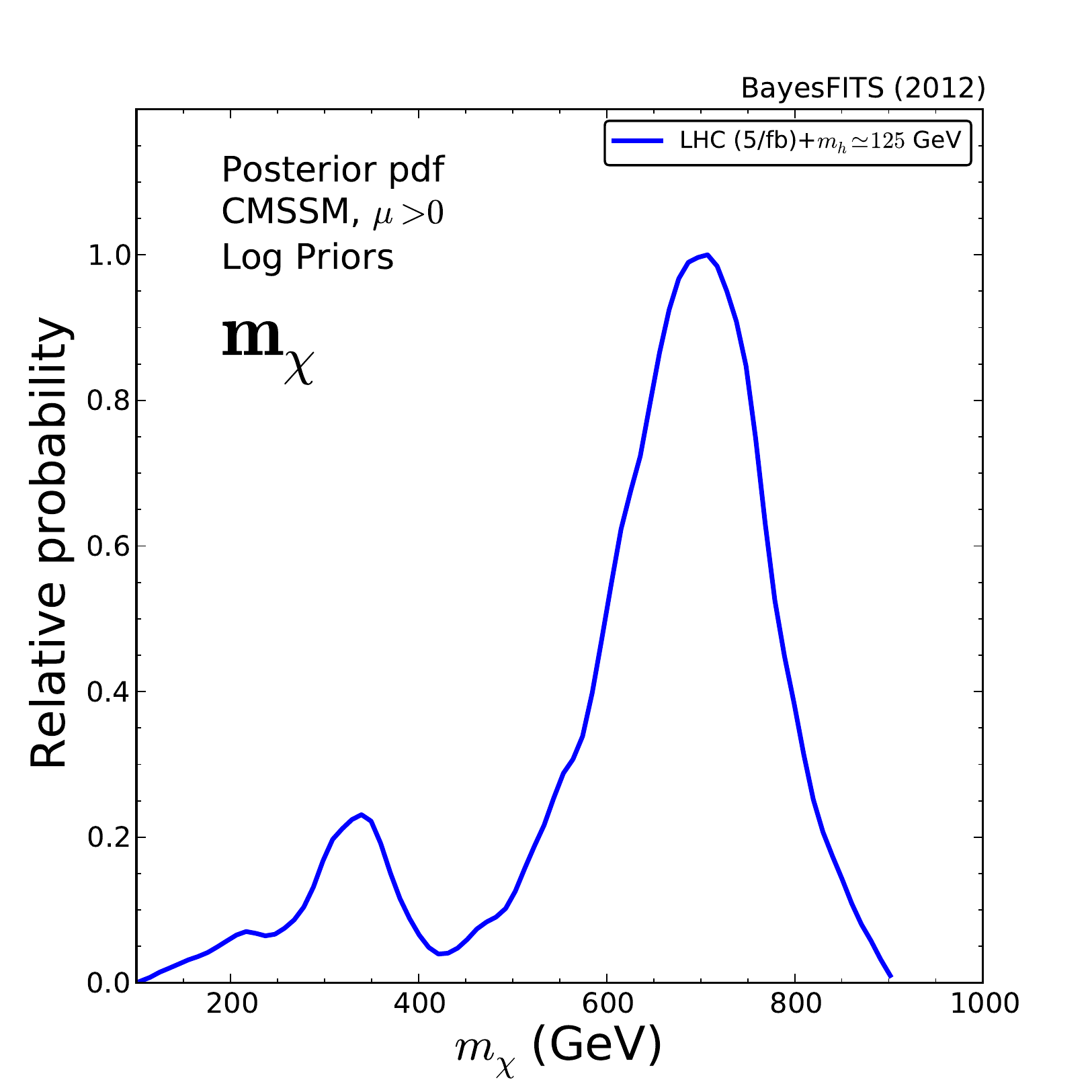}
}%
\caption[]{One-dimensional marginalized posterior pdf 
for the mass of: \subref{fig:4masses-a} the lightest stop, \subref{fig:4masses-b} 
the ${\widetilde u}_L$ squark, \subref{fig:4masses-c} the gluino, and
\subref{fig:4masses-d} the lightest neutralino in the CMSSM
 constrained by the experiments listed in Table~\ref{tab:exp_constraints}. }%
\label{fig:cmssm_4masses}%
\end{figure}
%%%%%%%%%%%%%%%%%%%%%%%%%%%%%%%%%%%%%%%%%%%%%%%%%%%%%%%%%%%%%%%%%%%%%%%%%%%%%%%%

%%%%%%%%%%%%%%%%%%%%%%%%%   F   I   G   U   R   E   %%%%%%%%%%%%%%%%%%%%%%%%%%%%
% 
\begin{figure}[p]
\centering
\includegraphics[width=0.49\textwidth]{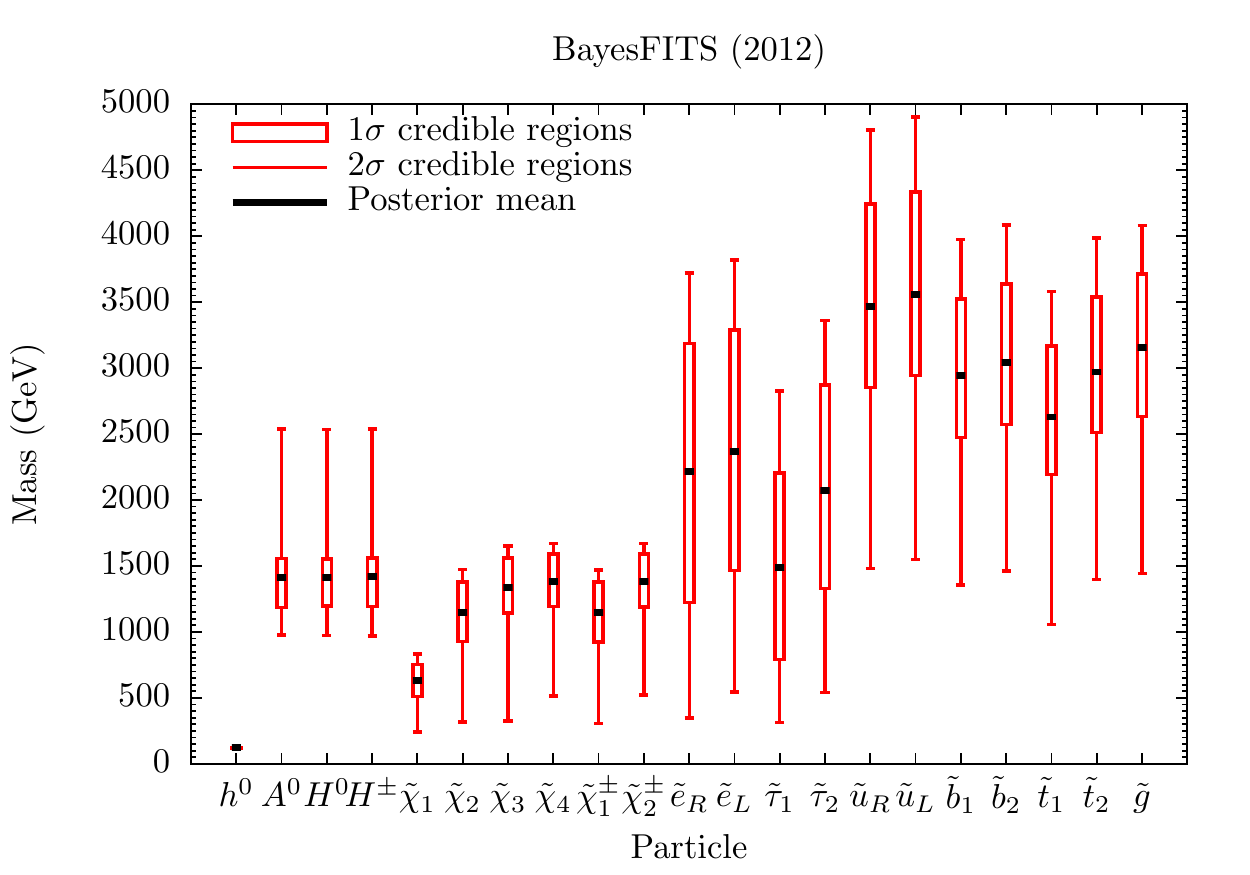}
\caption[]{One-dimensional marginalized posterior pdf for the supersymmetric spectrum
 constrained by the experiments listed in
 Table~\ref{tab:exp_constraints}.}
\label{fig:spectra}
\end{figure} 
%%%%%%%%%%%%%%%%%%%%%%%%%%%%%%%%%%%%%%%%%%%%%%%%%%%%%%%%%%%%%%%%%%%%%%%%%%%%%%%%

A similar bimodal behavior of the marginalized posterior can be
observed in
Fig.~\ref{fig:cmssm_params_lhc_mhl125}\subref{fig:Comparison_a0tanb_lhc}.
The large $1\sigma$ credible region at $\tanb\sim 45-55$ corresponds
to the large $1\sigma$ region in the $A$-funnel of the (\mzero,
\mhalf) plane. Conversely, the $2\sigma$ region at $\azero\sim 0$ and
$\tanb\lesssim 30$ can be mapped back to the \stau-coannihilation
region of the (\mzero, \mhalf) plane.
In Refs.\cite{Fowlie:2011mb,Roszkowski:2012uf} we could observe a wide
1$\sigma$ credible region at intermediate $\tanb$, whose statistical
relevance has now decreased. It corresponds to the FP/HB region of the
(\mzero, \mhalf) plane, now disfavored by the new LHC constraints on
the Higgs mass.

Since \brbsmumu\ is proportional to $\tan^6\beta/m_A^4$, one could
have naively expected to see small values of \tanb\ favored by the new
upper bound from LHCb. As we can see in
Fig.~\ref{fig:cmssm_params_lhc_mhl125}\subref{fig:Comparison_a0tanb_lhc},
this is actually not the case. This is because some other
constraints favor large \tanb. One is \deltagmtwomususy, even though
at the end it is poorly satisfied. The other is a combination of the
relic density favoring also larger \mha\ with the light Higgs mass
close to 125\gev, both of which can be more easily achieved at large
\mhalf.  The end result is that $m_A$ is now required to be larger
than in the past, which is consistent with the observed
prevalence of the $A$-funnel region over the FP/HB region.

The 1D relative marginalized posteriors for the masses of selected superpartners are shown
in the four panels of Fig.~\ref{fig:cmssm_4masses}.
In Fig.~\ref{fig:cmssm_4masses}\subref{fig:4masses-a} one can see the posterior for the lightest stop mass; in Fig.~\ref{fig:cmssm_4masses}\subref{fig:4masses-b} the one for the heaviest squark, $\tilde{u}_L$; Figs~\ref{fig:cmssm_4masses}\subref{fig:4masses-c} and~\ref{fig:cmssm_4masses}\subref{fig:4masses-d} show the gluino and lightest neutralino, respectively. As we mentioned in the previous section, the razor method will translate a lower bound on
$M_R$ and $R^2$ into a lower bound on squark and gluino masses. At small \mzero, where the cross section for $pp\rightarrow\tilde{q}\tilde{q}$ is dominant, this translates into
$m_{\tilde{t}_1}\gsim 800\gev$ and $m_{\tilde{u}_L}\gsim 1200\gev$; at large \mzero, where $pp\rightarrow\tilde{g}\tilde{g}$ dominates, the razor sets the limit $m_{\tilde{g}}\gsim 800\gev$. 

The highest peaks, indicating the values most favored by the present
constraints, are located at $\mtone\sim 2.5\tev$, $\mul\sim 3.2\tev$,
$\mglu\sim 3.2\tev$, and $\mchi\sim 700\gev$. The relative probability
of the peaks obtained in the $A$-funnel region is higher than the
probability of the peaks obtained in the \stau-coannihilation
region. We show in Fig.~\ref{fig:spectra} the one-dimensional
posteriors for all particles in the supersymmetric spectrum.

%%%%%%%%%%%%%%%%%%%%%%%%%   F   I   G   U   R   E   %%%%%%%%%%%%%%%%%%%%%%%%%%%%
% 
\begin{figure}[t]
\centering
\subfloat[]{%
\label{fig:Comparison_A0mh}%
\includegraphics[width=0.39\textwidth]{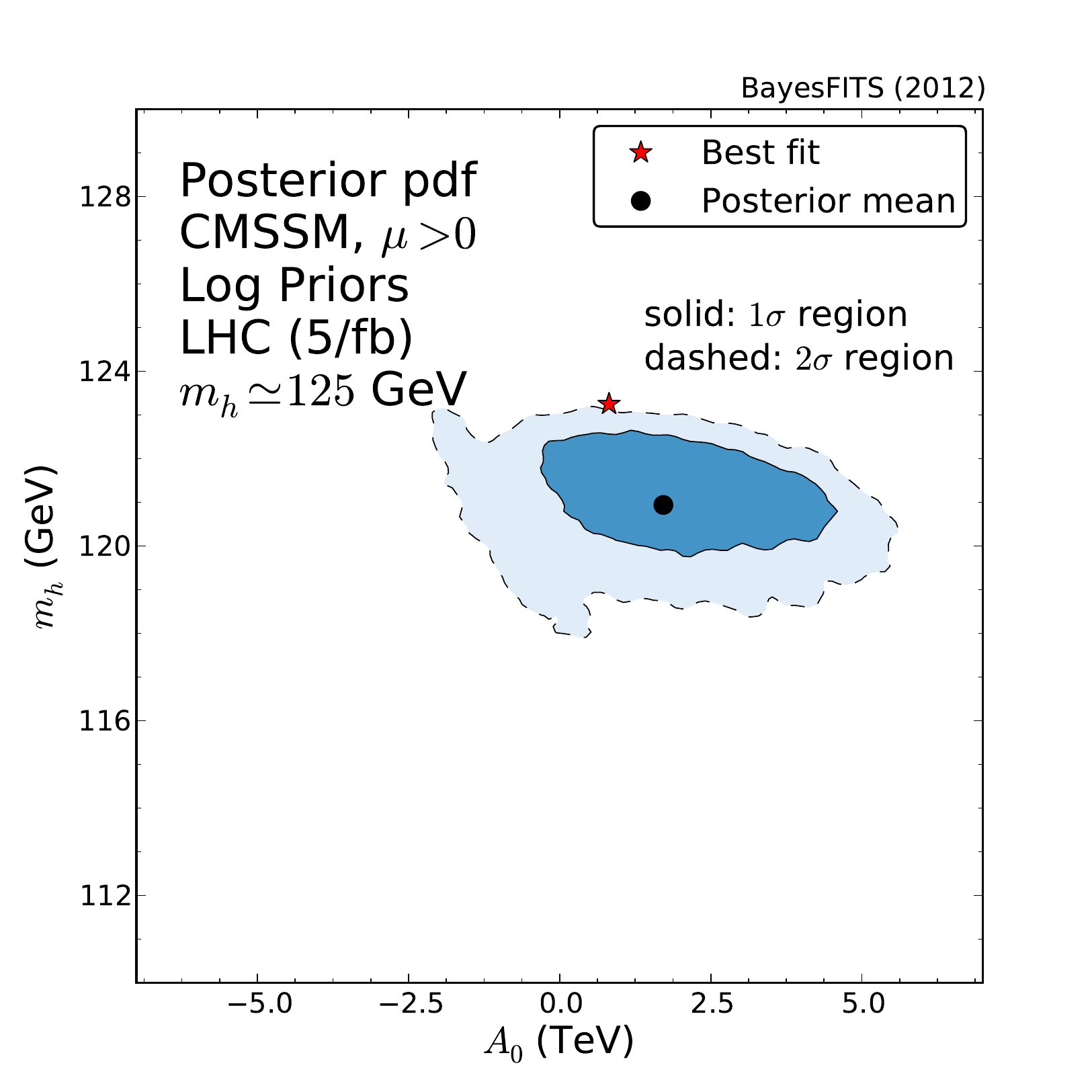}
}%
\hspace{1pt}%
\subfloat[]{%
\label{fig:Comparison_mAtanbeta}%
\includegraphics[width=0.39\textwidth]{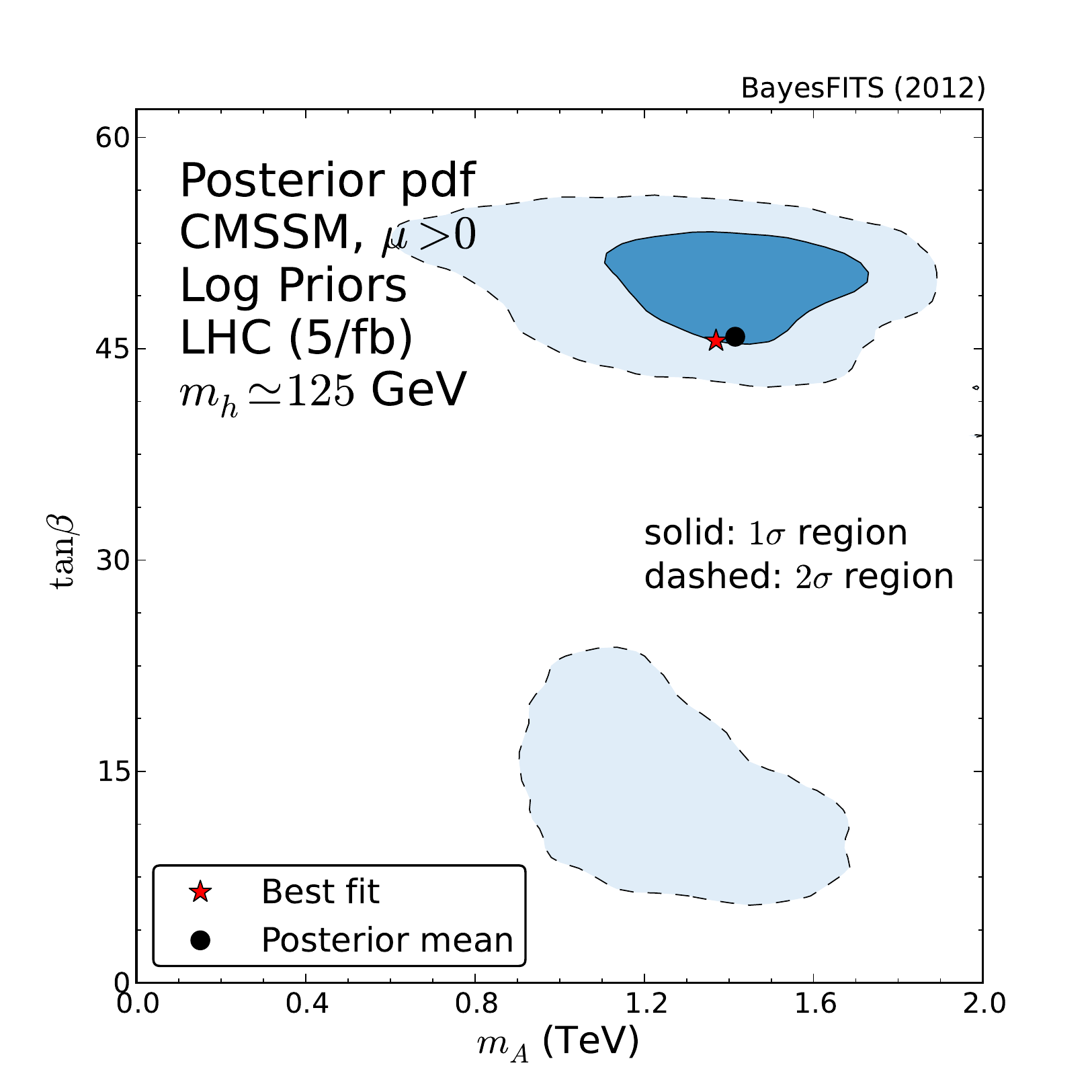}
}%
\caption[]{\subref{fig:Comparison_A0mh} Marginalized posterior pdf in the (\azero, \mhl) plane, in the CMSSM
 constrained by the experiments listed in
 Table~\ref{tab:exp_constraints}. \subref{fig:Comparison_mAtanbeta} Marginalized posterior pdf in the ($m_A$, \tanb) plane for the same constraints.
}%
\label{fig:cmssm_comparison_mAtanbeta}
\end{figure} 
%%%%%%%%%%%%%%%%%%%%%%%%%%%%%%%%%%%%%%%%%%%%%%%%%%%%%%%%%%%%%%%%%%%%%%%%%%%%%%%%

Figure~\ref{fig:cmssm_comparison_mAtanbeta}\subref{fig:Comparison_A0mh}
shows the two-dimensional posterior in the (\azero, \mhl) plane. It
presents an interesting behavior, not often pointed out in the
literature. Given the experimental and theoretical uncertainties on
the Higgs mass determination, the Bayesian fit to all constraints
favors positive values of \azero, although we confirm the known fact
that $\mhl>123\gev$ can be more easily obtained in the CMSSM only for
negative values of \azero.

Figure~\ref{fig:cmssm_comparison_mAtanbeta}\subref{fig:Comparison_mAtanbeta} shows the posterior in the ($m_A$, \tanb) plane. As mentioned above, the combined effect of the new Higgs constraints and \brbsmumu\ now favors larger values of both parameters. Notice that the ($m_A$, \tanb) range encompassed by the high posterior probability contours safely place the model in the decoupling regime (Sec.~\ref{subsec:higgs}) and thus justify our assumption of a SM-like Higgs.

%%%%%%%%%%%%%%%%%%%%%%%%%%%%%%%%%%%%%%%%%%%%%%%%%%%%%%%%%%%%%%%%%%%%%%%%%%%%%%%%
% Plots in 2 by 2 subfigure environment
% non-LHC + alphaT 1.1/fb + Xenon
% Various observables plotted against each other
\begin{figure}[b]
\centering
\subfloat[t][]{%
\label{fig:2d_g2_bsm-a}%
{\includegraphics[width=0.39\textwidth]{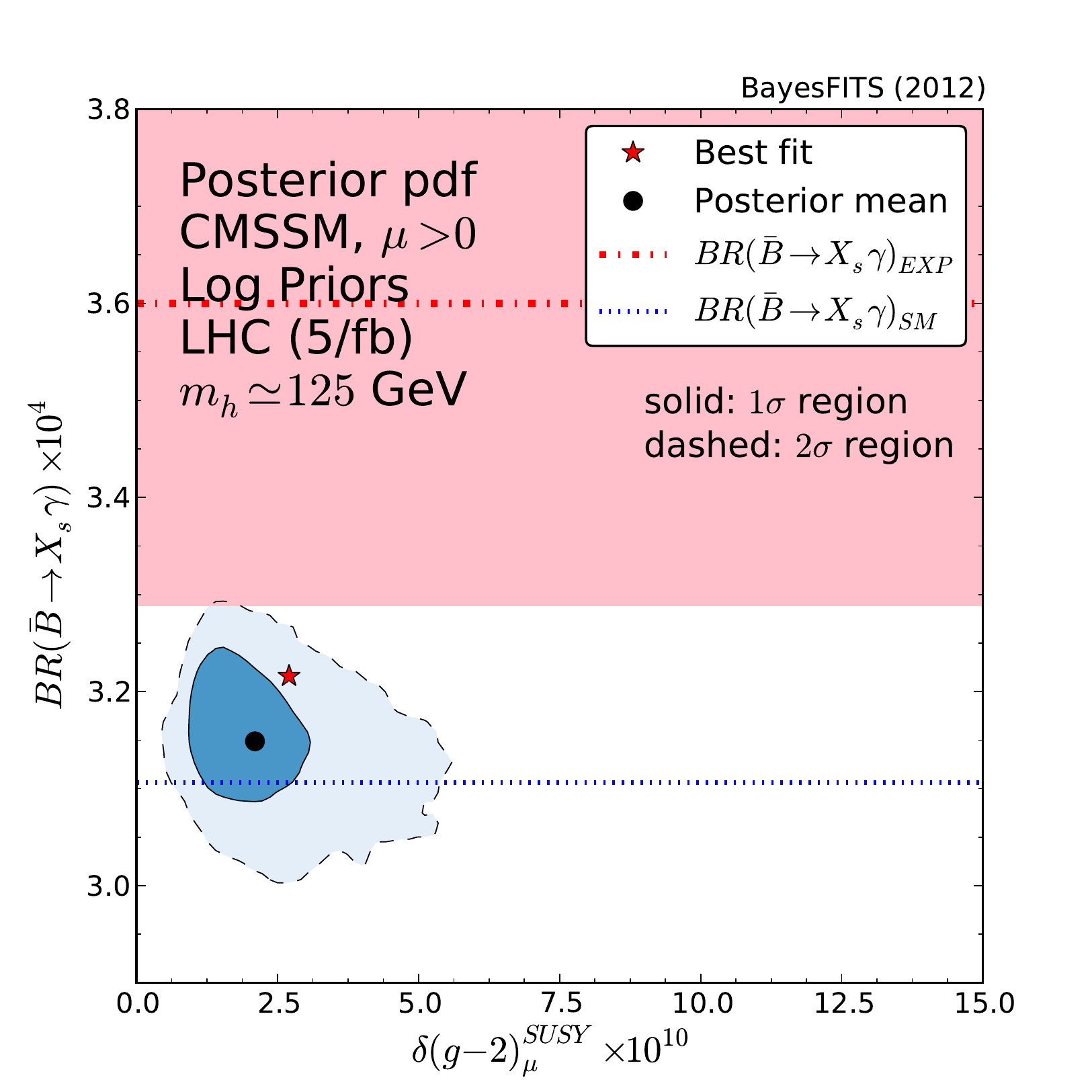}}
}%
%\hspace{1pt}%
%\subfloat[t][\brbsmumu\ against \brbxsgamma.]{%
%\label{fig:2d_bsg_g2_bsmm_oh2-b}%
%{\includegraphics[width=0.49\textwidth]{CMSSM-CONV_bsg_bsmumu.pdf}}
%}
\hspace{1pt}%
\subfloat[t][]{%
\label{fig:2d_g2_bsm-b}%
{\includegraphics[width=0.39\textwidth]{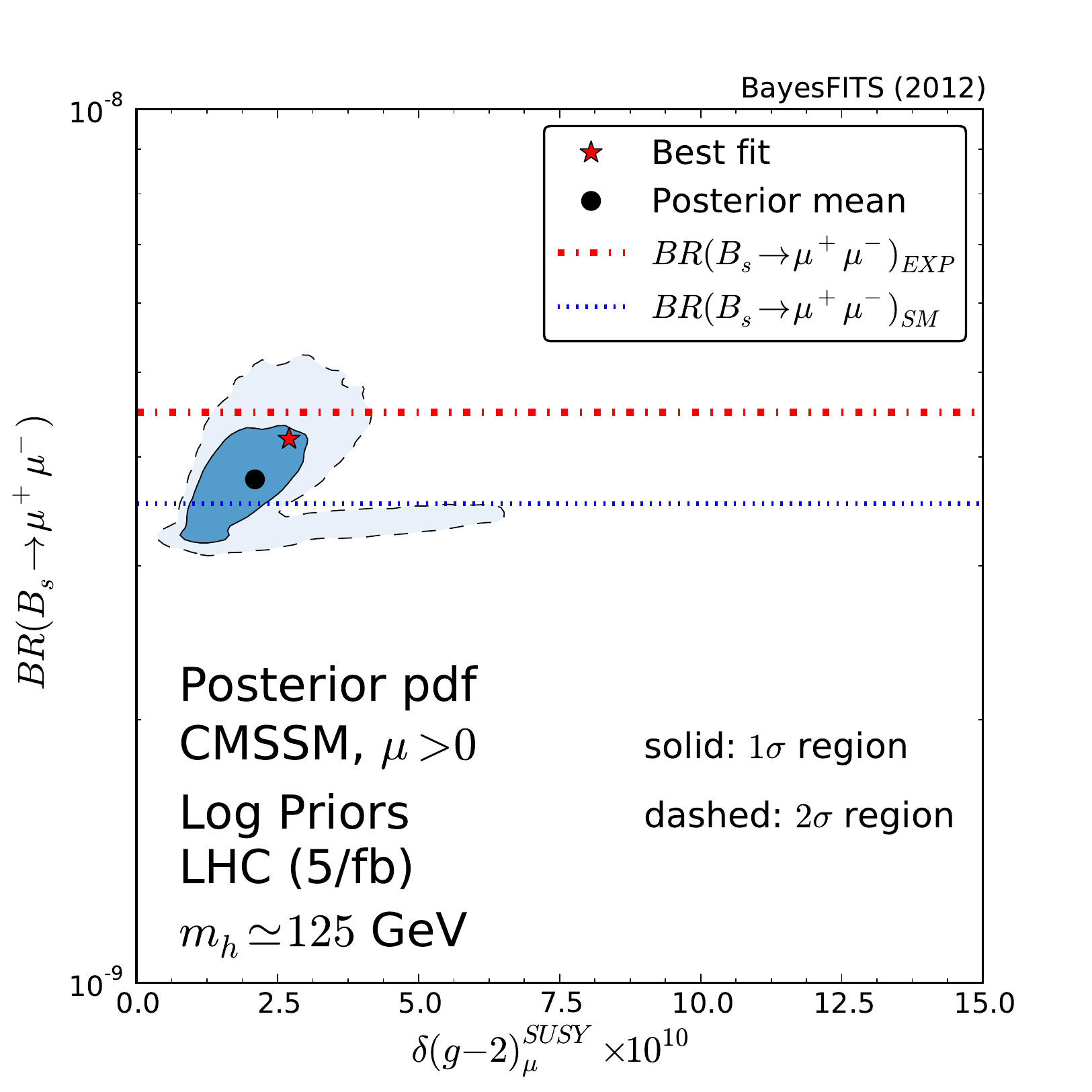}}
}
%\hspace{1pt}%
%\subfloat[t][\brbxsgamma\ against \abund.]{%
%\label{fig:2d_bsg_g2_bsmm_oh2-d}%
%\includegraphics[width=0.49\textwidth]{CMSSM-CONV_oh2_bsg.pdf}
%}\\%
\caption[]{\subref{fig:2d_g2_bsm-a} Marginalized posterior pdf of the
  experimental observables \deltagmtwomususy\ vs. \brbxsgamma\  in the
  CMSSM constrained by the experiments listed in Table~\ref{tab:exp_constraints}.
  \subref{fig:2d_g2_bsm-b} Marginalized posterior pdf of the
  experimental observables \deltagmtwomususy\ vs. \brbsmumu\ under the same constraints.
}%
\label{fig:cmssm_2d_g2_bsm}%
\end{figure}
%%%%%%%%%%%%%%%%%%%%%%%%%%%%%%%%%%%%%%%%%%%%%%%%%%%%%%%%%%%%%%%%%%%%%%%%%%%%%%%%

In Fig.~\ref{fig:cmssm_2d_g2_bsm}\subref{fig:2d_g2_bsm-a} we show the
2D posterior in the \deltagmtwomususy\ vs. \brbxsgamma\ plane for
$\mu>0$.  The \gmtwo\ constraint is applied. The red horizontal line
(dot-dashed) shows the experimental value of \brbxsgamma, and the pink
shaded region highlights the experimental uncertainties at
$1\sigma$. The blue horizontal line (dotted) shows the SM value, as
calculated by SuperISO.  One can see that the 68\% and 95\% Bayesian
credible regions are consistent with the experimental value of
\brbxsgamma\ at the $2\sigma$ level, while \deltagmtwomususy\ shows a
poor fit, as was noticed in many previous global scans of the CMSSM;
see, e.g.,
\cite{*deAustri:2006pe,*Akrami:2009hp,*Strege:2011pk,*Bechtle:2012zk,Buchmueller:2011sw,Buchmueller:2011ab}. In
particular, for $\mu>0$, a slightly better fit to \deltagmtwomususy\
is obtained in the \stau-coannihilation region, which implies values
of \brbxsgamma\ closer to the SM value, which lies $\sim 1.5\sigma$
away from the measured one\cite{Roszkowski:2007fd}.  On the other
hand, the best-fit point lies in the $A$-funnel region, where it is
harder to satisfy \gmtwo\ but easier to satisfy \brbxsgamma.

Figure~\ref{fig:cmssm_2d_g2_bsm}\subref{fig:2d_g2_bsm-b} shows that a
similar tension exists between the \brbsmumu\ and \gmtwo\
constraints. The red line (dot-dashed) shows the new LHCb 95\%~CL
upper bound, while the blue line (dotted) shows the SM value for
\brbsmumu\ that we use in our calculations. In an attempt to better
fit the \gmtwo\ constraint, a narrow 95\% credible region shows up
along the SM values of \brbsmumu, which lie in the
\stau-coannihilation region where \tanb\ is smaller. However, the
best-fit point is situated in the $A$-funnel region, where the \gmtwo\
constraint is overcome by the one due to \brbsmumu, which is now free to assume a
broader range of values.

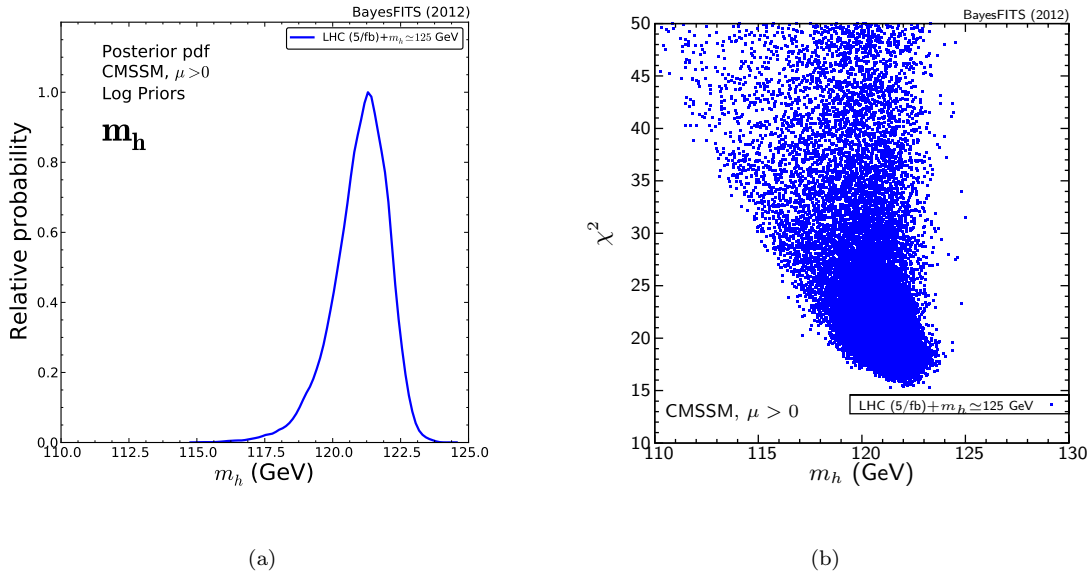
\begin{figure}[t]
\centering
\subfloat[]{%
\label{fig:-a}%
\raisebox{5.5mm}{\includegraphics[width=0.39\textwidth]{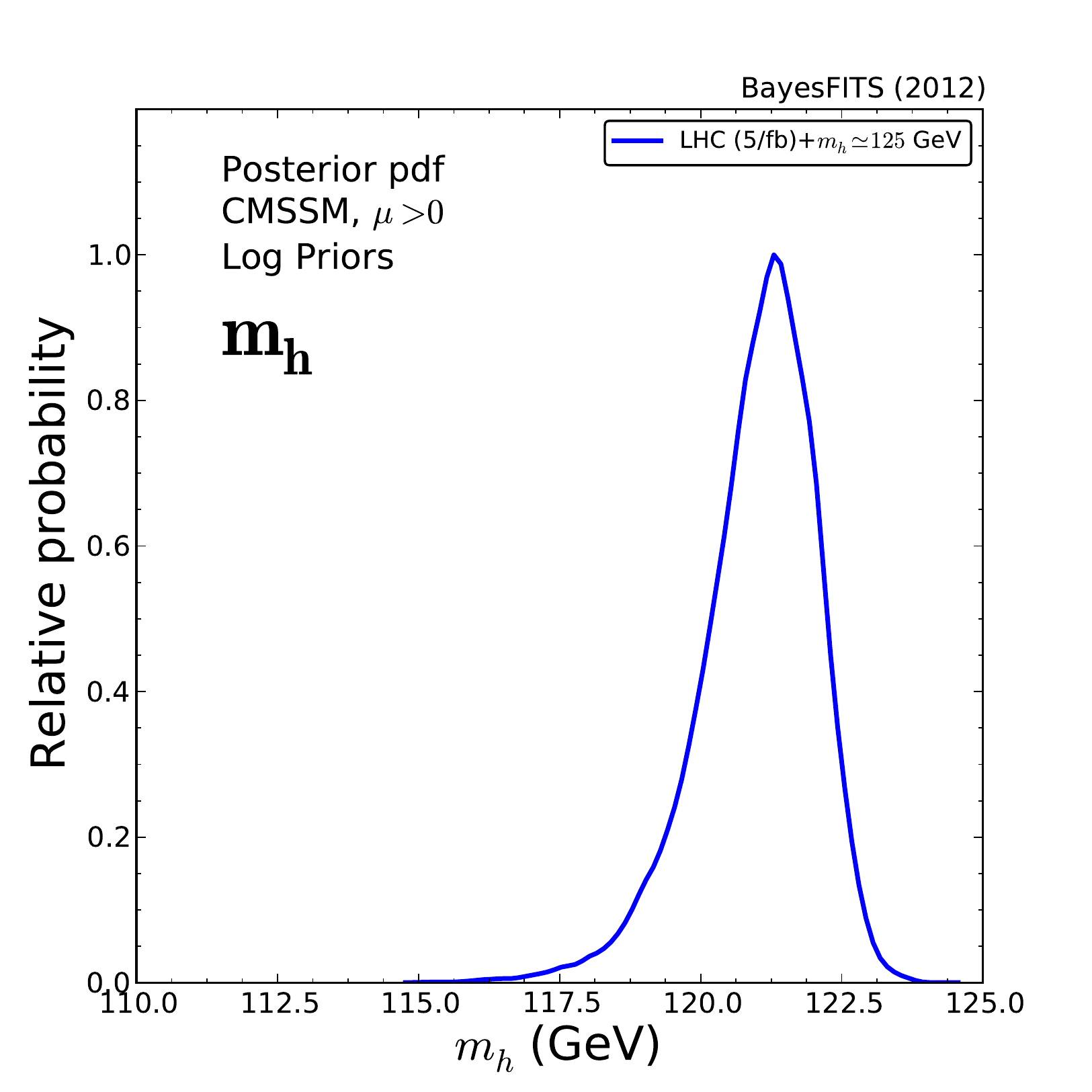}}
}%
%\hspace{1pt}%
\subfloat[]{%
\label{fig:-b}%
  \setlength{\unitlength}{0.0500bp}%
  \begin{picture}(4464.00,4132.80)%
%    \gplgaddtomacro\gplbacktext{%
      \csname LTb\endcsname%
      \put(914,704){\makebox(0,0)[r]{\strut{} \scriptsize{\textsf{10}}}}%
      \put(914,1100){\makebox(0,0)[r]{\strut{} \scriptsize{\textsf{15}}}}%
      \put(914,1495){\makebox(0,0)[r]{\strut{} \scriptsize{\textsf{20}}}}%
      \put(914,1891){\makebox(0,0)[r]{\strut{} \scriptsize{\textsf{25}}}}%
      \put(914,2286){\makebox(0,0)[r]{\strut{} \scriptsize{\textsf{30}}}}%
      \put(914,2682){\makebox(0,0)[r]{\strut{} \scriptsize{\textsf{35}}}}%
      \put(914,3077){\makebox(0,0)[r]{\strut{} \scriptsize{\textsf{40}}}}%
      \put(914,3473){\makebox(0,0)[r]{\strut{} \scriptsize{\textsf{45}}}}%
      \put(914,3868){\makebox(0,0)[r]{\strut{} \scriptsize{\textsf{50}}}}%
      \put(946,624){\makebox(0,0){\strut{} \scriptsize{\textsf{110}}}}%
      \put(1726,624){\makebox(0,0){\strut{} \scriptsize{\textsf{115}}}}%
      \put(2507,624){\makebox(0,0){\strut{} \scriptsize{\textsf{120}}}}%
      \put(3287,624){\makebox(0,0){\strut{} \scriptsize{\textsf{125}}}}%
      \put(4067,624){\makebox(0,0){\strut{} \scriptsize{\textsf{130}}}}%
      \put(550,2286){\rotatebox{-270}{\makebox(0,0){\strut{}$\chi^2$}}}%
      \put(2506,464){\makebox(0,0){\strut{}$m_h$ \textsf{(GeV)}}}%
      \put(3271,3947){\makebox(0,0)[l]{\strut{}\tiny{\textsf{BayesFITS (2012)}}}}%
      \put(1024,941){\makebox(0,0)[l]{\strut{}\scriptsize{\textsf{CMSSM}, $\mu>0$}}}%
%    }%
%    \gplgaddtomacro\gplfronttext{%
%      \csname LTb\endcsname%
     \put(3803,997){\makebox(0,0)[r]{\strut{}\tiny{\textsf{LHC (5/fb)$+m_h$$\simeq$125 \textsf{GeV}}}}}%
%      \csname LTb\endcsname%
%      \put(3803,854){\makebox(0,0)[r]{\strut{}\tiny{\textsf{LHC (5/fb)}}}}%
%    }%
%    \gplbacktext
    \put(0,0){\includegraphics{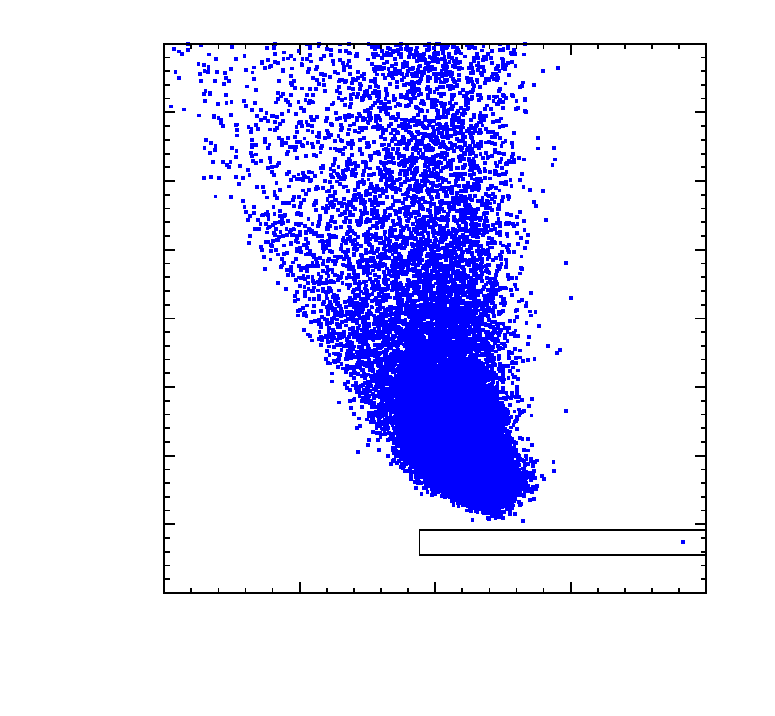}}%
%    \gplfronttext
  \end{picture}%
}%
\caption[]{\subref{fig:-a} Marginalized 1D posterior pdf of \mhl\ in the CMSSM constrained by the experiments listed in Table~\ref{tab:exp_constraints}. \subref{fig:-b} Scatter plot showing the distribution of the total $\chi^2$ of the points in our chain versus the Higgs mass.   
 }%
\label{fig:cmssm_mhchisq}
\end{figure}

The probability distribution of the lightest Higgs mass is shown in
Fig.~\ref{fig:cmssm_mhchisq}\subref{fig:-a}. The present constraints
highly favor Higgs masses centered around $\mhl\sim 122\gev$. Points
having $\mhl>124\gev$ are difficult to achieve in the CMSSM with the
prior ranges we consider ($\mzero\lesssim 4\tev$, $\mhalf\lesssim
2\tev$), as is well known. They are, nonetheless, present in our chain
in appreciable number but they are disfavored by the global
constraints. This point is made clear in
Fig.~\ref{fig:cmssm_mhchisq}\subref{fig:-b}, where we show a scatter
plot of the total $\chi^2$ versus the Higgs mass. Points giving Higgs
masses as large as 125\gev\ are generated, but their global fit to all
constraints is generally poor.

%%%%%%%%%%%%%%%%%%%%%%%%%   F   I   G   U   R   E   %%%%%%%%%%%%%%%%%%%%%%%%%%%%
% 2 by 1: left: plot of 1d pdf of mhl,
% right: chi2 vs mhl
%%%%%%%%%%%%%%%%%%%%%%%%%%%%%%%%%%%%%%%%%%%%%%%%%%%%%%%%%%%%%%%%%%%%%%%%%%%%%%%%

%%%%%%%%%%%%%%%%%%%%%%%%%   F   I   G   U   R   E   %%%%%%%%%%%%%%%%%%%%%%%%%%%%
% Plots in 2 by 2 subfigure environment
% 1D plots
\begin{figure}[t]
\centering
\subfloat[]{%
%\label{fig:-a}% bsgamma
%\includegraphics[width=0.49\textwidth]{CMSSM_1D_bsg.pdf}
%}%1, 0.9942, 0.9922, 0.9883
%\hspace{1pt}%
%\subfloat[]{%
\label{fig:-a}% g-2
  \setlength{\unitlength}{0.0500bp}%
  \begin{picture}(4464.00,4132.80)%
%    \gplgaddtomacro\gplbacktext{%
      \csname LTb\endcsname%
      \put(958,594){\makebox(0,0)[r]{\strut{} \scriptsize{\textsf{0}}}}%
      \put(958,1224){\makebox(0,0)[r]{\strut{} \scriptsize{\textsf{500}}}}%
      \put(958,1853){\makebox(0,0)[r]{\strut{} \scriptsize{\textsf{1000}}}}%
      \put(958,2483){\makebox(0,0)[r]{\strut{} \scriptsize{\textsf{1500}}}}%
      \put(958,3112){\makebox(0,0)[r]{\strut{} \scriptsize{\textsf{2000}}}}%
      \put(958,3742){\makebox(0,0)[r]{\strut{} \scriptsize{\textsf{2500}}}}%
      \put(990,494){\makebox(0,0){\strut{} \scriptsize{\textsf{0}}}}%
      \put(1375,494){\makebox(0,0){\strut{} \scriptsize{\textsf{500}}}}%
      \put(1759,494){\makebox(0,0){\strut{} \scriptsize{\textsf{1000}}}}%
      \put(2144,494){\makebox(0,0){\strut{} \scriptsize{\textsf{1500}}}}%
      \put(2529,494){\makebox(0,0){\strut{} \scriptsize{\textsf{2000}}}}%
      \put(2913,494){\makebox(0,0){\strut{} \scriptsize{\textsf{2500}}}}%
      \put(3298,494){\makebox(0,0){\strut{} \scriptsize{\textsf{3000}}}}%
      \put(3682,494){\makebox(0,0){\strut{} \scriptsize{\textsf{3500}}}}%
      \put(4067,494){\makebox(0,0){\strut{} \scriptsize{\textsf{4000}}}}%
      \put(486,2231){\rotatebox{-270}{\makebox(0,0){\strut{}$m_{1/2}$ \textsf{(GeV)}}}}%
      \put(2528,304){\makebox(0,0){\strut{}$m_0$ \textsf{(GeV)}}}%
      \put(3105,3944){\makebox(0,0)[l]{\strut{}\tiny{\textsf{BayesFITS (2012)}}}}%
      \put(1105,3729){\makebox(0,0)[l]{\strut{}\footnotesize{\textsf{Light Higgs mass $m_h$}}}}%
      \put(1105,3541){\makebox(0,0)[l]{\strut{}\scriptsize{\textsf{CMSSM}, $\mu>0$}}}%
      \put(1105,3377){\makebox(0,0)[l]{\strut{}\scriptsize{\textsf{with $\delta(g-2)_\mu$}}}}%
      \put(1105,3226){\makebox(0,0)[l]{\strut{}\tiny{\textsf{LHC (5/fb)$+m_h$$\simeq$125 \textsf{GeV}}}}}%
%    }%
%    \gplgaddtomacro\gplfronttext{%
      \csname LTb\endcsname%
      \put(3803,3619){\makebox(0,0)[r]{\strut{}\tiny{$m_h$: 81 - 117 \textsf{GeV}}}}%
      \csname LTb\endcsname%
      \put(3803,3487){\makebox(0,0)[r]{\strut{}\tiny{117 - 119 \textsf{GeV}}}}%
      \csname LTb\endcsname%
      \put(3803,3355){\makebox(0,0)[r]{\strut{}\tiny{119 - 122 \textsf{GeV}}}}%
      \csname LTb\endcsname%
      \put(3803,3223){\makebox(0,0)[r]{\strut{}\tiny{122 - 128 \textsf{GeV}}}}%
%    }%
%    \gplbacktext
    \put(0,0){\includegraphics{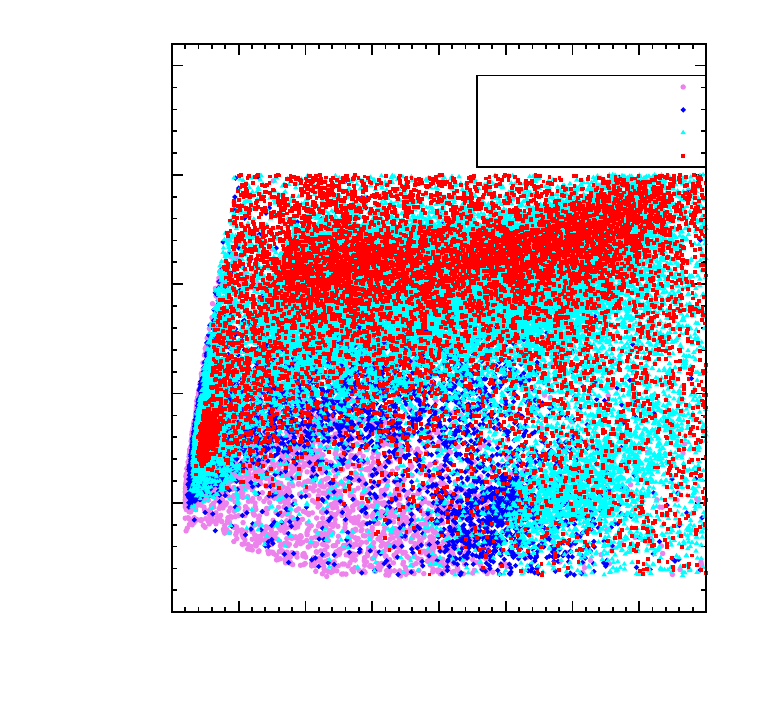}}%
%    \gplfronttext
  \end{picture}%
}
\subfloat[]{%
%\label{fig:-c}% oh2
%\includegraphics[width=0.49\textwidth]{CMSSM_1D_oh2.pdf}
%}%
%\hspace{-3pt}%
\label{fig:-b}% bsmm
\raisebox{3.1mm}{\includegraphics[width=0.39\textwidth]{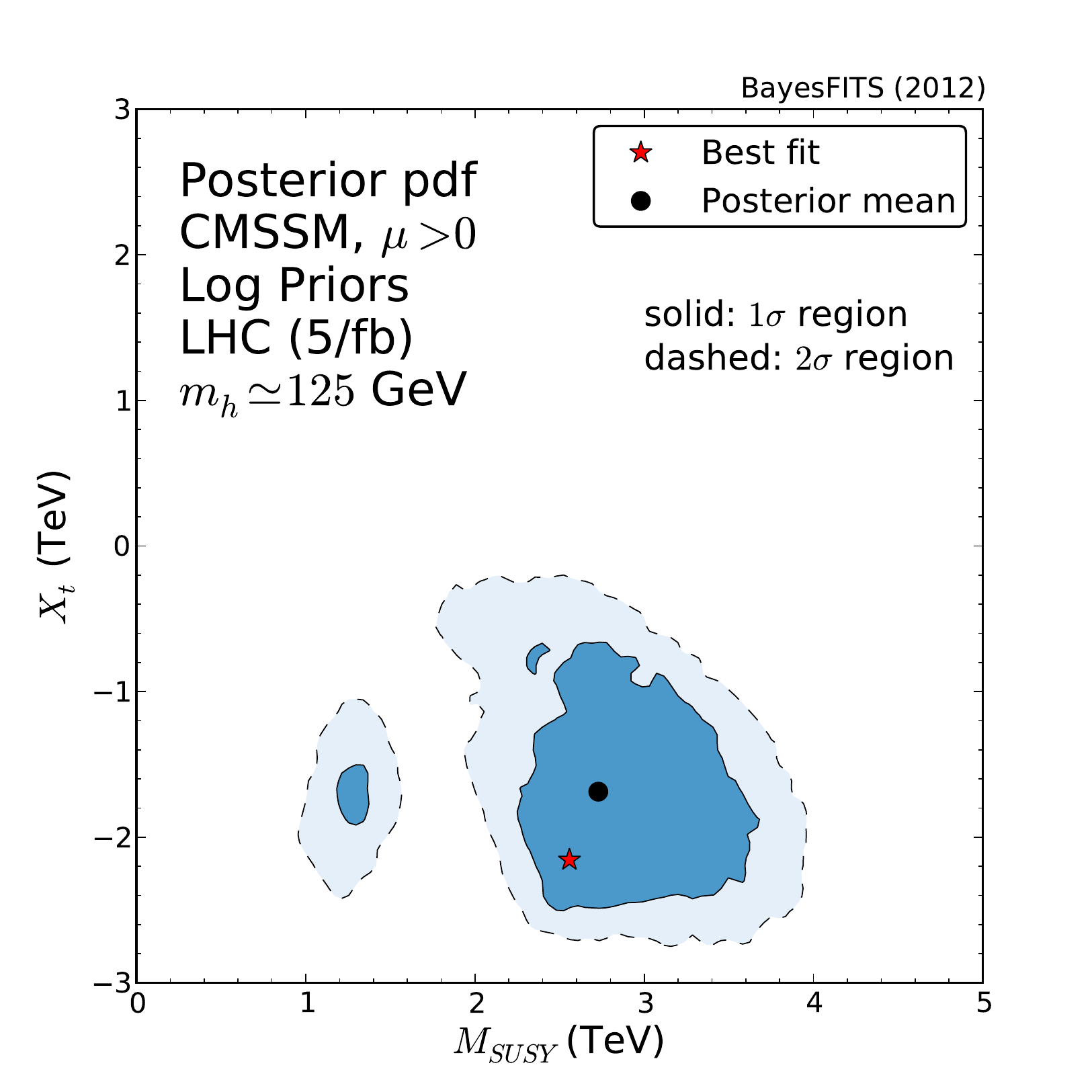}}
}%
\caption[]{\subref{fig:-a} Scatter plot showing the value of \mhl\ in the (\mzero, \mhalf) plane of the CMSSM. \subref{fig:-b} Marginalized posterior pdf in the parameters $X_t$ vs $\msusy$, relevant for the loop corrections to the Higgs mass.} 
\label{fig:2d_Higgs_loop_corrections}
\end{figure}
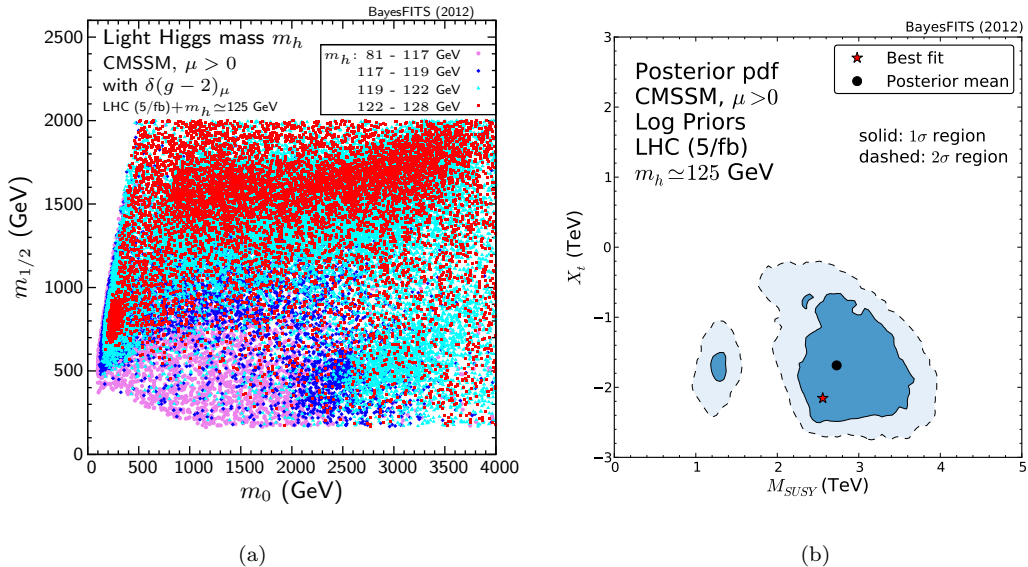
%%%%%%%%%%%%%%%%%%%%%%%%%%%%%%%%%%%%%%%%%%%%%%%%%%%%%%%%%%%%%%%%%%%%%%%%%%%%%%%%

The reason for so strongly disfavoring larger values of \mhl\ is the
tension between the Higgs mass above 124 \gev\ and the correct value
of the relic density. This tension manifests itself both in the
$A$-funnel and in the FP/HB region, though its origin in each of those
regions is different. In the $A$-funnel the main mechanism that allows
to obtain the correct value of the relic density is the resonance
annihilation of neutralinos through the pseudoscalar $A$ boson. To
allow such a process, an approximate relation $m_A\sim 2\mchi$ should
hold. However, for cases where $\mhl>124\gev$ the mass of the pseudoscalar $m_A$
exceeds significantly the doubled mass of the neutralino, and
annihilation at the resonance cannot take place.

In the FP/HB region the correct relic density is achieved in another
way. Because of the relatively small values of $|\mu|$ the lightest
neutralino becomes more Higgsino-like, and the annihilation cross
section is enhanced. However, as we have already stated, in the
CMSSM the lightest Higgs boson with mass larger than 124\gev\ can be
much more easily obtained for large ($\lesssim -1\tev$) negative
values of $A_0$ at the GUT scale. After running down to the
electroweak scale, negative values for \azero\ yield even larger
negative $A_t$, which is one of the conditions to obtain large Higgs
boson masses, as it will appear clear below.  On the other hand,
taking into account the minimalization condition for the scalar
potential, large negative $A_0$ do not allow the parameter $\mu$ to be
small enough to enhance the Higgsino-like component of
the neutralino. That creates the tension between the relic density and the
Higgs mass above $124\gev$.

In Fig.~\ref{fig:2d_Higgs_loop_corrections}\subref{fig:-a} we show a
scatter plot representing the distribution of the lightest Higgs mass
over the (\mzero, \mhalf) plane. One can see that Higgs masses
compatible with 125\gev\ at $1\sigma$ can be obtained in large numbers
across the whole plane. Particularly, the mass distribution presented
in Fig.~\ref{fig:2d_Higgs_loop_corrections}\subref{fig:-a} has one
interesting aspect. The one-loop contribution to the Higgs mass in the
decoupling limit ($m_A\gg\mz$) for moderate-to-large \tanb\ is given
by\cite{oneloophiggs}
\begin{equation}
  \Delta\mhl^2\propto\ln\frac{\msusy^2}{m_t^2}+\frac{X_t^2}{\msusy^2}\left(1-\frac{X_t^2}{12\msusy^2}\right),\label{1loopmh}
\end{equation}
where $m_t$ is the top quark mass, $\msusy$ is the geometrical average
of the physical stop masses, and $X_t=A_t-\mu\cot\beta$. While the
presence of a relatively heavy Higgs is not a surprise in the
$A$-funnel region, where the one-loop contribution to \mhl\ is driven
up by a large SUSY scale, it is more striking in the \stauc\
region. As anticipated above, to ensure such a heavy Higgs mass in the
region of low \mzero\ and \mhalf, the contribution from the $X_t$
factor in Eq.~(\ref{1loopmh}) should be significant. ($X_t\sim A_t$
almost throughout the whole parameter space.) In fact, it turns out
that the \stauc\ region is the only region of parameter space where
the factor $|X_t|/\msusy$ reaches values close to $\sim 2.5$, the
maximal contribution from the stop-mixing.

The interplay between $\msusy$ and $X_t$ just described is often
claimed in the literature to be an indication of
fine-tuning\cite{Chankowski:1997zh,*Barbieri:1998uv,*Kane:1998im},
thus making the CMSSM a less natural model than, for instance, the
Next-to-Minimal Supersymmetric Standard
Model\cite{Ellwanger:2011aa}. We plot in
Fig.~\ref{fig:2d_Higgs_loop_corrections}\subref{fig:-b} the
two-dimensional marginalized posterior in the $(\msusy,X_t)$
plane. One can see two separate high-probability regions. The one on
the right corresponds to the $A$-funnel region, where the best-fit
point lies, while the one on the left, smaller in size, to the \stauc\
region. We gather that, even if the model might be intrinsically
fine-tuned, given the present status of experimental and theoretical
uncertainties, our global set of constraints favors $2\sigma$ credible
regions that span an area of $\sim10\tev^2$, thus allowing a broad
range of values for these parameters. Moreover, it appears clear that
the present set of constraints highly favors negative values of $X_t$.

%%%%%%%%%%%%%%%%%%%%%%%%%   F   I   G   U   R   E   %%%%%%%%%%%%%%%%%%%%%%%%%%%%
% Likelihood maps
\begin{figure}[t]
\centering
\subfloat[$\mu>0$ and no \gmtwo.]
{% 
\label{fig:LikeMap-a}%
% razor 4.4/fb SB profile likelihood
\includegraphics[width=0.39\linewidth]{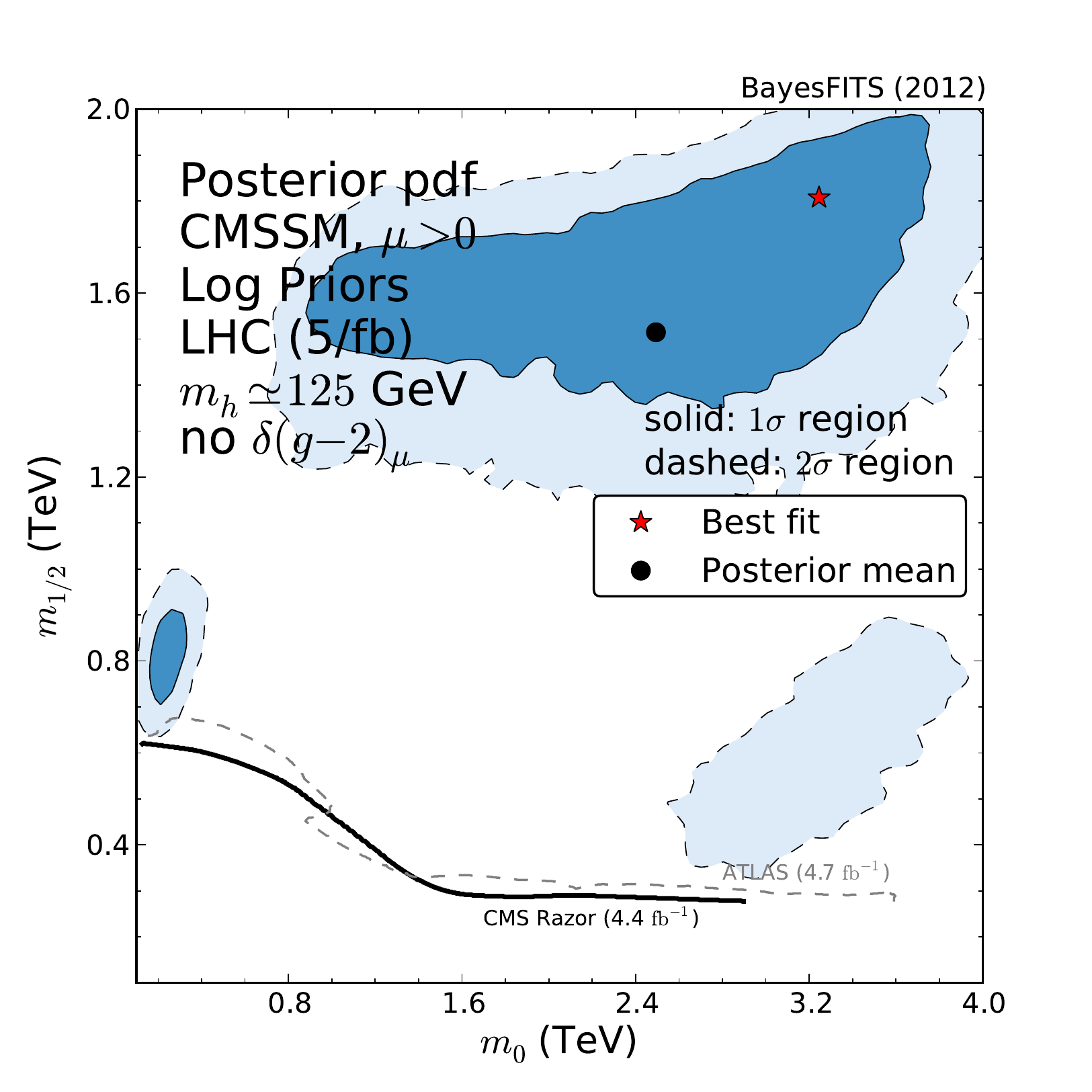}
}%
\hspace{8pt}
\subfloat[$\mu<0$ and no \gmtwo.]
{% 
\label{fig:LikeMap-b}%
% razor 4.4/fb SB profile likelihood
\includegraphics[width=0.39\linewidth]{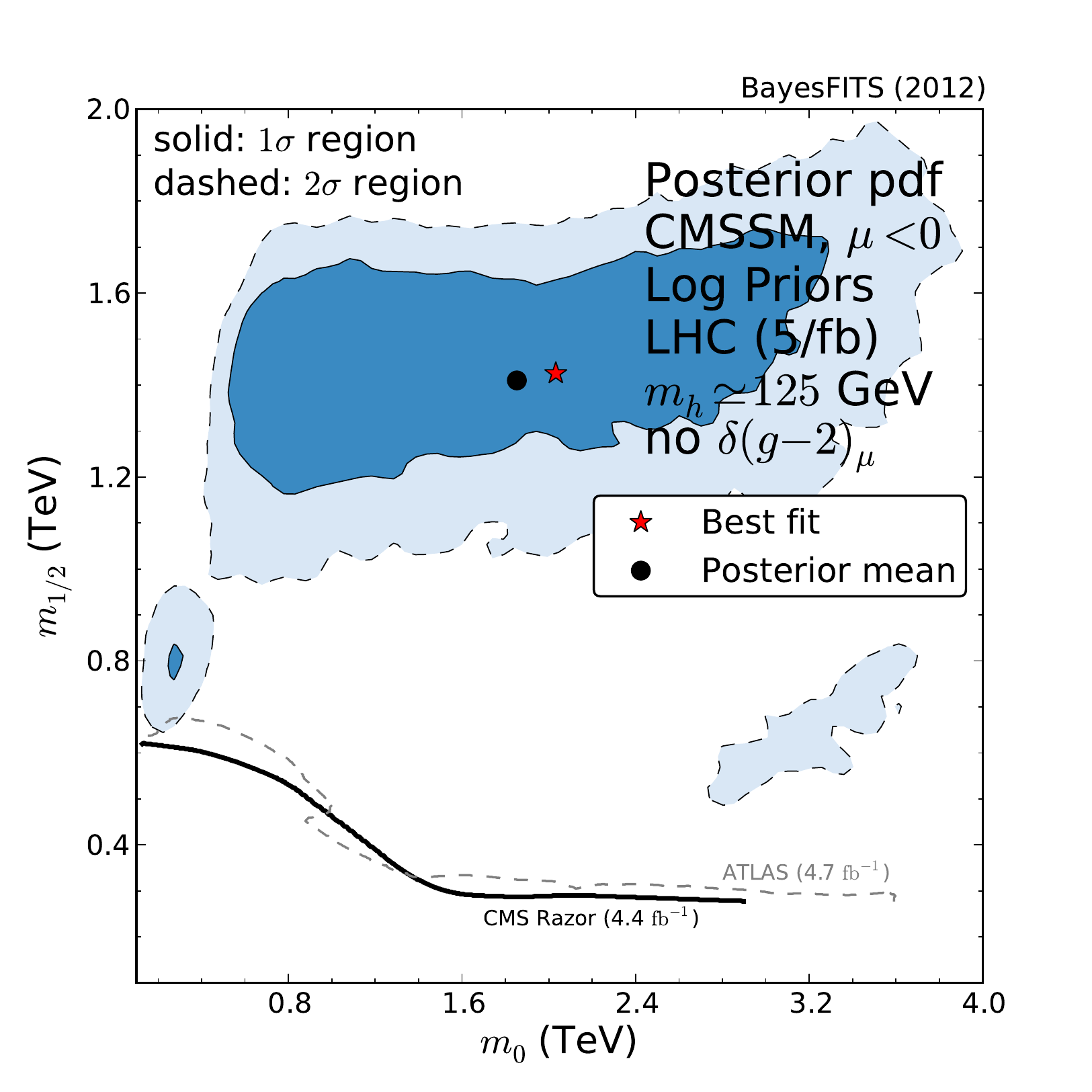}
}%
\hspace{8pt}
\subfloat[$\mu>0$ and no \gmtwo.]
{%
\label{fig:LikeMap-c}%
% razor 4.4/fb likelihood plot
\includegraphics[width=0.39\linewidth]{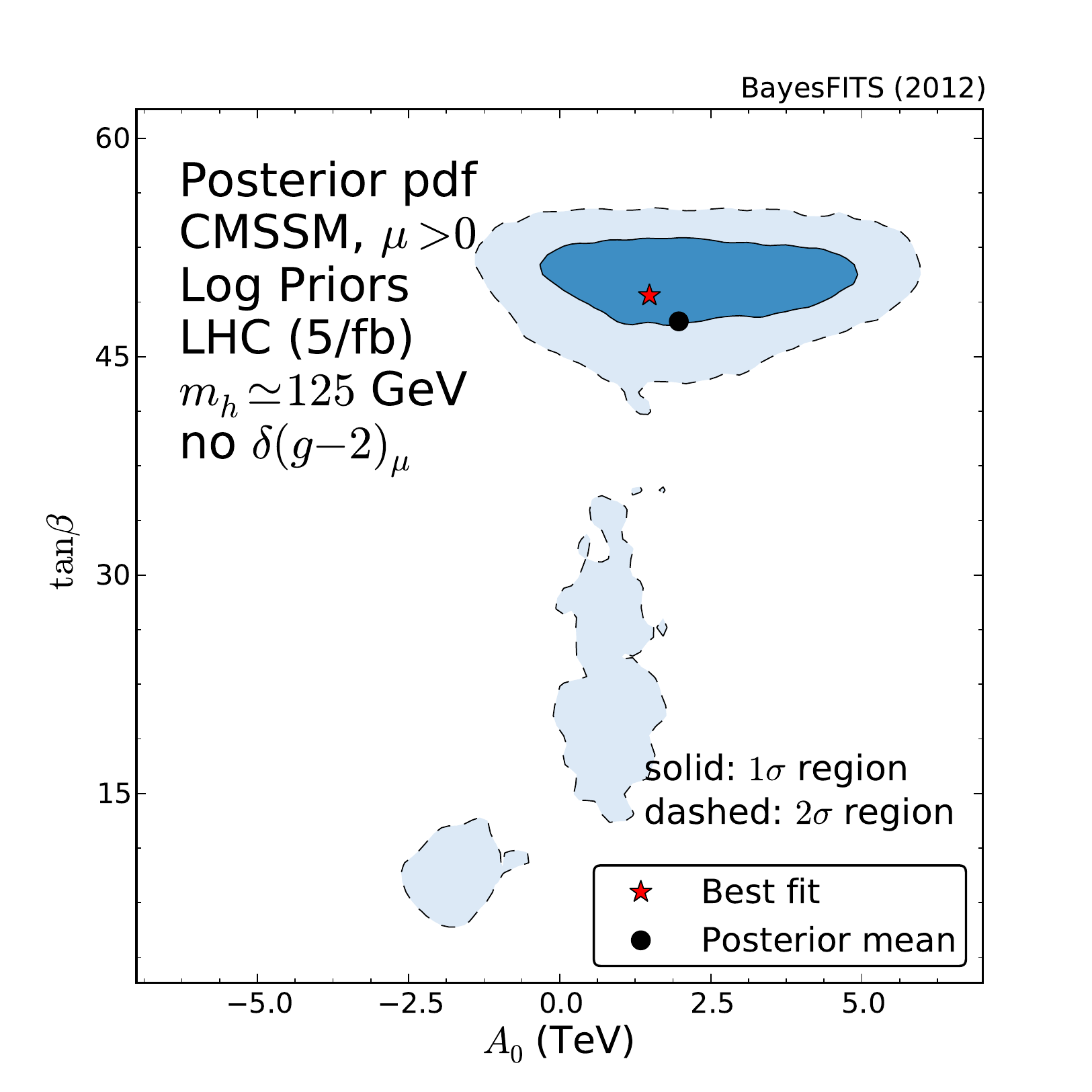}
}
\hspace{8pt}
\subfloat[$\mu<0$ and no \gmtwo.]
{%
\label{fig:LikeMap-d}%
% razor 4.4/fb likelihood plot
\includegraphics[width=0.39\linewidth]{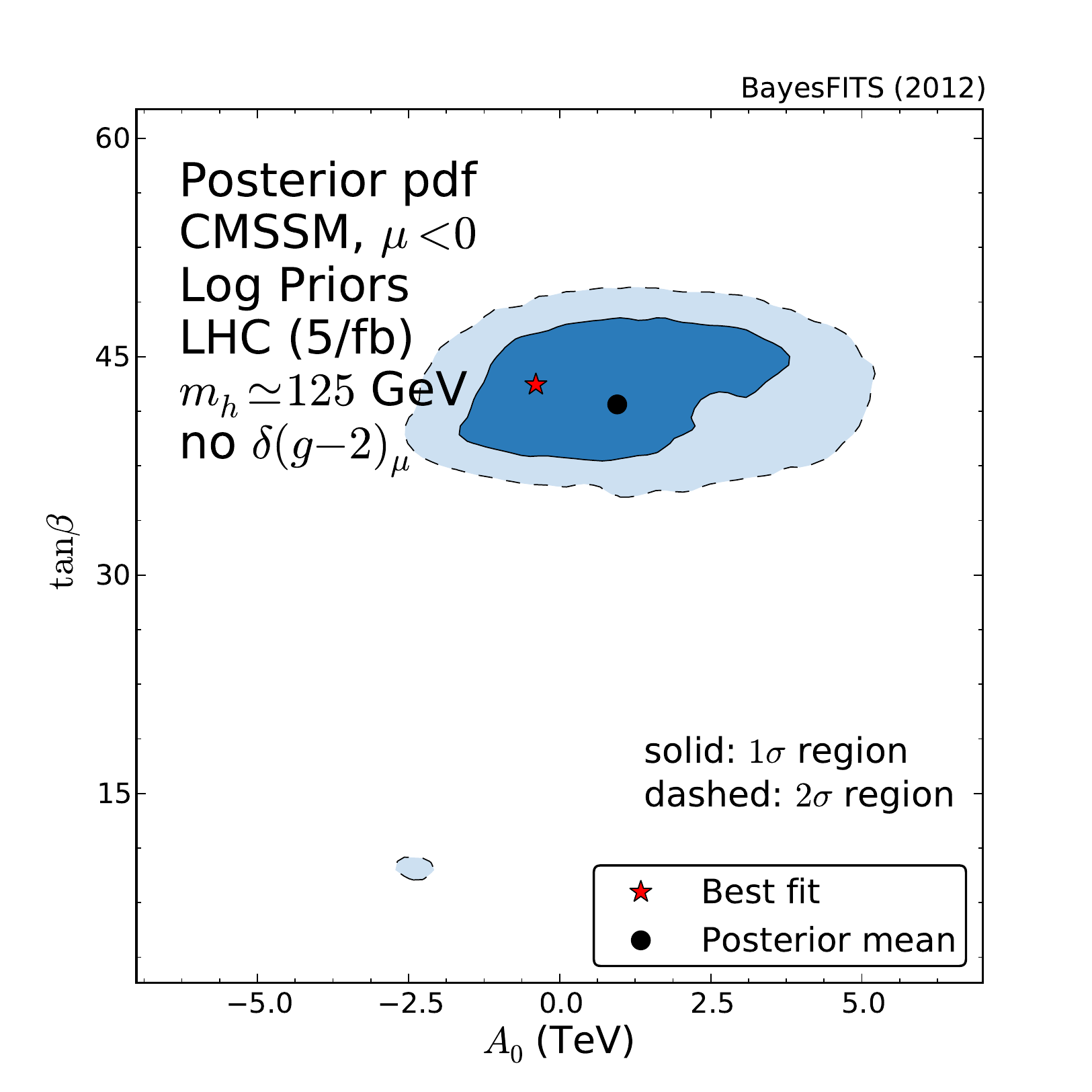}
}
\caption[]{\subref{fig:LikeMap-a} Marginalized posterior pdf in the (\mzero, \mhalf) plane for the constraints listed in Table~\ref{tab:exp_constraints} except \gmtwo, for $\mu>0$. \subref{fig:LikeMap-b} Marginalized posterior pdf in the (\mzero, \mhalf) plane for the same constraints as in \subref{fig:LikeMap-a} and $\mu<0$. \subref{fig:LikeMap-c} Marginalized posterior pdf in the (\azero, \tanb) plane for the same constraints as in \subref{fig:LikeMap-a}.\subref{fig:LikeMap-d} Marginalized posterior pdf in the (\azero, \tanb) plane for the same constraints as in \subref{fig:LikeMap-a} and $\mu<0$.}
\label{fig:likemap_gmtwo}
\end{figure} 
%%%%%%%%%%%%%%%%%%%%%%%%%%%%%%%%%%%%%%%%%%%%%%%%%%%%%%%%%%%%%%%%%%%%%%%%%%%%%%%%%%%%%%%%%%

%%%%%%%%%%%%%%%%%%%%%%%%%%%%%%%%%%%%%%%%%%%%%%%%%%%%%%%%%%%%%%%%%%%%%%%%%%%%%%%
\subsection{Impact of \gmtwo\ and the case $\mu<0$}
%%%%%%%%%%%%%%%%%%%%%%%%%%%%%%%%%%%%%%%%%%%%%%%%%%%%%%%%%%%%%%%%%%%%%%%%%%%%%%%

Since the poor global fit is mainly a result of including the \gmtwo\
constraint in the likelihood, and the SM prediction is to this day
still marred by substantial theoretical uncertainties, we have also
performed scans without the \gmtwo\ constraint. In this case there is
no reason anymore to assume $\signmu=+1$, as the main reason for such
choice was to improve the fit to this particular observable. For this
reason we will not show the case with \gmtwo\ and $\mu<0$ because the
global fit worsens, although actually not so much. We will summarize the goodness of all the fits in
Table~\ref{tab:bestfit_lnlike}.

Before we move to the case with no \gmtwo, a few remarks on the effect
of reversing \signmu\ while maintaining the \gmtwo\ constraint are in
order.  Even with \gmtwo\ taken into account, we checked that the main
effect of taking $\mu<0$ would impact on the value and location of the
best-fit point, rather than the posterior distribution. The
probability distributions obtained in this case are very similar to
the ones depicted in Fig.~\ref{fig:cmssm_params_lhc_mhl125}, but the
best-fit point is now pushed up to larger \mzero\ and \mhalf.
Clearly, when $\mu$ is negative, the \gmtwo\ constraint tends to favor
large mass scales, since it tends to minimize the (now negative)
contribution. On the other hand, the change in the sign of $\mu$
allows almost all points in the scan to satisfy \bsmumu, and this
provides a significant difference from the cases with positive $\mu$,
where a relatively wide region of parameter space at small \mzero\ and
\mhalf\ was disfavored under the new LHCb limit. These two contrasting
effects can be thought as balancing out, thus producing a similar
posterior distribution.

Let us now analyze the effects of lifting the \gmtwo\ constraint. The
case $\mu>0$ is shown in
Fig.~\ref{fig:likemap_gmtwo}\subref{fig:LikeMap-a}, where we plot the
two-dimensional posterior in the (\mzero, \mhalf) plane, and in
Fig.~\ref{fig:likemap_gmtwo}\subref{fig:LikeMap-c} where the
distribution in the (\azero, \tanb) plane is shown. The plots do not
show much difference from the cases with \gmtwo\ included.  The
best-fit point moves towards larger \mzero, and one can notice the
slightly increased relevance of the FP/HB region.  The near
independence of the global posterior distribution of the \gmtwo\
constraint for $\mu>0$ was to be expected.  As one can see in
Table~\ref{tab:bestfit_lnlike}, the contribution to the total $\chi^2$
of the best-fit point due to this constraint is by far the largest,
thus making it the observable most poorly fit.  When all other constraints
pull in a different direction, the pdf becomes insensitive to this
constraint, in all effects treating it as an outlier.

In fact, when \gmtwo\ is ignored, the lowest $\chi^2$ for all the four
cases we have studied, is obtained with negative
$\mu$. We show the marginalized posterior for this case in the
(\mzero, \mhalf) plane in
Fig.~\ref{fig:likemap_gmtwo}\subref{fig:LikeMap-b}. One can see that
the area of parameter space corresponding to the $A$-resonance region
extends to values of \mzero\ lower than in the other cases; the
\stauc\ and FP/HB regions are instead reduced. As described above,
$\mu<0$ allows to satisfy \brbsmumu\ in broader regions of parameter
space. Moreover, it appears that the Higgs mass constraint can be
satisfied better in the low \mzero\ region for $\mu<0$. When it comes
to the marginalized posterior in the (\azero, \tanb) plane [shown in
Fig.~\ref{fig:likemap_gmtwo}\subref{fig:LikeMap-d}] one can see that
low values of \tanb\ are nearly excluded, and the $1\sigma$ credible
region has shifted down, to values around $\tanb\sim 40-45$.

%%%%%%%%%%%%%%%%%%%%%%%%%   F   I   G   U   R   E   %%%%%%%%%%%%%%%%%%%%%%%%%%%%
% 
\begin{figure}[t]
\centering
\subfloat[]{%
\label{fig:Comparison_mAtanbeta-a}%
\includegraphics[width=0.39\textwidth]{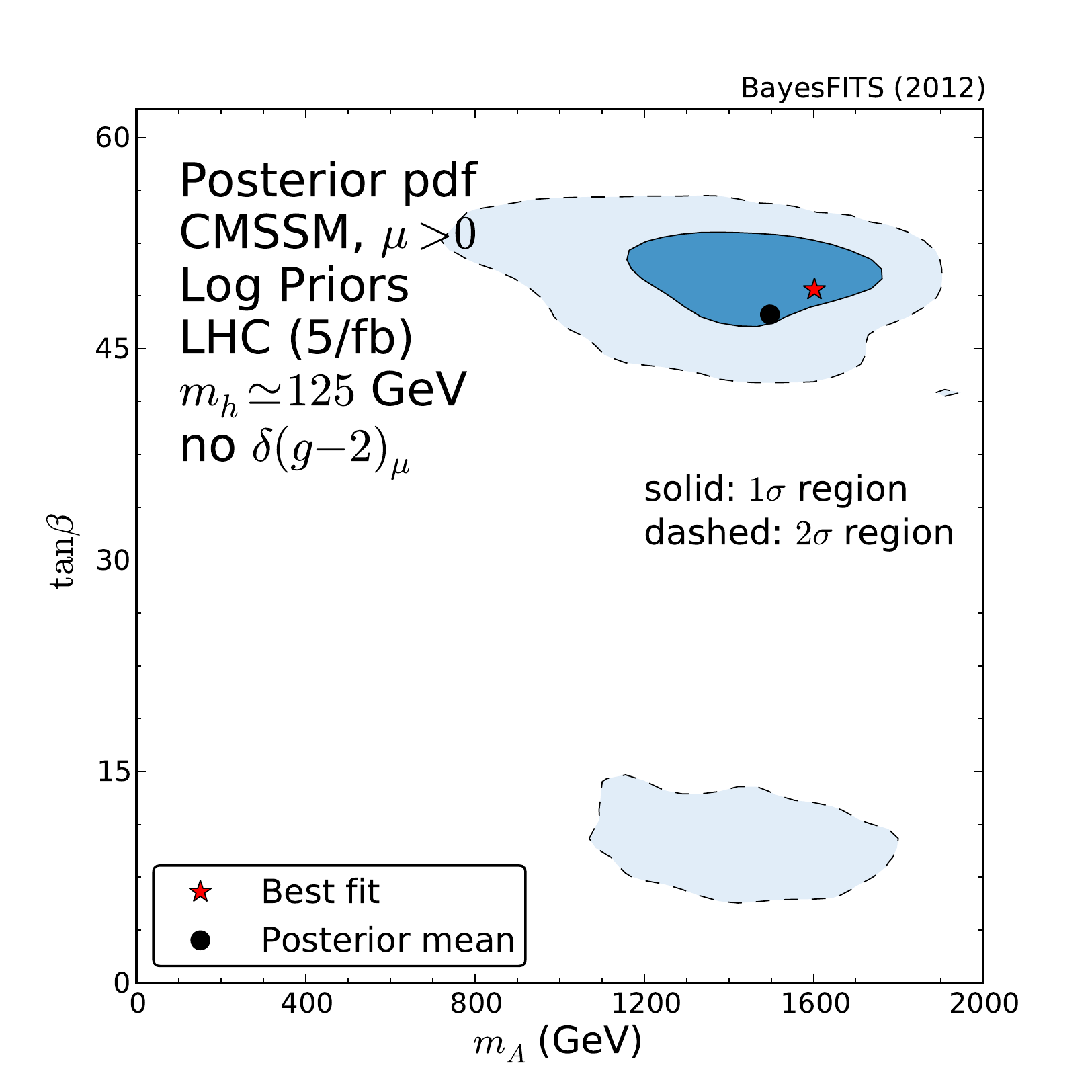}
}%
\hspace{1pt}%
\subfloat[]{%
\label{fig:Comparison_mAtanbeta-b}%
\includegraphics[width=0.39\textwidth]{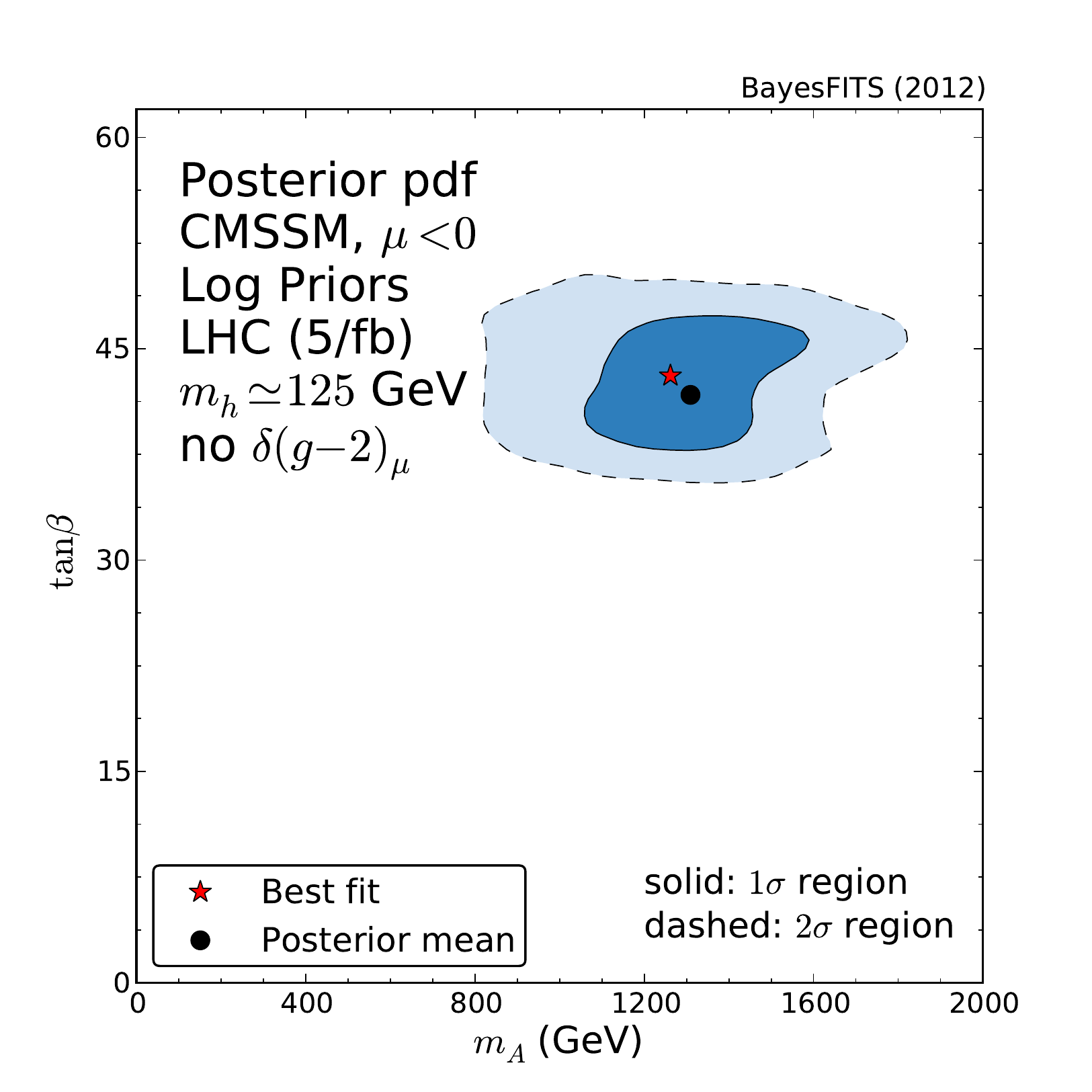}
}%
\caption[]{Marginalized posterior pdf in the (\mha, \tanb) plane, in the CMSSM
 constrained by all the experiments listed in
 Table~\ref{tab:exp_constraints} except \gmtwo.
 \subref{fig:Comparison_mAtanbeta-a} $\mu>0$, \subref{fig:Comparison_mAtanbeta-b} $\mu<0$. }%
\label{fig:nog2_mAtanbeta}
\end{figure} 
%%%%%%%%%%%%%%%%%%%%%%%%%%%%%%%%%%%%%%%%%%%%%%%%%%%%%%%%%%%%%%%%%%%%%%%%%%%%%%%%

%%%%%%%%%%%%%%%%%%%%%%%%%%%%%%%%%%%%%%%%%%%%%%%%%%%%%%%%%%%%%%%%%%%%%%%%%%%%%%%%
% Plots in 2 by 2 subfigure environment
% non-LHC + alphaT 1.1/fb + Xenon
% Various observables plotted against each other
\begin{figure}[p]
\centering
\subfloat[t][]{%
\label{fig:2d_g2_bsm-a}%
{\includegraphics[width=0.39\textwidth]{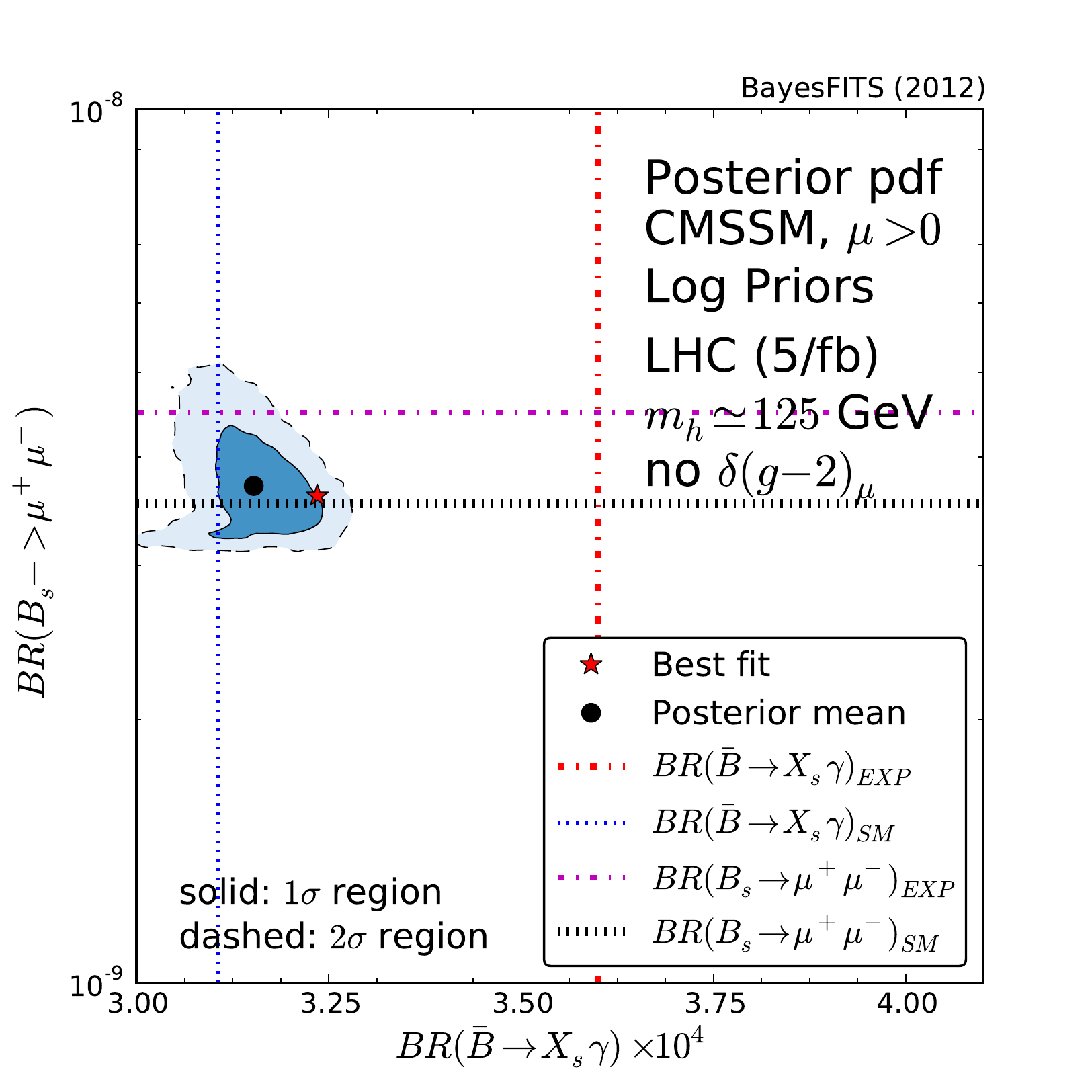}}
}%
%\hspace{1pt}%
%\subfloat[t][\brbsmumu\ against \brbxsgamma.]{%
%\label{fig:2d_bsg_g2_bsmm_oh2-b}%
%{\includegraphics[width=0.49\textwidth]{CMSSM-CONV_bsg_bsmumu.pdf}}
%}
\hspace{1pt}%
\subfloat[t][]{%
\label{fig:2d_g2_bsm-b}%
{\includegraphics[width=0.39\textwidth]{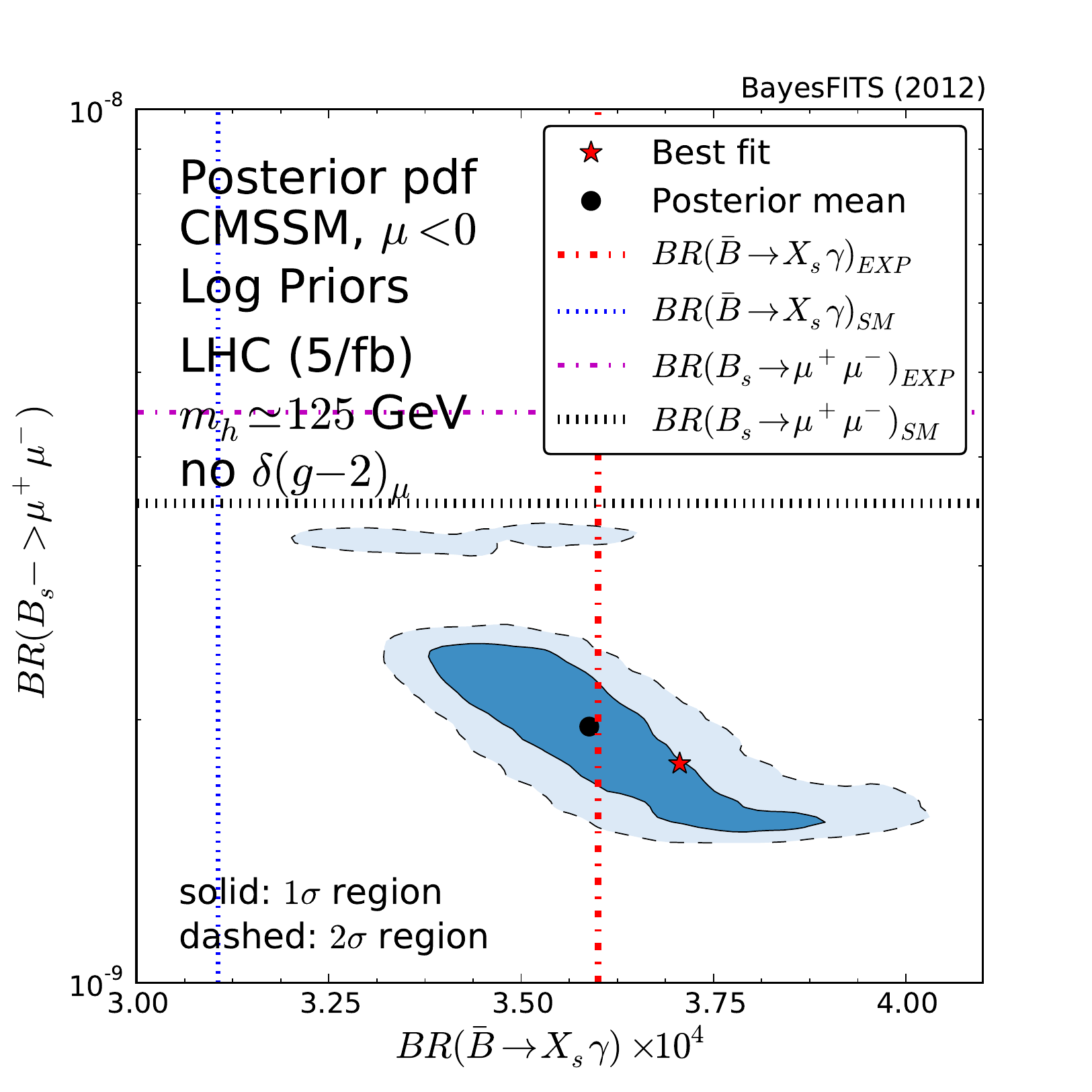}}
}
%\hspace{1pt}%
%\subfloat[t][\brbxsgamma\ against \abund.]{%
%\label{fig:2d_bsg_g2_bsmm_oh2-d}%
%\includegraphics[width=0.49\textwidth]{CMSSM-CONV_oh2_bsg.pdf}
%}\\%
\caption[]{Marginalized posterior pdf of the
  experimental observables \brbxsgamma\ vs. \brbsmumu\ in the
  CMSSM constrained by all the experiments listed in Table~\ref{tab:exp_constraints} except \gmtwo. \subref{fig:2d_g2_bsm-a} $\mu>0$, \subref{fig:2d_g2_bsm-b} $\mu<0$.}%
\label{fig:cmssm_2d_nog2_bsg_bsm}%
\end{figure}
%%%%%%%%%%%%%%%%%%%%%%%%%%%%%%%%s%%%%%%%%%%%%%%%%%%%%%%%%%%%%%%%%%%%%%%%%%%%%%%%%

In Fig.~\ref{fig:nog2_mAtanbeta}\subref{fig:Comparison_mAtanbeta-a} we
show the two-dimensional pdf in the $(m_A, \tanb)$ plane without the
\gmtwo\ constraint, and taking $\mu>0$. No visible difference appears
with the case which included \gmtwo. Significant differences appear
instead for $\mu<0$, as shown in
Fig.~\ref{fig:nog2_mAtanbeta}\subref{fig:Comparison_mAtanbeta-b}. Not
only can one notice the down-shifting of the preferred values for
\tanb\ mentioned above, but also lower values of $m_A$ than in the
positive $\mu$ case are now favored at large \tanb. The reason lies in
the improved fit to the $b$-physics observables, and in particular to
\brbsmumu.

This can be seen in Fig.~\ref{fig:cmssm_2d_nog2_bsg_bsm} where we show the two-dimensional
posterior for the observables \brbxsgamma\ vs. \brbsmumu\ for
\subref{fig:Comparison_mAtanbeta-a} $\mu>0$ and
\subref{fig:Comparison_mAtanbeta-b} $\mu<0$. The purple horizontal
line (dot-dashed) and the red vertical line (dot-dashed) show the
respective experimental values, while the horizontal gray line
(dotted) and the vertical blue line (dotted) show the respective SM
values.  One can see that, for $\mu>0$ the probability distribution
does not change significantly when we lift the \gmtwo\ constraint. It
confirms the fact that, given the poorness of the fit to
\deltagmtwomususy, the posterior is effectively insensitive to this
constraint.

%%%%%%%%%%%%%%%%%%%%%%%%%   F   I   G   U   R   E   %%%%%%%%%%%%%%%%%%%%%%%%%%%%
% 2 by 1: left: plot of 1d pdf of mhl,
% right: chi2 vs mhl
\begin{figure}[p]
\centering
\subfloat[]{%
\label{fig:-a}%
\raisebox{3.5mm}
{\includegraphics[width=0.39\textwidth]{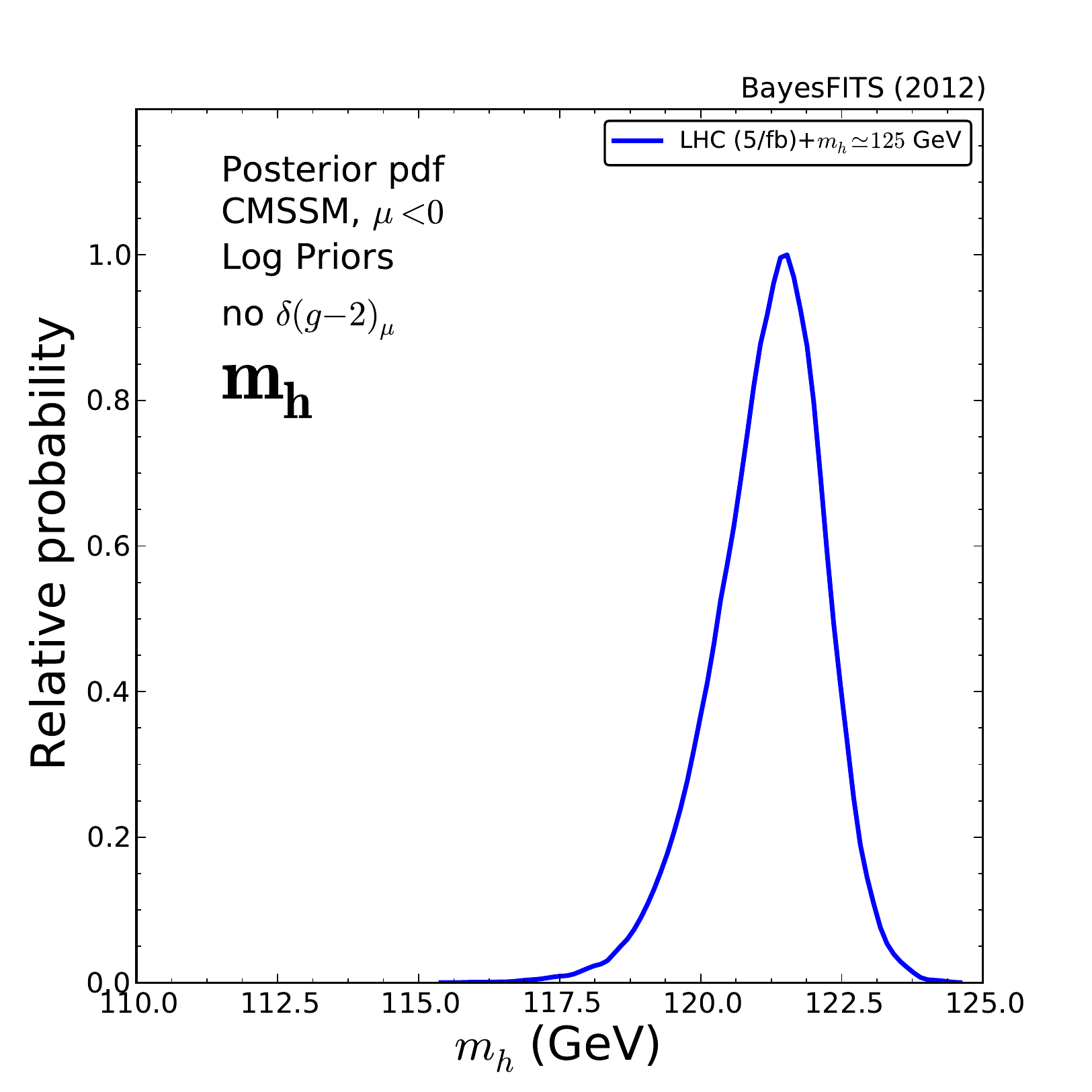}}
}%
%\hspace{1pt}%
\subfloat[]{%
\label{fig:-b}%
  \setlength{\unitlength}{0.0500bp}%
  \begin{picture}(5040.00,4032.00)%
%    \gplgaddtomacro\gplbacktext{%
      \csname LTb\endcsname%
      \put(958,594){\makebox(0,0)[r]{\strut{} \scriptsize{\textsf{0}}}}%
      \put(958,1204){\makebox(0,0)[r]{\strut{} \scriptsize{\textsf{500}}}}%
      \put(958,1814){\makebox(0,0)[r]{\strut{} \scriptsize{\textsf{1000}}}}%
      \put(958,2425){\makebox(0,0)[r]{\strut{} \scriptsize{\textsf{1500}}}}%
      \put(958,3035){\makebox(0,0)[r]{\strut{} \scriptsize{\textsf{2000}}}}%
      \put(958,3645){\makebox(0,0)[r]{\strut{} \scriptsize{\textsf{2500}}}}%
      \put(990,474){\makebox(0,0){\strut{} \scriptsize{\textsf{0}}}}%
      \put(1447,474){\makebox(0,0){\strut{} \scriptsize{\textsf{500}}}}%
      \put(1903,474){\makebox(0,0){\strut{} \scriptsize{\textsf{1000}}}}%
      \put(2360,474){\makebox(0,0){\strut{} \scriptsize{\textsf{1500}}}}%
      \put(2817,474){\makebox(0,0){\strut{} \scriptsize{\textsf{2000}}}}%
      \put(3273,474){\makebox(0,0){\strut{} \scriptsize{\textsf{2500}}}}%
      \put(3730,474){\makebox(0,0){\strut{} \scriptsize{\textsf{3000}}}}%
      \put(4186,474){\makebox(0,0){\strut{} \scriptsize{\textsf{3500}}}}%
      \put(4643,474){\makebox(0,0){\strut{} \scriptsize{\textsf{4000}}}}%
      \put(436,2180){\rotatebox{-270}{\makebox(0,0){\strut{}$m_{1/2}$ \textsf{(GeV)}}}}%
      \put(2816,274){\makebox(0,0){\strut{}$m_0$ \textsf{(GeV)}}}%
      \put(3501,3840){\makebox(0,0)[l]{\strut{}\scriptsize{\textsf{BayesFITS (2012)}}}}%
      \put(1127,3633){\makebox(0,0)[l]{\strut{}\footnotesize{\textsf{Light Higgs mass $m_h$}}}}%
      \put(1127,3450){\makebox(0,0)[l]{\strut{}\footnotesize{\textsf{CMSSM}, $\mu<0$}}}%
      \put(1127,3291){\makebox(0,0)[l]{\strut{}\footnotesize{\textsf{no $\delta(g-2)_\mu$}}}}%
      \put(1127,3120){\makebox(0,0)[l]{\strut{}\footnotesize{\textsf{LHC (5/fb)$+m_h$$\simeq$125 \textsf{GeV}}}}}%
%    }%
%    \gplgaddtomacro\gplfronttext{%
      \csname LTb\endcsname%
      \put(4379,3611){\makebox(0,0)[r]{\strut{}\scriptsize{$m_h$: 81 - 117 \textsf{GeV}}}}%
      \csname LTb\endcsname%
      \put(4379,3457){\makebox(0,0)[r]{\strut{}\scriptsize{117 - 119 \textsf{GeV}}}}%
      \csname LTb\endcsname%
      \put(4379,3303){\makebox(0,0)[r]{\strut{}\scriptsize{119 - 122 \textsf{GeV}}}}%
      \csname LTb\endcsname%
      \put(4379,3149){\makebox(0,0)[r]{\strut{}\scriptsize{122 - 128 \textsf{GeV}}}}%
%    }%
%    \gplbacktext
    \put(0,0){\includegraphics{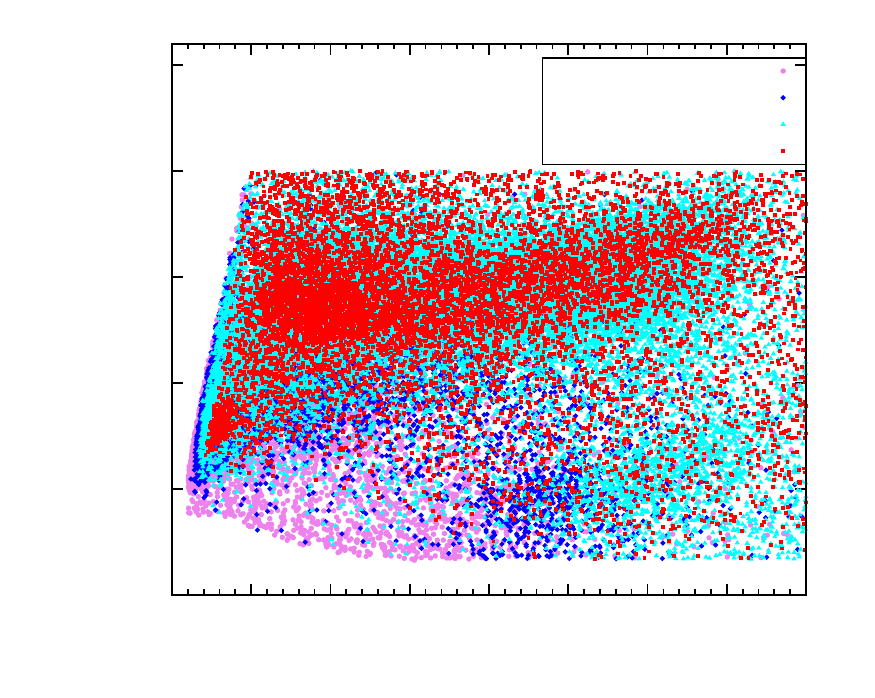}}%
%    \gplfronttext
  \end{picture}%
}%
\caption[]{\subref{fig:-a} Marginalized 1D posterior pdf of \mhl\ for $\mu<0$, constrained by the experiments listed in Table~\ref{tab:exp_constraints} except \gmtwo. \subref{fig:-b} Scatter plot distribution of the Higgs mass in the (\mzero, \mhalf) plane without the \gmtwo\ constraint, $\mu<0$.
 }%
\label{fig:higgs_mu_neg}
\end{figure}
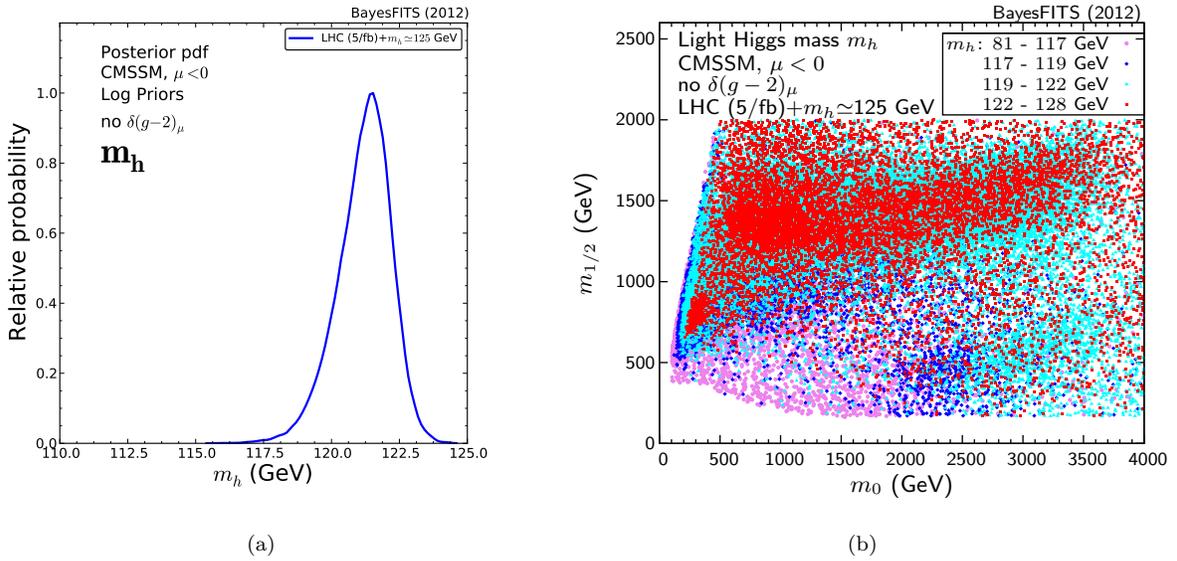 
%%%%%%%%%%%%%%%%%%%%%%%%%%%%%%%%%%%%%%%%%%%%%%%%%%%%%%%%%%%%%%%%%%%%%%%%%%%%%%%%

However, again, a significant difference arises for the
case with $\mu<0$ , shown in
Fig.~\ref{fig:cmssm_2d_nog2_bsg_bsm}\subref{fig:2d_g2_bsm-b}. The
contribution from the chargino-stop loop to \brbxsgamma\ changes sign
and now contributes positively to alleviate the discrepancy between
the experimental and the the SM value. As a consequence, the overall fit
to the experimental measurement improves, with the exception of the
region at small \mzero\ and \mhalf\, where the value becomes a bit too
high. \brbsmumu\ gets instead negative contributions that improve the
fit over all parameter space, even pushing the preferred value below
the SM calculation.

%%%%%%%%%%%%%%%%%%%%%%%%%   F   I   G   U   R   E   %%%%%%%%%%%%%%%%%%%%%%%%%%%%
% 
\begin{figure}[t]
\centering
\subfloat[$\mu>0$]{%
\label{fig:fig-a}%
\includegraphics[width=0.39\textwidth]{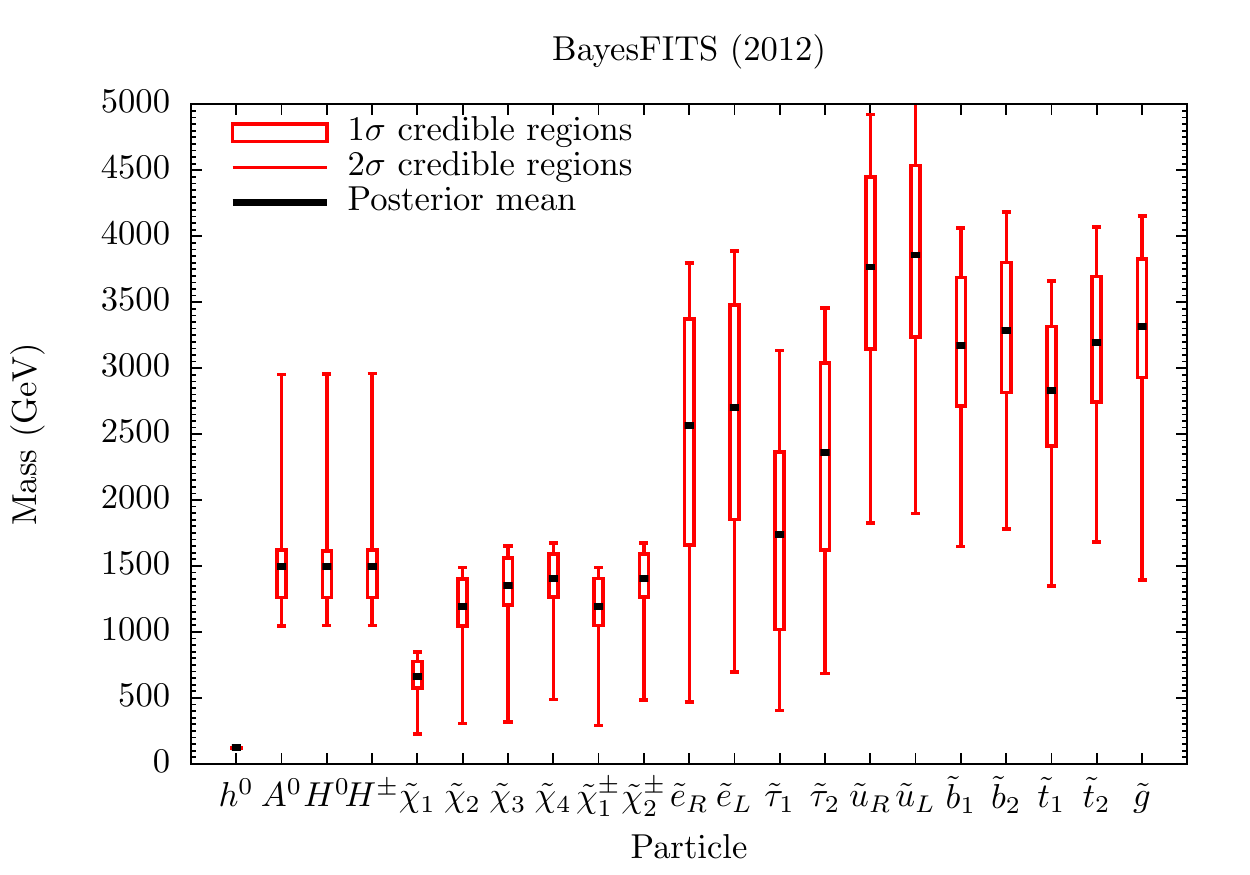}
}%
\hspace{1pt}%
\subfloat[$\mu<0$]{%
\label{fig:fig-b}%
\includegraphics[width=0.39\textwidth]{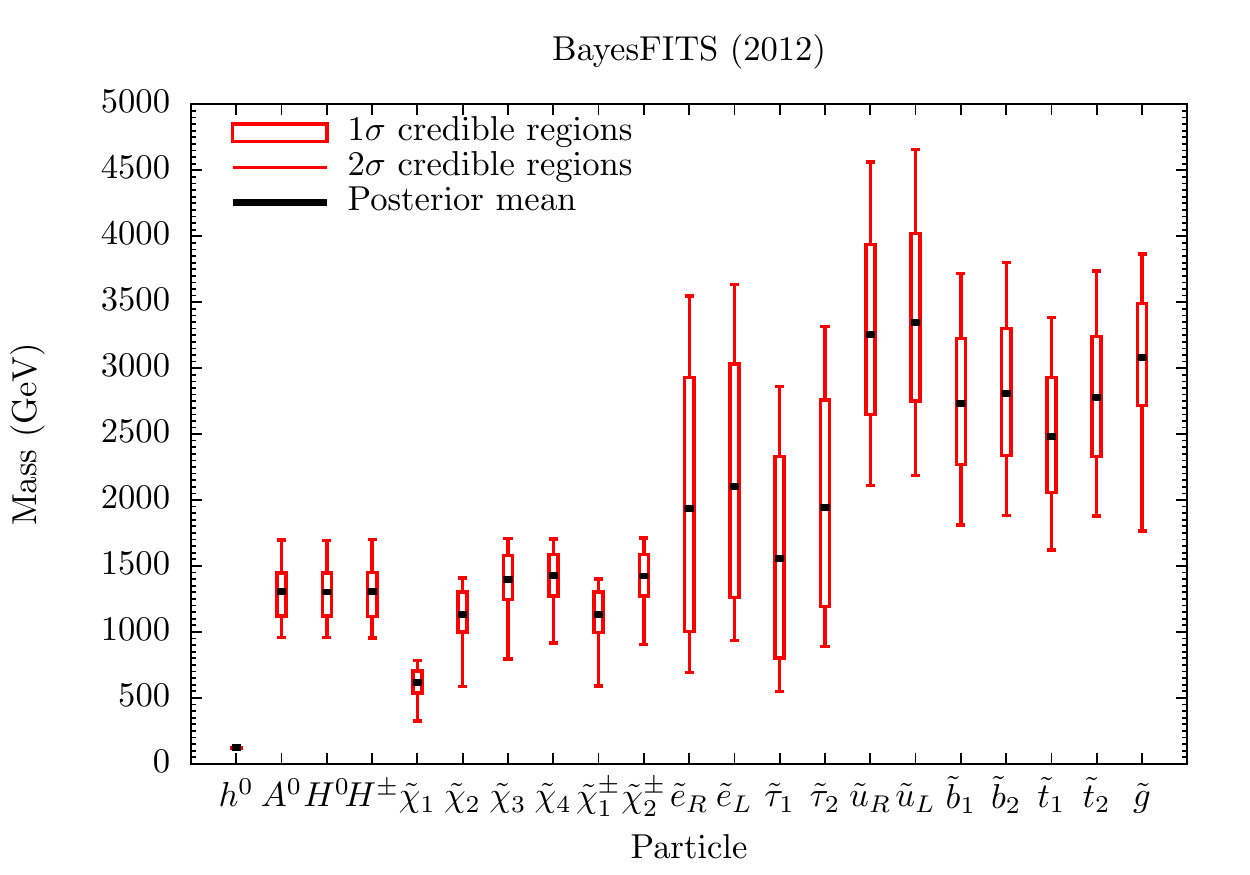}
}%
\caption[]{1D marginalized posterior pdf for the supersymmetric spectrum
 constrained by the experiments listed in
 Table~\ref{tab:exp_constraints} except \gmtwo. \subref{fig:fig-a} $\mu>0$. \subref{fig:fig-b} $\mu<0$.
}%
\label{fig:spectra_nog2}
\end{figure} 
%%%%%%%%%%%%%%%%%%%%%%%%%%%%%%%%%%%%%%%%%%%%%%%%%%%%%%%%%%%%%%%%%%%%%%%%%%%%%%%%

In Fig.~\ref{fig:higgs_mu_neg}\subref{fig:-a} we show the
one-dimensional marginalized posterior on Higgs mass distribution in
the case without \gmtwo\ and $\mu<0$. In
Fig.~\ref{fig:higgs_mu_neg}\subref{fig:-b} we show a scatter plot of
the distribution of Higgs masses over the (\mzero, \mhalf)
plane. Basically no difference in the distribution of the
Higgs mass is found for $\mu<0$. Finally, Fig.~\ref{fig:spectra_nog2}
shows the Bayesian credibility regions for the supersymmetric spectrum
when the \gmtwo\ constraint is lifted for $\mu>0$ \subref{fig:-a} and
$\mu<0$ \subref{fig:-b}.

%%%%%%%%%%%%%%%%%%%%%%%%%%%%%%%%%%%%%%%%%%%%%%%%%%%%%%%%%%%%%%%%%%%%%%%%%%%%%%%%
\subsection{Dark matter direct detection and $\mu$ combination}

%%%%%%%%%%%%%%%%%%%%%%%%%   F   I   G   U   R   E   %%%%%%%%%%%%%%%%%%%%%%%%%%%%
% 
\begin{figure}[b]
\centering
\subfloat[]{%
\label{fig:fig-a}%
\includegraphics[width=0.39\textwidth]{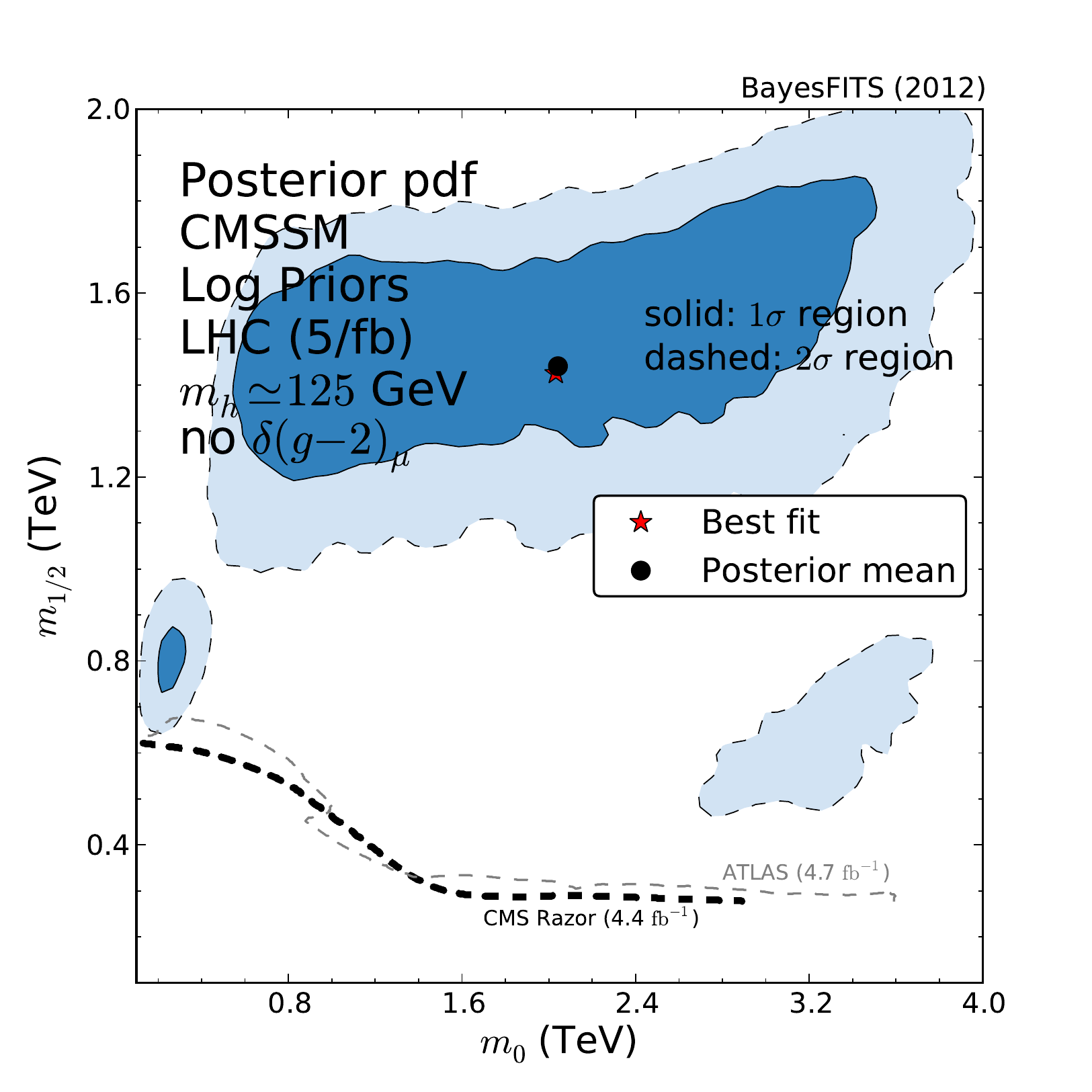}
}%
\hspace{1pt}%
\subfloat[]{%
\label{fig:fig-b}%
\includegraphics[width=0.39\textwidth]{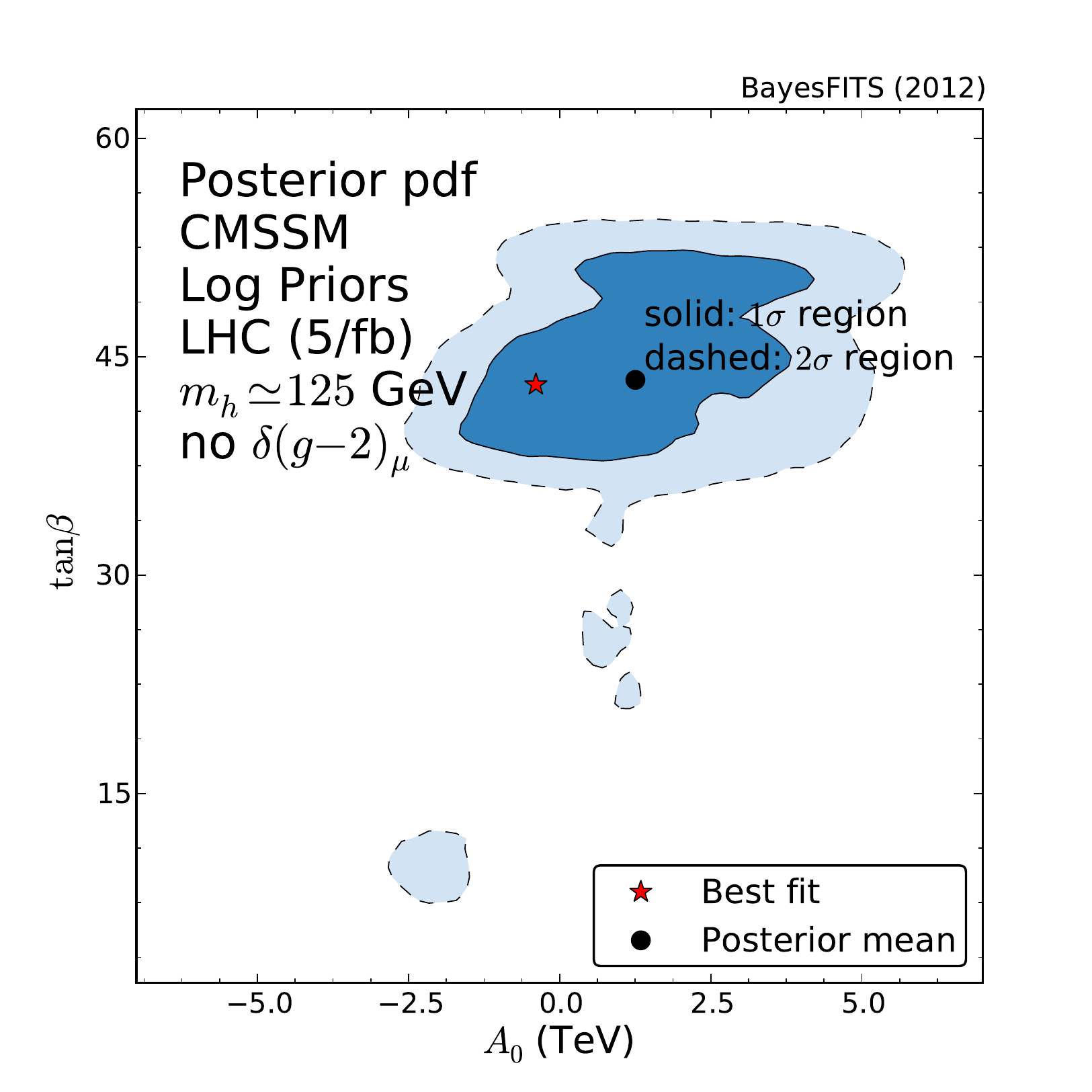}
}%
\caption[]{Marginalized posterior pdf in \subref{fig:fig-a} the (\mzero, \mhalf)
  plane and \subref{fig:fig-b} the (\azero, \tanb) plane of the CMSSM for $\mu>0$ and $\mu<0$ combined, constrained by the
  experiments listed in Table~\ref{tab:exp_constraints} except \gmtwo.
  The dashed black line shows the CMS razor 95\%~CL exclusion bound.
}
\label{fig:combine}
\end{figure} 
%%%%%%%%%%%%%%%%%%%%%%%%%%%%%%%%%%%%%%%%%%%%%%%%%%%%%%%%%%%%%%%%%%%%%%%%%%%%%%%%

In Fig.~\ref{fig:combine} we show a combination of the $\mu>0$ and
$\mu<0$ cases, without the \gmtwo\ constraint.  We concatenated the
two chains, with the appropriate statistical weights.  Each chain's
posterior pdf was multiplied by its own Bayesian evidence
$\mathcal{Z}=\int\mathcal{L}(m)\pi(m)dm$ and divided by the sum of
both evidences to normalize the resulting pdf to unity,
%%%
\begin{equation}
  p(m|d)_{\textrm{tot}} = p(m|d)_{\mu<0} \times
  \frac{\mathcal{Z}_{\mu<0}}{\mathcal{Z}_{\mu<0}+\mathcal{Z}_{\mu>0}}
  + p(m|d)_{\mu>0}   \times
  \frac{\mathcal{Z}_{\mu>0}}{\mathcal{Z}_{\mu<0}+\mathcal{Z}_{\mu>0}}\,.
\label{mucombination} 
\end{equation}

%%%%%%%%%%%%%%%%%%%%%%%%%   F   I   G   U   R   E   %%%%%%%%%%%%%%%%%%%%%%%%%%%%
% 2 by 1 plot of  pdf sigsip vs mchi, nonLHC and with LHC
\begin{figure}[t]
\centering
\subfloat[
%lr The unknown quantities relevant to direct detection experiments
%constrained by the \alphaTexp\ and the non-LHC experiments.
]{% 
\label{fig:Comparison_sigsipmchi-a}%
\includegraphics[width=0.39\textwidth]{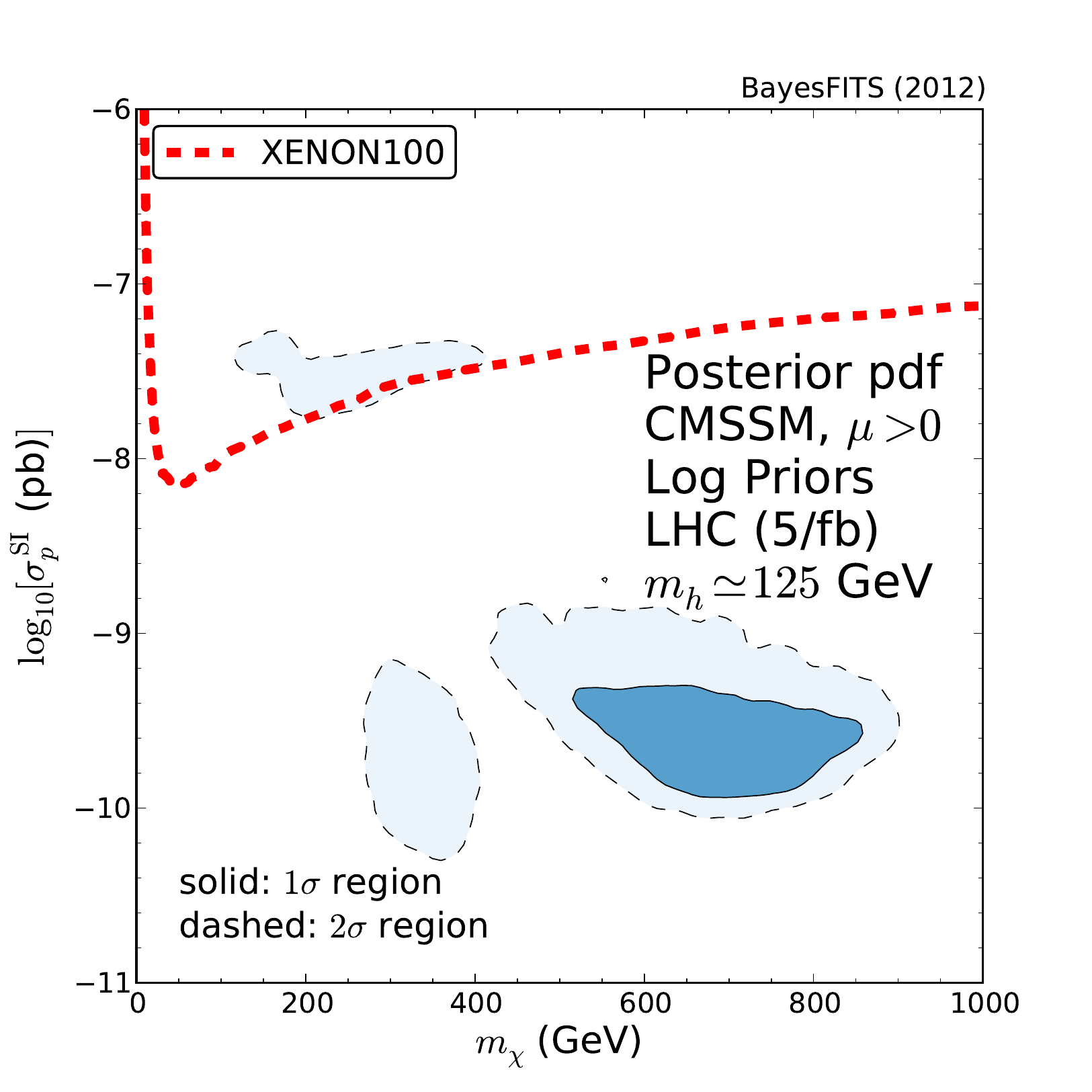}
}%
\hspace{1pt}%
\subfloat[
%lr The unknown quantities relevant to direct detection experiments
%constrained by \xenon, the \alphaTexp\ and the non-LHC experiments.
]{% 
\label{fig:Comparison_sigsipmchi-b}%
\includegraphics[width=0.39\textwidth]{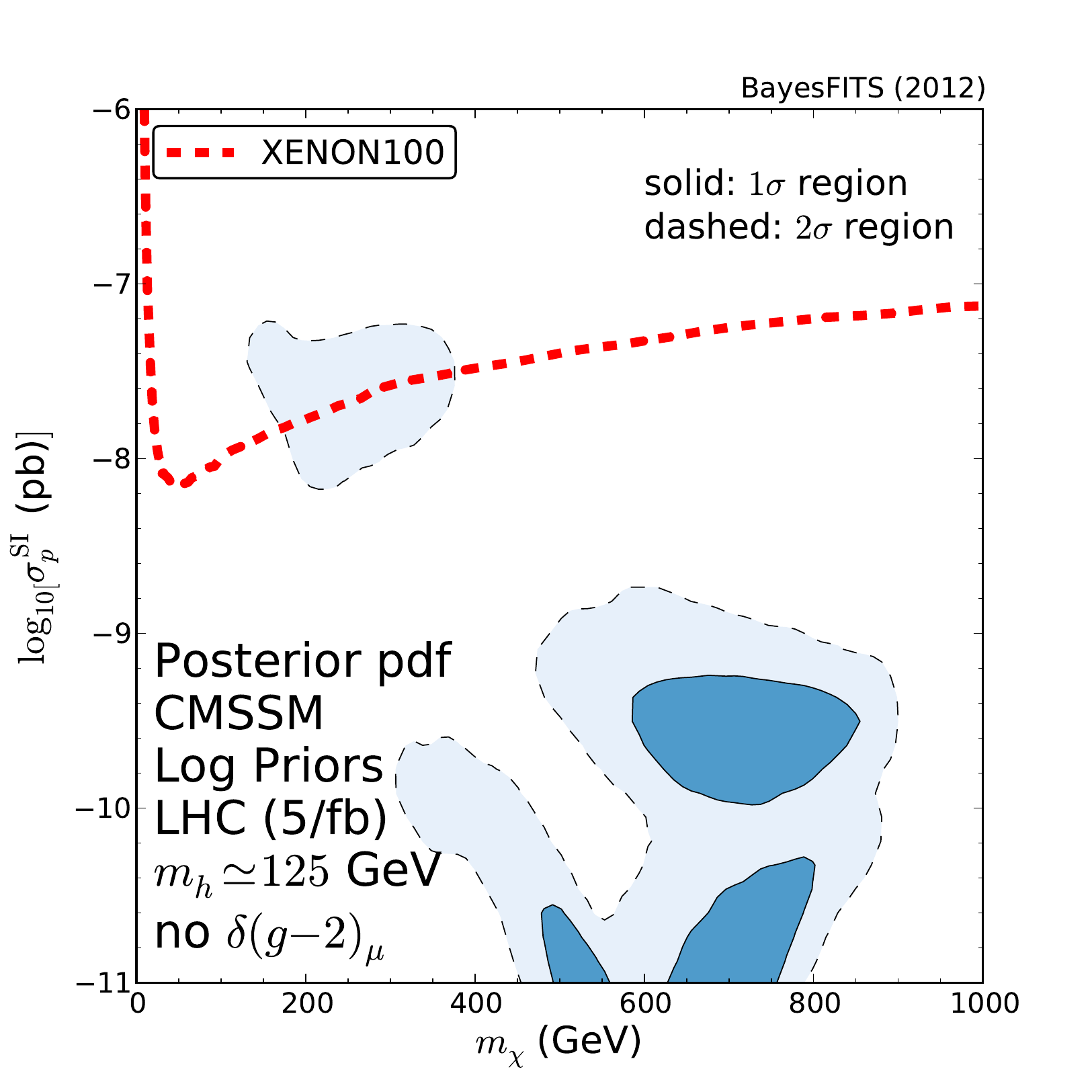}
}
\caption[]{Marginalized posterior pdf on the ($m_\neutone$, \sigsip)
  plane in the CMSSM constrained by the experiments listed in
  Table~\ref{tab:exp_constraints}, for the case
  \subref{fig:Comparison_sigsipmchi-a} with \gmtwo\ and positive
  $\mu$, and \subref{fig:Comparison_sigsipmchi-b} without \gmtwo\ and
  with a combination of the $\mu>0$ and $\mu<0$.  
}

\label{fig:cmssm_comparison_sigsipmchi}
\end{figure}
%%%%%%%%%%%%%%%%%%%%%%%%%%%%%%%%%%%%%%%%%%%%%%%%%%%%%%%%%%%%%%%%%%%%%%%%%%%%%%%%

In Fig.~\ref{fig:cmssm_comparison_sigsipmchi} we present the results
of our scan in the (\mchi, \sigsip) plane in the most popular case of
positive $\mu$ with the $\gmtwo$ constraint in the likelihood
\subref{fig:Comparison_sigsipmchi-a}, and in the case where we combine
both positive and negative $\mu$ scans done without the $\gmtwo$
constraint in the likelihood \subref{fig:Comparison_sigsipmchi-b}.
Differently from our previous studies of the
CMSSM\cite{Fowlie:2011mb,Roszkowski:2012uf}, we do not include the
XENON100\cite{Aprile:2011hi} limit in the likelihood function, due to
the large theoretical uncertainties which render the impact on CMSSM
parameters of the present experimental bounds from DM direct detection
considerably weaker than the limits obtained from the LHC.

In
Fig.~\ref{fig:cmssm_comparison_sigsipmchi}\subref{fig:Comparison_sigsipmchi-a},
the FP/HB region, which is just above the XENON100 90\%~CL upper bound
on \sigsip, has the potential to be ruled out with the sensitivity
planned for future XENON1T detector\cite{Aprile:2011hi}.  We checked
that the posterior distribution in the case without the \gmtwo\
constraint and $\mu>0$ is similar.  With respect to what was observed in
our previous studies\cite{Fowlie:2011mb,Roszkowski:2012uf}, we note
that the 68\% credible region corresponding to the \stauc\ region
($200\gev\lsim\mchi\lsim400\gev$) has been washed out. On the other hand, the
$A$-resonance region ($\mchi\gsim 400\gev$ and $\sigsip\lsim 10^{-9}\pb$) is not
likely to be further constrained by the new spin-independent cross
section measurements planned for the next year.

However, the $\mu$-combined case shows a very different shape for the
posterior, due to the total effective coupling being reduced by
negative $\mu$. Hence, the FP/HB region yields a slightly lower
$\sigsip$ than in the $\mu>0$
case. Figure~\ref{fig:cmssm_comparison_sigsipmchi}\subref{fig:Comparison_sigsipmchi-b}
shows that the FP/HB region still remains partially below the XENON100
bound. It can be tested with the future XENON1T sensitivity.

\section{\label{sec:statistics} Statistical Discussion}

We dedicate this section to some further statistical considerations. In
Sec.~\ref{sec:statistics}A we analyze in detail the individual
contributions to the minimum $\chi^2$ of our scans, and try to derive
some conclusions on the goodness of the global fit of the
CMSSM. Note that Bayesian scans are by definition \textit{not}
optimized for calculating the best-fit points to the highest accuracy,
because their results are dependent on the choice of priors, while the
best-fit point is entirely determined by the likelihood
function. Nevertheless, we think that the conclusions presented in
this section are general, as they are based on the properties of our
likelihood functions over a broad range of parameters. In
Sec.~\ref{sec:statistics}B we perform a Bayesian model comparison of
the model with $\mu>0$ and $\mu<0$, based on the relative evidence. We
find that both the frequentist and Bayesian approaches favor the case
of  $\mu<0$  and without the \gmtwo\ constraint.

%%%%%%%%%%%%%%%%%%%%%%%%%%%%%%%%%%%%%%%%%%%%%%%%%%%%%%%%%%%%%%%%%%%%%%%%%%%%%%%%
\subsection{\label{sec:bestfit}The $\chi^2$ and the best-fit point}
%%%%%%%%%%%%%%%%%%%%%%%%%%%%%%%%%%%%%%%%%%%%%%%%%%%%%%%%%%%%%%%%%%%%%%%%%%%%%%%%

%%%%%%%%%%%%%%%%%%%%%%%%%%%%   T   A   B   L   E   %%%%%%%%%%%%%%%%%%%%%%%%%%%%
% Table showing likelihoods for best-fit point from tan beta = 10 map,  with log priors
\begin{table}[t]
%%%%%%%%%%%%%%%%%%%%
%\footnotesize{
%%%%%%%%%%%%%%%%%%%%%%%%%%%%%%%%%%%%%%%%%%%%%%%%%%%%%%%%%%%%%%%%%%%%%%%%%%%%%%%%
    \begin{tabular}{|c|c|c|c|c|c|c|c|c|c|c|c|c|c|c|c|c|}
%lr \begin{tabular}{|l|l|l|l|l|l|l|l|l|l|l|l|l|l|l|l|l|}
\hline %\toprule
%%%%%%%%%%%%%%%%%%%%%%%%%%%%%%%%%%%%%%%%%%%%%%%%%%%%%%%%%%%%%%%%%%%%%%%%%%%%%%%%
 & Contribution to $\chi^2_{min}$ & \abundchi & \mh & \bxsgamma & \bsmumu & \sinsqeff & \mw & \deltagmtwomususy &  \butaunu & \delmbs & \razor &  Total\\ \hline %\midrule
%%%%%%%%%%%%%%%%%%%%%%%%%%%%%%%%%%%%%%%%%%%%%%%%%%%%%%%%%%%%%%%%%%%%%%%%%%%%%%%%
1 & with \gmtwo, $\mu>0$& 0.10 & 0.38 & 1.52 & 0.70 & 1.07 & 0.13 & 10.40 & 0.85 & 0.12 & 0.14 & 15.42\\ \hline
2 & with \gmtwo, $\mu<0$ & 0.06 & 0.70 & 0.00004 & 0 & 0.21 & 0.14 & 13.93 & 0.91 & 0.46 & 0.14 & 16.56\\ \hline
3 & w/o \gmtwo, $\mu>0$ & 0.15 & 0.74 & 1.37 & 0.08 &0.05 & 0.44 & - & 0.84 & 0.16 & 0.14 & 3.97\\ \hline
4 & w/o \gmtwo, $\mu<0$ & 0.15 & 0.33 & 0.12 & 0 & 0.31 & 0.06& - & 0.93 & 0.70 & 0.14 & 2.74\\ \hline
%%%%%%%%%%%%%%%%%%%%%%%%%%%%%%%%%%%%%%%%%%%%%%%%%%%%%%%%%%%%%%%%%%%%%%%%%%%%%%%%
%\bottomrule
\end{tabular}
%}
%%%%%%%%%%%%%%%%%%%%
%lr \input{./Tables/bestfit_lnlike}
\caption{Breakdown of all contributions to the $\chi^2$ of the
  best-fit points of our four different CMSSM likelihood scans.
}
\label{tab:bestfit_lnlike}
\end{table}
%%%%%%%%%%%%%%%%%%%%%%%%%%%%%%%%%%%%%%%%%%%%%%%%%%%%%%%%%%%%%%%%%%%%%%%%%%%%%%%%

%%%%%%%%%%%%%%%%%%%%%%%%%   F   I   G   U   R   E   %%%%%%%%%%%%%%%%%%%%%%%%%%%%
% 
%
%\begin{figure}%[\pos]
%\centering
%\label{fig:cmssm_massbars}% 3d scatter of mh in m0,m12 without mh125
%\includegraphics[width=0.45\textwidth]{CMSSM_masses.pdf}
%\caption[]{}
%\end{figure}

\begin{figure}[b]
\centering
\label{fig:chi2bar_bfpoint}% 3d scatter of mh in m0,m12 with mh125
\includegraphics[width=0.65\textwidth]{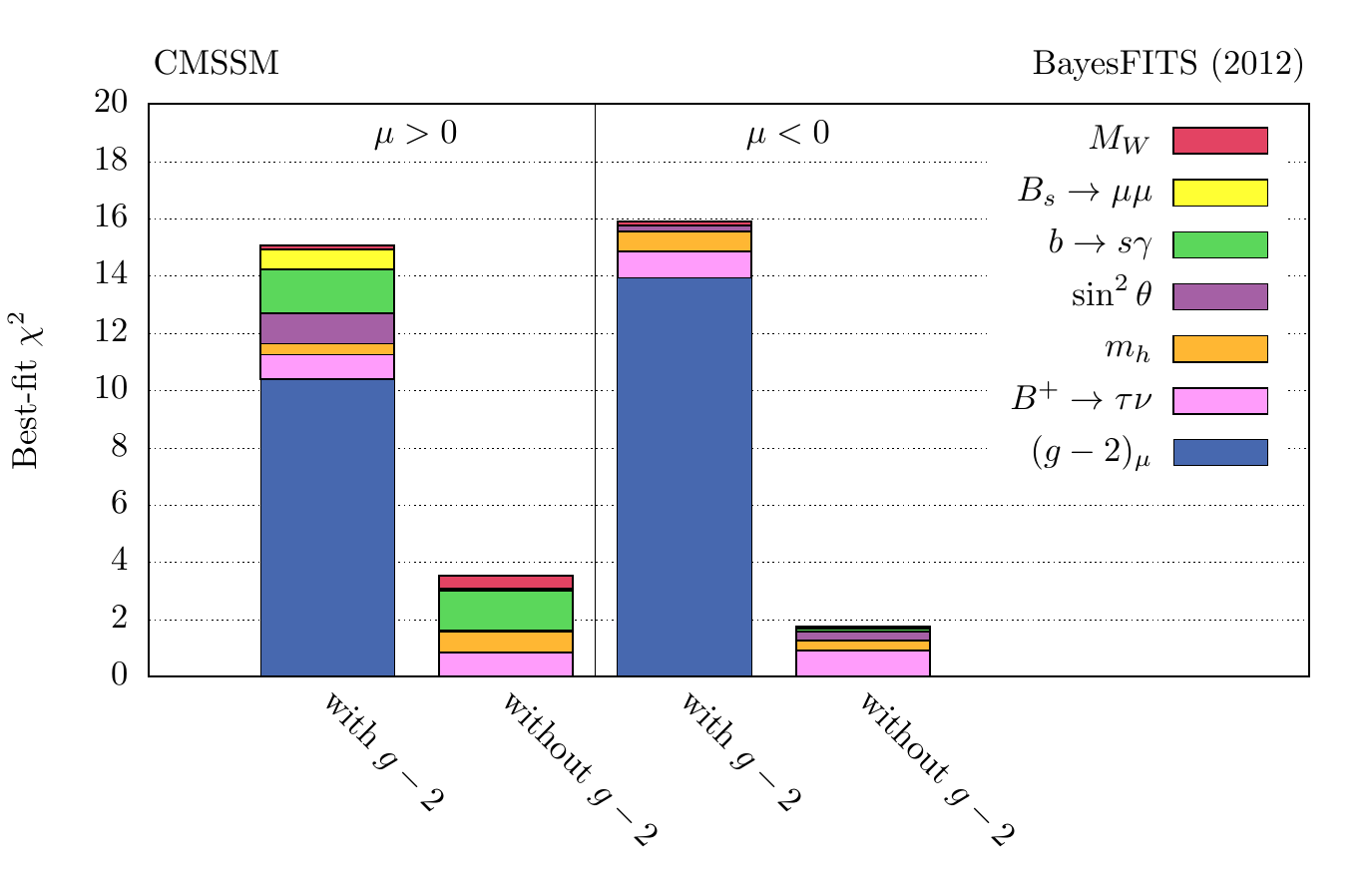}
\caption[]{A bar chart showing the breakdown of the main contributions to the $\chi^2$ of the
  best-fit points of our four different likelihood scans.
}
\label{fig:chi2bar_bfpoint}
\end{figure} 
%%%%%%%%%%%%%%%%%%%%%%%%%%%%%%%%%%%%%%%%%%%%%%%%%%%%%%%%%%%%%%%%%%%%%%%%%%%%%%%%

%%%%%%%%%%%%%%%%%%%%%%%%%%%%   T   A   B   L   E   %%%%%%%%%%%%%%%%%%%%%%%%%%%%
% Table showing parameters for best-fit points from various scans
\begin{table}[t]
%%%%%%%%%%%%%%%%%%%%%%%%%%%%%%%%%%%%%%%%%%%%%%%%%%%%%%%%%%%%%%%%%%%%%%%%%%%%%%%%
\begin{tabular}{|c|c|c|c|c|c|c|}
\hline %\toprule
%%%%%%%%%%%%%%%%%%%%%%%%%%%%%%%%%%%%%%%%%%%%%%%%%%%%%%%%%%%%%%%%%%%%%%%%%%%%%%%%
 			& \mzero & \mhalf & \azero & \tanb & \mhl & $\chi^2_{min}$ \\
\hline %\midrule
%%%%%%%%%%%%%%%%%%%%%%%%%%%%%%%%%%%%%%%%%%%%%%%%%%%%%%%%%%%%%%%%%%%%%%%%%%%%%%%%
with \gmtwo, $\mu>0$ & 945 & 1570 & 817 & 45.6 & 123.3 & 15.42 \\ \hline
with \gmtwo, $\mu<0$ & 2430 & 1480 & 1413 & 43.7 & 122.6 & 16.56 \\ \hline
w/o \gmtwo, $\mu>0$ & 3245 & 1808 & 1485 & 49.2 & 122.6 & 3.97 \\ \hline
w/o \gmtwo, $\mu<0$ & 2032 & 1425 & -393 & 43.1 & 123.4 & 2.74 \\ \hline
%%%%%%%%%%%%%%%%%%%%%%%%%%%%%%%%%%%%%%%%%%%%%%%%%%%%%%%%%%%%%%%%%%%%%%%%%%%%%%%%
\hline %\bottomrule
\end{tabular}
%%%%%%%%%%%%%%%%%%%%
\caption{
CMSSM parameters and Higgs masses for the
best-fit points of our four different likelihood scans. Masses and $A_0$ are in \gev. 
}
\label{tab:bestfit_params}
\end{table}
%%%%%%%%%%%%%%%%%%%%%%%%%%%%%%%%%%%%%%%%%%%%%%%%%%%%%%%%%%%%%%%

In Table~\ref{tab:bestfit_lnlike} we present the breakdown of the
individual constraint contributions to the total $\chi^2$ of our
best-fit points, for the scans performed in this analysis. (We define
the test statistic as $\chi^2=-2\ln\mathcal{L}$.) A bar chart showing
the main individual contributions to the minimum $\chi^2$ is given in
Fig.~\ref{fig:chi2bar_bfpoint}. In Table~\ref{tab:bestfit_params} we
present the best-fit points' CMSSM parameters and the corresponding
Higgs mass. As one could have expected, the largest contribution is
due to the \gmtwo\ constraint which is very poorly fitted
in the CMSSM after the low-mass region has been excluded by the
increasingly constraining LHC limits.

We refrain from calulating $p$-values for our best-fit points in this
paper, given the highly non-Gaussian nature of the distribution of the
uncertainties. Nonetheless, we point out that, given the number of
constraints we employ, $\chisqmin\simeq 15-16$ seem to
indicate that the present status of the global fit to all the
constraints, especially \gmtwo, is poor.

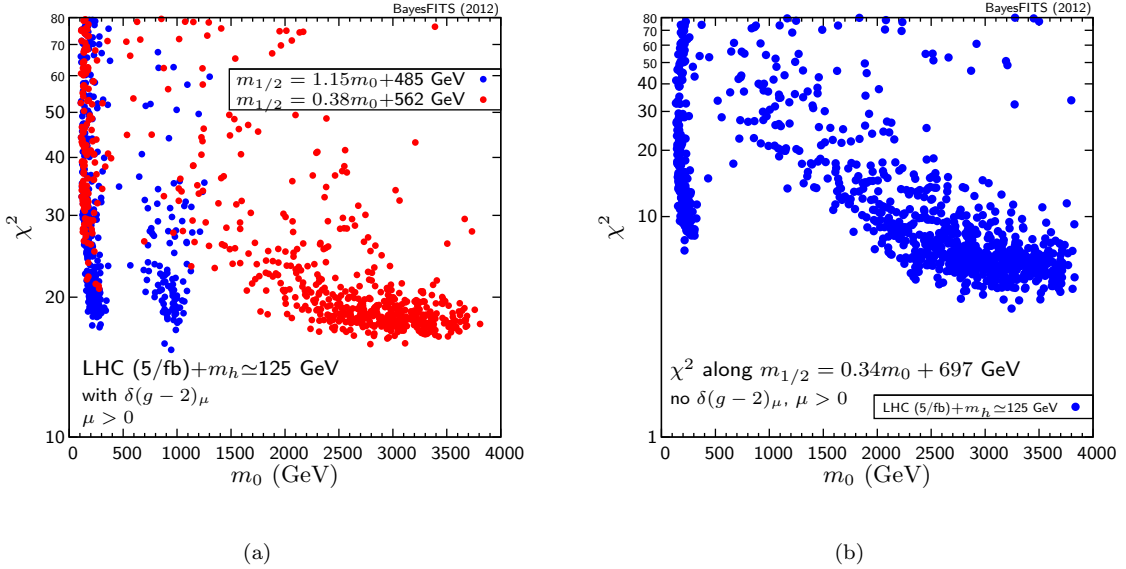
\begin{figure}[t]
\centering
%\hspace{-1.3cm}%
\subfloat[]{%
\label{fig:-a}%
  \setlength{\unitlength}{0.0500bp}%
  \begin{picture}(4464.00,4132.80)%
%    \gplgaddtomacro\gplbacktext{%
      \csname LTb\endcsname%
      \put(782,704){\makebox(0,0)[r]{\strut{}\scriptsize{\textsf{10}}}}%
      \put(782,1759){\makebox(0,0)[r]{\strut{}\scriptsize{\textsf{20}}}}%
      \put(782,2376){\makebox(0,0)[r]{\strut{}\scriptsize{\textsf{30}}}}%
      \put(782,2813){\makebox(0,0)[r]{\strut{}\scriptsize{\textsf{40}}}}%
      \put(782,3153){\makebox(0,0)[r]{\strut{}\scriptsize{\textsf{50}}}}%
      \put(782,3430){\makebox(0,0)[r]{\strut{}\tiny{\textsf{60}}}}%
      \put(782,3665){\makebox(0,0)[r]{\strut{}\tiny{\textsf{70}}}}%
      \put(782,3868){\makebox(0,0)[r]{\strut{}\tiny{\textsf{80}}}}%
      \put(814,584){\makebox(0,0){\strut{} \scriptsize{\textsf{0}}}}%
      \put(1221,584){\makebox(0,0){\strut{} \scriptsize{\textsf{500}}}}%
      \put(1627,584){\makebox(0,0){\strut{} \scriptsize{\textsf{1000}}}}%
      \put(2034,584){\makebox(0,0){\strut{} \scriptsize{\textsf{1500}}}}%
      \put(2441,584){\makebox(0,0){\strut{} \scriptsize{\textsf{2000}}}}%
      \put(2847,584){\makebox(0,0){\strut{} \scriptsize{\textsf{2500}}}}%
      \put(3254,584){\makebox(0,0){\strut{} \scriptsize{\textsf{3000}}}}%
      \put(3660,584){\makebox(0,0){\strut{} \scriptsize{\textsf{3500}}}}%
      \put(4067,584){\makebox(0,0){\strut{} \scriptsize{\textsf{4000}}}}%
      \put(450,2286){\rotatebox{-270}{\makebox(0,0){\strut{}$\chi^2$}}}%
      \put(2440,404){\makebox(0,0){\strut{}$m_0$ (GeV)}}%
      \put(3254,3942){\makebox(0,0)[l]{\strut{}\tiny{\textsf{BayesFITS (2012)}}}}%
      \put(912,1216){\makebox(0,0)[l]{\strut{}\footnotesize{\textsf{LHC~(5/fb)$+m_h$$\simeq$125~\textsf{GeV}}}}}%
      \put(912,1019){\makebox(0,0)[l]{\strut{}\scriptsize{\textsf{with $\delta(g-2)_\mu$}}}}%
      \put(912,849){\makebox(0,0)[l]{\strut{}\scriptsize{\textsf{$\mu>0$}}}}%
%    }%
%    \gplgaddtomacro\gplfronttext{%
%      \csname LTb\endcsname%
      \put(3803,3403){\makebox(0,0)[r]{\strut{}\scriptsize{\textsf{$m_{1/2}=1.15m_0$+485~\textsf{GeV}}}}}%
%      \csname LTb\endcsname%
      \put(3803,3249){\makebox(0,0)[r]{\strut{}\scriptsize{\textsf{$m_{1/2}=0.38m_0$+562~\textsf{GeV}}}}}%
%    }%
%    \gplbacktext
    \put(0,0){\includegraphics{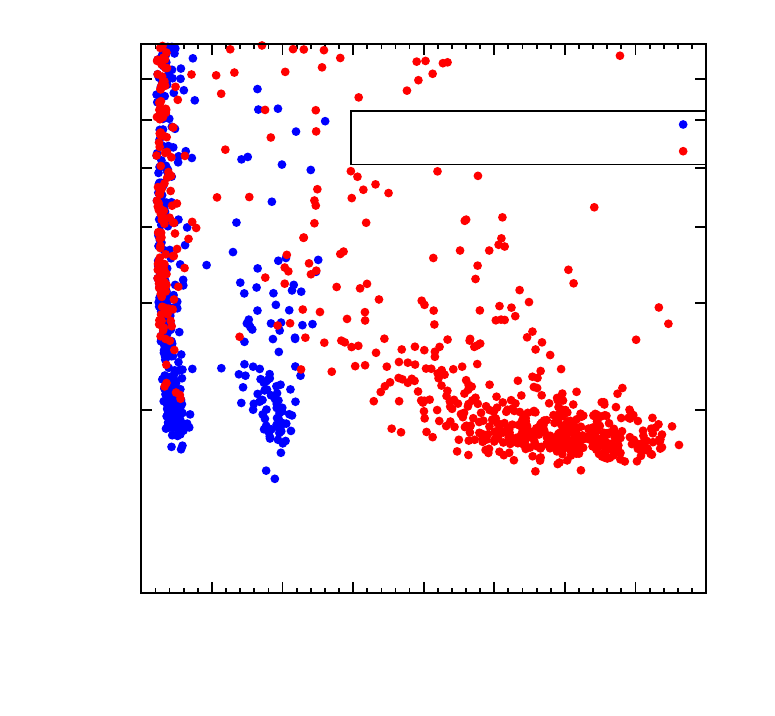}}%
%    \gplfronttext
  \end{picture}%

}%
%\hspace{-0.5cm}%
\subfloat[]{%
\label{fig:-b}%
  \setlength{\unitlength}{0.0500bp}%
  \begin{picture}(4464.00,4132.80)%
%    \gplgaddtomacro\gplbacktext{%
      \csname LTb\endcsname%
      \put(782,704){\makebox(0,0)[r]{\strut{}\scriptsize{\textsf{1}}}}%
      \put(782,2367){\makebox(0,0)[r]{\strut{}\scriptsize{\textsf{10}}}}%
      \put(782,2867){\makebox(0,0)[r]{\strut{}\scriptsize{\textsf{20}}}}%
      \put(782,3160){\makebox(0,0)[r]{\strut{}\scriptsize{\textsf{30}}}}%
      \put(782,3368){\makebox(0,0)[r]{\strut{}\scriptsize{\textsf{40}}}}%
      \put(782,3529){\makebox(0,0)[r]{\strut{}\tiny{\textsf{50}}}}%
      \put(782,3660){\makebox(0,0)[r]{\strut{}\tiny{\textsf{60}}}}%
      \put(782,3772){\makebox(0,0)[r]{\strut{}\tiny{\textsf{70}}}}%
      \put(782,3868){\makebox(0,0)[r]{\strut{}\tiny{\textsf{80}}}}%
      \put(814,584){\makebox(0,0){\strut{} \scriptsize{\textsf{0}}}}%
      \put(1221,584){\makebox(0,0){\strut{} \scriptsize{\textsf{500}}}}%
      \put(1627,584){\makebox(0,0){\strut{} \scriptsize{\textsf{1000}}}}%
      \put(2034,584){\makebox(0,0){\strut{} \scriptsize{\textsf{1500}}}}%
      \put(2441,584){\makebox(0,0){\strut{} \scriptsize{\textsf{2000}}}}%
      \put(2847,584){\makebox(0,0){\strut{} \scriptsize{\textsf{2500}}}}%
      \put(3254,584){\makebox(0,0){\strut{} \scriptsize{\textsf{3000}}}}%
      \put(3660,584){\makebox(0,0){\strut{} \scriptsize{\textsf{3500}}}}%
      \put(4067,584){\makebox(0,0){\strut{} \scriptsize{\textsf{4000}}}}%
      \put(450,2286){\rotatebox{-270}{\makebox(0,0){\strut{}$\chi^2$}}}%
      \put(2440,404){\makebox(0,0){\strut{}$m_0$ (GeV)}}%
      \put(3254,3949){\makebox(0,0)[l]{\strut{}\tiny{\textsf{BayesFITS (2012)}}}}%
      \put(879,1204){\makebox(0,0)[l]{\strut{}\footnotesize{\textsf{$\chi^2$ along $m_{1/2}=0.34m_0+697$ GeV}}}}%
      \put(879,997){\makebox(0,0)[l]{\strut{}\scriptsize{\textsf{no $\delta(g-2)_\mu$, $\mu>0$}}}}%
%   }%
%    \gplgaddtomacro\gplfronttext{%
%      \csname LTb\endcsname%
      \put(3803,932){\makebox(0,0)[r]{\strut{}\tiny{\textsf{LHC (5/fb)$+m_h$$\simeq$125 \textsf{GeV}}}}}%
%    }%
%    \gplbacktext
    \put(0,0){\includegraphics{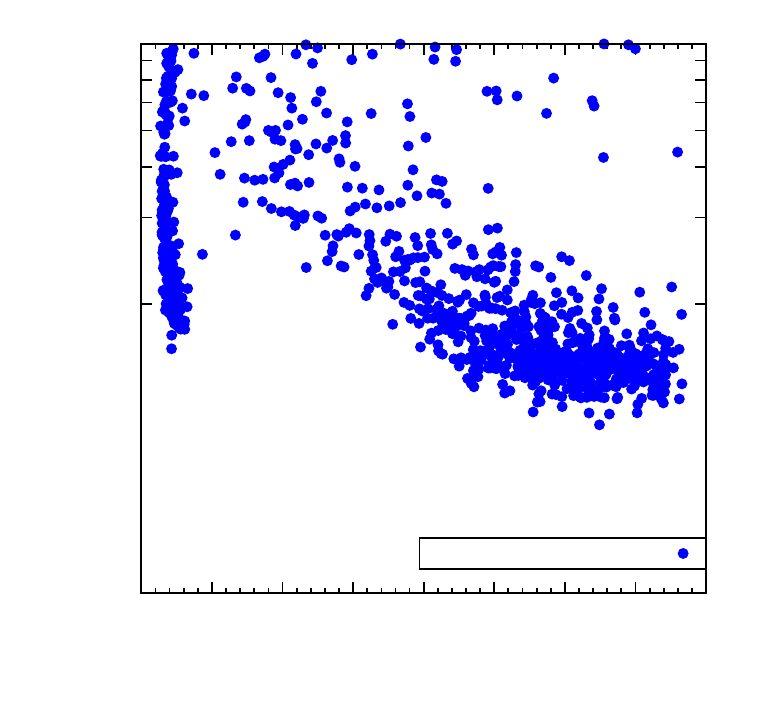}}%
%    \gplfronttext
  \end{picture}%
}%
\caption[]{
\subref{fig:-a} Scatter plot distribution of the total $\chi^2$ vs \mzero\ for the points along two narrow strips of the (\mzero, \mhalf) plane parametrized by 
(in blue) $\mhalf=1.15\mzero+485\gev$ and (in red) $\mhalf=0.38\mzero+562\gev$, with $\mu>0$, \gmtwo. \subref{fig:-b} $\mu>0$, no \gmtwo. \mzero, \mhalf\ parametrized by $\mhalf=0.34\mzero+697\gev$.}
\label{fig:scatter_m0_chi2}
\end{figure}

One important issue should be emphasized when trying to find the
position of the best-fit point in a global fit to the CMSSM. 
At present none of the experimental constraints, including
positive measurements of the Higgs mass and the DM relic density, have
a strongly constraining effect on the CMSSM parameters. As a result,
fairly similar values of \chisq\ can be achieved over large ranges of
the model's parameters. To illustrate the point we show in
Fig.~\ref{fig:scatter_m0_chi2}\subref{fig:-a} the combined
distribution of the total $\chi^2$, as a function of $m_0$, for the
points that lie along two narrow strips of the (\mzero, \mhalf) plane,
with $\mu>0$ and \deltagmtwomususy\ taken into account. The first
strip cuts through the \stauc\ region to reach the best-fit point
in the $A$-funnel region. It is parametrized by
$\mhalf=1.15\mzero+485\gev$, and the points lying along the strip are
indicated in blue. The second strip also crosses the \stauc\ region
with a different inclination, so to reach the $A$-funnel region at
large \mzero. It is parametrized by $\mhalf=0.38\mzero+562\gev$, and
the relative points are shown in red. Both lines cross the $1\sigma$
credibility intervals in the \stauc\ region and the $A$-resonance
region. The $\chi^2$ distribution shows a plateau that extends across
the $A$-resonance region, with approximately the same $\chi^2$ values
as those obtained in the \stauc\ region.  Thus $\chi^2$ analyses can
be very sensitive to minor changes in the adopted methodology
(scanning procedure, modeling of the likelihood for different
observables, etc) and, as a consequence, the position of the best-fit
point can also undergo dramatic changes.  This should be kept in
mind when comparing results of different groups.

%\begin{figure}[b]
%\centering
%%\label{fig:plateau}% 3d scatter of mh in m0,m12 with mh125
%\includegraphics[width=0.65\textwidth]{BF_regions.pdf}
%%\caption[]{The plateau.
%}
%\label{fig:plateau}
%\end{figure} 
%

Nevertheless, some general conclusions can be drawn by briefly
analyzing the main individual contributions to the best-fit point. For $\mu>0$, a
tension between the Higgs mass at 125\gev\ and \gmtwo\ is expected, as
is predicted theoretically by the fact that $\msusy$ should be large enough
to obtain the correct mass of the Higgs, but small enough to fit the \gmtwo\
constraint. So, naively one would expect that if the latter were
released, \mhl\ would show a better fit in the $A$-funnel region. This
is not the case for some parts of the $A$-funnel region,
particularly the one where the best-fit point is located. As
Table~\ref{tab:bestfit_lnlike} shows, even in the presence of the
\gmtwo\ constraint the contribution of the Higgs to the fit is
relatively good.

As we discussed in Sec.~\ref{sec:results}, lifting the \gmtwo\
constraint allows a better fit to the $b$-physics observables, which
can be seen, particularly in the case of \brbsmumu, by comparing the
first and third rows in Table~\ref{tab:bestfit_lnlike}.  However, even when we keep \gmtwo\ in place, an even better fit to $b$-physics can be
obtained for $\mu<0$. Since the SUSY contribution to
\deltagmtwomususy\ is proportional to $\mu$, it becomes negative when
$\mu<0$, and in that case high supersymmetric masses are required in
order to suppress it. As we mentioned in Sec.~\ref{sec:results}, for
$\mu<0$ heavy SUSY masses are also required to suppress the
chargino-stop contribution to \brbxsgamma. Thus the \bsgamma\ and
\gmtwo\ constraints add a same-sign ``pull" to the minimum
$\chi^2$. Moreover, for the negative $\mu$ case, the best-fit point
shows an excellent fit to \brbsmumu.

%%%%%%%%%%%%%%%%%%%%%%%%%   F   I   G   U   R   E   %%%%%%%%%%%%%%%%%%%%%%%%%%%%
% 2 by 1: left: plot of 1d pdf of mhl,
% right: chi2 vs mhl
%%%%%%%%%%%%%%%%%%%%%%%%%%%%%%%%%%%%%%%%%%%%%%%%%%%%%%%%%%%%%%%%%%%%%%%%%%%%%%%%

%Let us now spend a few more words on the issue of the accurate
%position of the best-fit point, which we introduced at the beginning
%of this section. To exemplify the problem, we performed an additional
%scan (with all observable constraints unchanged \kk{Scan is for thee
%  bounds case}), where, *** as an approximation to the razor limit, we
%used a somewhat more crude likelihood map . \lr{can you give a  bit more
%  details here?} In this case the best-fit point for the scan was
%found in the \stauc\ region. This not only outlines the problems with
%proper determination of the best-fit point, but also emphasizes the
%necessity for a correct treatment of the exclusion limits coming from
%the LHC SUSY searches. When a hard-cut or a crude smear bound is
%applied, the \razor\ contribution to the $\chi^2$ for a given point is
%either large (when the point lies below the 95\%~CL exclusion bound,
%so that it is rejected by the scan) or very close to zero (in the
%allowed region). In our likelihood map approach we move one step
%further and we model a slope decrease in the likelihood taking into
%account the appropriate statistical and systematic errors. Given the
%$\chi^2$ plateau we encounter over most of the parameter space this
%seems to be the only viable way of obtaining meaningful information.

Figure~\ref{fig:scatter_m0_chi2}\subref{fig:-b} shows the $\chi^2$
distribution when the \gmtwo\ constraint is lifted. When neglecting
the observable that has been most difficult to fit, one finds a
more informative distribution of the $\chi^2$. The scan clearly favors the regions at
large mass scales, as described in the previous
sections. Unfortunately, those regions will be  much more difficult to
probe at the LHC and in DM searches.

In conclusion, one can notice a rather striking improvement of the
global fits when the \gmtwo\ constraint is lifted, exemplified by the
drop of more than ten units of $\chi^2$ for one less constraint.

%%%%%%%%%%%%%%%%%%%%%%%%%%%%%%%%%%%%%%%%%%%%%%%%%%%%%%%%%%%%%%%%%%%%%%%%%%%%%%%%%%%%%%%%%%%%%%%
\subsection{\label{sec:evidence} Comparison between $\mu>0$ and $\mu<0$ without $\gmtwo$}
%%%%%%%%%%%%%%%%%%%%%%%%%%%%%%%%%%%%%%%%%%%%%%%%%%%%%%%%%%%%%%%%%%%%%%%%%%%%%%%%%%%%%%%%%%%%%%%

In this subsection, we compare the Bayesian evidences given in our
scans to see if either the $\mu>0$ or $\mu<0$ case is favored by the
experiments, according to Bayesian statistics.

In Table~\ref{tab:MuComparison} we show the log-evidence for our four
scans. Two of them include the $\gmtwo$ constraint. In this case we
expect $\mu>0$ to be slightly favored, in agreement with what we
found for the $\chi^2$ analysis. The other two scans do not include
$\gmtwo$.

The Bayesian evidence favors (it is larger for) $\mu>0$ when we
include $\gmtwo$, but favors $\mu<0$ when we omit the
constraint. Without $\gmtwo$, the Bayes factor (or evidence ratio)
yields $2.5:1$ in favor of the $\mu<0$ case. This reads ``barely worth
mentioning'' (1:1 to 3:1) on Jeffrey's scale\cite{Jeffery:1998}, which
measures the so-called ``strength of evidence''. With $\gmtwo$
included, the Bayes factor yields $2.9:1$ in favor of the $\mu>0$
case, which also reads ``barely worth mentioning'' on the Jeffrey's
scale.

We conclude that both the minimum $\chi^2$ and Bayesian approaches
indicate that, when the \gmtwo\ constraint is lifted, the fit for the
CMSSM is better for negative $\mu$.

%%%%%%%%%%%%%%%%%%%%%%%%%%%%   T   A   B   L   E   %%%%%%%%%%%%%%%%%%%%%%%%%%%%%
\begin{table}[h]
\begin{tabular}{|c||c|c||c|c|}
\hline%------------------------------------------------------------------------%
 & with \gmtwo, $\mu > 0$ & with \gmtwo, $\mu < 0$ & w/o \gmtwo, $\mu >
0$ & w/o \gmtwo, $\mu < 0$\\
\hline%------------------------------------------------------------------------%
$\ln\mathcal{Z}$       & -18.8 & -19.8 & -13.5 & -12.6 \\
\hline%------------------------------------------------------------------------%
\end{tabular}

\caption{Bayesian evidences found for $\mu < 0$ and $\mu >0$ with and
without \gmtwo.}
\label{tab:MuComparison}
\end{table}
%%%%%%%%%%%%%%%%%%%%%%%%%%%%%%%%%%%%%%%%%%%%%%%%%%%%%%%%%%%%%%%%%%%%%%%%%%%%%%%%

%%%%%%%%%%%%%%%%%%%%%%%%%%%%%%%%%%%%%%%%%%%%%%%%%%%%%%%%%%%%%%%%%%%%%%%%%%%%%%%%
\section{\label{sec:summary}Summary and Conclusions}
%%%%%%%%%%%%%%%%%%%%%%%%%%%%%%%%%%%%%%%%%%%%%%%%%%%%%%%%%%%%%%%%%%%%%%%%%%%%%%%%
In this paper we have performed an updated global statistical analysis
of the CMSSM. In terms of new experimental inputs that we incorporated into
the likelihood function in an approximate but accurate way, new
stringent limits from the \cms\ razor analysis of 4.4/fb of data on the
mass parameters \mzero\ and \mhalf, as well as the new limit from LHCb
on \brbsmumu. We also considered the impact of the SM-like light Higgs 
with mass being close to 125\gev.

A combination of these new inputs with other usual constraints, most
notably from $b$-physics, electroweak observables and dark matter relic
density, as well as from \deltagmtwomususy, generally pushes the favored
ranges of posterior probability beyond the 1\tev\ scale for \mhalf\
and above $\sim0.8\tev$ for \mzero, into the \ha-resonance region
where $\mha\sim 2\mchi$. As for the other two CMSSM
parameters: large \tanb\ remains favored, with $\tanb\sim50$, while
\azero\ remains poorly constrained and can take both signs.

With \deltagmtwomususy\ included in the likelihood, the overall fit in
terms of \chisqmin, for $\mu>0$, remains poor (compare
Table~\ref{tab:bestfit_lnlike}; see also \eg,\cite{Fowlie:2011mb}),
invariably primarily due to the high mass scales of the CMSSM causing
SUSY to generate only about a tenth of a needed contribution to the
variable. This has prompted us to consider the case of negative $\mu$,
where we found that \chisqmin\ is not significantly worse.

On the other hand, when we relaxed the \gmtwo\ constraint (since the
CMSSM fails to satisfy it anyway), overall we found a much better fit,
with $\mu<0$ being actually somewhat favored (again compare
Table~\ref{tab:bestfit_lnlike}). In particular, \brbxsgamma\ is now
reproduced much better, as well as \brbsmumu. This calls for a new
serious look at the phenomenology for negative $\mu$.

One concrete observable of interest that is strongly affected by the
sign of $\mu$ is the spin-independent cross section on DM neutralino
$\sigsip$. While for both signs of $\mu$ its high-probability ranges
have now dropped at least an order of magnitude below the XENON100
limit, for negative $\mu$ it can become even much lower (compare
Fig.~\ref{fig:cmssm_comparison_sigsipmchi}).

The light Higgs of about 125\gev\ remains a challenge for the CMSSM.
On the other hand, it is true that physical values of \mhl\ within a
\gev\ or so from 125\gev\ can only be achieved at the expense of poor
$\chi^2$ (compare Fig.~\ref{fig:cmssm_mhchisq}\subref{fig:fig-b}) and
also for negative \azero.

Finding a stable location of the best-fit point in the CMSSM parameter space
is a real challenge because of an extended ``plateau'' of comparable,
low values of $\chi^2$, which we have pointed out for $\mu>0$ and the
\gmtwo\ constraint included. (Compare also Ref.\cite{Fowlie:2011mb}.)

In contrast, high posterior probability regions remain relatively
robust, but unfortunately now favoring superpartner mass ranges which
will be even more difficult to test at the LHC than before, and
similarly for DM searches. Thus the CMSSM is now favoring new
territories whose experimental exploration may be a real challenge for
the next few years.

%%%%%%%%%%%%%%%%%%%%%%%%%%%%%%%%%%%%%%%%%%%%%%%%%%%%%%%%%%%%%%%%%%%%%%%%%%%%%%%%%
%\section{\label{sec:summary}Summary}
%%%%%%%%%%%%%%%%%%%%%%%%%%%%%%%%%%%%%%%%%%%%%%%%%%%%%%%%%%%%%%%%%%%%%%%%%%%%%%%%%
%
%In this paper we performed an updated global Bayesian analysis of the CMSSM. With respect to our previous analyses, we implemented two novel features in the likelihood function:
%
%$\bullet$ We approximated the CMS razor hadronic search with 4.4/fb. By taking into account signal production, detector efficiencies and background uncertainties, we created a likelihood function in the (\mzero,\mhalf) plane which reproduces exactly the razor 95\%~CL exclusion bound.
%
%$\bullet$ We constructed likelihood functions for the Higgs mass in two ways: with improved ATLAS and CMS limits on the SM-like Higgs boson mass, and assuming a possible Higgs signal at 125 \gev.
%
%We performed global scans assuming both signs of the CMSSM parameter $\mu$. We also investigated the effects of lifting the \gmtwo\ experiment
%al contraint, 
    
\bigskip

\textbf{Note added:} On July 4th, 2012, the discovery at $4.9\sigma$ by CMS and at $5.0\sigma$ by ATLAS\cite{HiggsFiveSig}
of a boson consistent with the SM Higgs, with mass near 125\gev, was announced. Particularly, the mass claimed by CMS, $\mhl=125.3\pm0.6\gev$, is very close in central value and experimental error to the signal case considered in this paper. Subsequent to the announcement, we post-processed our chains with a likelihood function modified to incorporate the updated result. We found no changes in the posterior distribution and location of the best-fit point for the putative signal case presented here.  
    
\bigskip
%%%%%%%%%%%%%%%%%%%%%%%%%%%%%%%%%%%%%%%%%%%%%%%%%%%%%%%%%%%%%%%%%%%%%%%%%%%%%%%%
\begin{acknowledgments}

We would like to thank Maurizio Pierini, Christopher Rogan and Maria Spiropulu for valuable discussions and inputs. 
E.M.S. would like to thank Azar Mustafayev for discussions on the impact of the \gmtwo\ constraint.

  This work has been funded in part by the Welcome Programme
  of the Foundation for Polish Science. A.J.F.~is funded by the Science Technology and Facilities
  Council.
  K.K. is also supported in part by the EU and MSHE Grant No. POIG.02.03.00-00-013/09.
  L.R. is also supported in part by the Polish National Science Centre Grant No. N202 167440, an STFC
  consortium grant of Lancaster, Manchester and Sheffield Universities
  and by the EC 6th Framework Programme MRTN-CT-2006-035505. 
\end{acknowledgments}
%%%%%%%%%%%%%%%%%%%%%%%%%%%%%%%%%%%%%%%%%%%%%%%%%%%%%%%%%%%%%%%%%%%%%%%%%%%%%%%%

%%%%%%%%%%%%%%%%%%% down to here dotad %%%%%%%%%%%%%%%%%%%%%!!!!!!
\bibliography{myref}
\end{document}